\documentclass[prd,amsmath,amssymb,superscriptaddress,preprintnumbers,twocolumn,nofootinbib,10pt]{revtex4-1}%superscriptaddress,showpacs,

\pdfoutput=1

\usepackage{graphicx}
\usepackage{dcolumn}
\usepackage{bm}
\usepackage{amssymb}
\usepackage{latexsym}
\usepackage{booktabs}
\usepackage{amsmath}
\usepackage{multirow}
\usepackage{url}

\usepackage{float}
\usepackage[colorlinks=true, linkcolor=red, citecolor=blue]{hyperref}

\usepackage[normalem]{ulem}
\usepackage{color}
\usepackage{array}
\usepackage{enumerate}

\begin{document}

\title{Prospects for searching for sterile neutrinos in dynamical dark energy cosmologies using joint observations of gravitational waves and $\gamma$-ray bursts}

\author{Lu Feng}\thanks{These authors contributed equally to this paper.}
\affiliation{College of Physical Science and Technology, Shenyang Normal University, Shenyang 110034, China}
\affiliation{Key Laboratory of Cosmology and Astrophysics (Liaoning) \& College of Sciences, Northeastern University, Shenyang 110819, China}
\author{Tao Han}\thanks{These authors contributed equally to this paper.}%\footnote{Lu Feng and Tao Han contributed equally to this work}
\affiliation{Key Laboratory of Cosmology and Astrophysics (Liaoning) \& College of Sciences, Northeastern University, Shenyang 110819, China}
\author{Jing-Fei Zhang}
\affiliation{Key Laboratory of Cosmology and Astrophysics (Liaoning) \& College of Sciences, Northeastern University, Shenyang 110819, China}
\author{Xin Zhang}\thanks{Corresponding author}
\email{zhangxin@mail.neu.edu.cn}
\affiliation{Key Laboratory of Cosmology and Astrophysics (Liaoning) \& College of Sciences, Northeastern University, Shenyang 110819, China}
\affiliation{Key Laboratory of Data Analytics and Optimization for Smart Industry (Ministry of Education), Northeastern University, Shenyang 110819, China}
\affiliation{National Frontiers Science Center for Industrial Intelligence and Systems Optimization, Northeastern University, Shenyang 110819, China}
%\affiliation{Key Laboratory of Data Analytics and Optimization
%for Smart Industry, (Northeastern University), Ministry of Education, Shenyang 110819, China}
%\affiliation{Center for High Energy Physics, Peking University, Beijing 100080, China}

\begin{abstract}

In the era of third-generation (3G) gravitational-wave (GW) detectors, GW standard siren observations from binary neutron star mergers provide a powerful tool for probing the expansion history of the universe. Since sterile neutrinos can influence cosmic evolution by modifying the radiation content and suppressing structure formation, GW standard sirens offer promising prospects for constraining sterile neutrino properties within a cosmological framework.
Building on this, we investigate the prospects for detecting sterile neutrinos in dynamical dark energy (DE) models using joint observations from 3G GW detectors and a future short gamma-ray burst detector, such as a THESEUS-like telescope. We consider three DE models: the $w$CDM, holographic DE (HDE), and Chevallier–Polarski–Linder (CPL) models. Our results show that the properties of DE can influence the constraints on sterile neutrino parameters. Moreover, the inclusion of GW data significantly improves constraints on both sterile neutrino parameters and other cosmological parameters across all three models, compared to the current limits derived from CMB+BAO+SN (CBS) observations. 
When GW data are included into the CBS dataset, a preference for $\Delta N_{\rm eff} > 0$ emerges at approximately the $1\sigma$ level in the $w$CDM and CPL models, and reaches about $3\sigma$ in the HDE model. Moreover, the upper limits on $m_{\nu,{\rm sterile}}^{\rm eff}$ are reduced by approximately 13\%, 75\%, and 3\% in the $w$CDM, HDE, and CPL models, respectively.

\end{abstract}

\maketitle

\section{Introduction}
On 17 August 2017, the signal of a gravitational wave (GW) produced by the merger of a binary neutron star system (BNS) was detected for the first time \cite{LIGOScientific:2017vwq}, and the electromagnetic (EM) signals generated by the same transient source were also observed subsequently, marking the beginning of the era of GW multi-messenger astronomy. GWs can act as ``standard sirens'' \cite{Schutz:1986gp,Holz:2005df} because the GW waveform encodes the absolute luminosity distance to the source. If the redshift of the EM counterpart of the GW source can be determined, a distance-redshift relation can be established, which can be used to probe the expansion history of the universe (see, e.g., Refs. \cite{Dalal:2006qt,Nissanke:2009kt,Cutler:2009qv,Camera:2013xfa,Cai:2016sby,Cai:2017cbj,Cai:2017aea,Chen:2017rfc,Wang:2018lun,Vitale:2018wlg,Belgacem:2019tbw,DAgostino:2019hvh,Howlett:2019mdh,Zhao:2019gyk,Borhanian:2020vyr,Qi:2021iic,Yu:2021nvx,Zhu:2021bpp,Zheng:2022gfi,Ezquiaga:2022zkx,Jin:2022tdf,LISACosmologyWorkingGroup:2022jok,Hou:2022rvk,Song:2022siz,Zhu:2023jti,Muttoni:2023prw,Branchesi:2023mws,Chen:2023dgw,Han:2023exn,Li:2023gtu,Yu:2023ico,Dong:2024bvw,Feng:2024lzh,Zheng:2024mbo,Xiao:2024nmi,Feng:2024mfx,Han:2024sxm,Han:2025fii,Jin:2025dvf} for related discussions). 

In the future, third-generation (3G) ground-based GW detectors, such as the Einstein Telescope (ET) \cite{ET-web,Punturo:2010zz} and the Cosmic Explorer (CE) \cite{CE-web,Evans:2016mbw}, are expected to detect a large number of standard sirens from BNS merger events. These observations can be used to constrain cosmological parameters with high precision, making GW standard sirens a powerful new cosmological probe. In particular, GW standard sirens can help break the parameter degeneracies generated by the current EM cosmological observations, thereby improving constraints on both active and sterile neutrino parameters; see, e.g., Refs. \cite{Wang:2018lun,Jin:2022tdf,Feng:2024lzh,Feng:2024mfx}.

The phenomenon of neutrino oscillation observed in solar and atmospheric neutrino experiments has shown that neutrinos are massive and that there is significant mixing among the three active neutrino flavors (see Ref. \cite{Lesgourgues:2006nd} for a review). 
However, anomalies observed in short-baseline neutrino oscillation experiments have hinted at the existence of additional light neutrino species \cite{LSND:2001aii,Mention:2011rk,Conrad:2012qt,Giunti:2012tn,Kopp:2013vaa,Giunti:2013aea,Gariazzo:2013gua,MiniBooNE:2018esg}, which, if they exist, would be sterile neutrinos.
Explaining these anomalous results appears to require the existence of at least one sterile neutrino with mass at the eV scale.
{However, this result is not favored by several other neutrino experiments, such as the Daya Bay \cite{DayaBay:2016lkk,DayaBay:2024nip} and IceCube  \cite{IceCube:2016rnb,IceCubeCollaboration:2024nle}. Consequently, further detailed investigations are necessary to explore the potential existence of sterile neutrinos. 
Meanwhile, laboratory experiments (such as JUNO, MINOS, and IceCube)  provide relatively model-independent and direct tests but are limited by restricted sensitivity ranges, energy resolution, and statistical uncertainties \cite{JUNO:2015zny,Forero:2021lax,JUNO:2024jaw,MINOS:2013utc,IceCube-Gen2:2019fet}. By contrast, cosmological observations can yield much tighter constraints on sterile neutrino parameters through global fits to the evolution of the universe. Therefore, cosmological observations serve as an independent and complementary avenue for exploring the properties of sterile neutrinos.}
Extensive studies on neutrinos in cosmology can be found in Refs. \cite{Hu:1997mj,deHolanda:2010am,Li:2012spm,Wang:2012vh,Palazzo:2013me,Wyman:2013lza,Battye:2013xqa,Zhang:2014dxk,Dvorkin:2014lea,Archidiacono:2014apa,Zhang:2014nta,DayaBay:2014fct,Zhang:2014ifa,Zhang:2015rha,Geng:2015haa,Zhang:2015uhk,Huang:2015wrx,Chen:2016eyp,Giusarma:2016phn,Wang:2016tsz,Xu:2016ddc,Vagnozzi:2017ovm,Li:2017iur,Yang:2017amu,Feng:2017nss,Feng:2017mfs,Wang:2017htc,Feng:2017usu,Vagnozzi:2018jhn,Knee:2018rvj,Feng:2019mym,Feng:2019jqa,Liu:2020vgn,Yang:2020tax,Zhang:2020mox,Yang:2020ope,Feng:2021ipq,DiValentino:2021rjj,Tanseri:2022zfe,Chernikov:2022mdn,Pang:2023joc,Pan:2023frx}.

Recently, the baryon acoustic oscillation (BAO) measurements from the second data release (DR2) of the Dark Energy Spectroscopic Instrument (DESI), derived from three years of observations, have been published. When combined with cosmic microwave background (CMB) and type Ia supernovae (SN) data, these measurements indicate a 2.8$\sigma$--4.2$\sigma$ preference for dynamical dark energy (DE) within the Chevallier–Polarski–Linder (CPL) model \cite{DESI:2025zgx}, providing stronger evidence than that obtained from the first data release (DR1) \cite{DESI:2024mwx}.
These significant deviations from the cosmological constant have drawn considerable attention within the cosmology community, motivating extensive studies on various aspects of cosmological physics, including the nature of DE and the properties of neutrinos~\cite{Giare:2024smz,Wang:2024dka,Li:2024qso,Giare:2024gpk,Dinda:2024ktd,Sabogal:2024yha,Escamilla:2024ahl,Li:2024qus,Li:2024bwr,Li:2025owk,Huang:2025som,Pang:2025lvh,You:2025uon,Pan:2025qwy,Wu:2025wyk,Cheng:2025lod,Yang:2025boq,Ling:2025lmw,Barua:2025ypw,Li:2025ula,Ozulker:2025ehg,Gialamas:2025pwv,Li:2025ops,Liu:2025myr,Qiang:2025cxp,Li:2025htp,Li:2025eqh,Li:2025dwz,Wu:2025vfs,Du:2024pai,Jiang:2024viw,Du:2025iow,Feng:2025mlo,RoyChoudhury:2025dhe,Du:2025xes}.
In particular, Ref.~\cite{Du:2025iow} has recently reported that, in the $\Lambda$CDM model,  when BAO data from DESI DR1 are combined with SN data from DESY5 and CMB data from Planck and ACT, the cosmological upper limits on sterile neutrino parameters are constrained to $N_{\rm eff} < 3.42$ and $m_{\nu,{\rm sterile}}^{\rm eff} < 0.217$~eV at the 2$\sigma$ level. In the CPL model, the corresponding constraints change to $N_{\rm eff} < 3.37$ and $m_{\nu,{\rm sterile}}^{\rm eff} < 0.350$~eV,  indicating that the properties of DE can influence  the cosmological constraints on sterile neutrino parameters.

In parallel,  Ref.~\cite{Feng:2024mfx} conducted a preliminary investigation of sterile neutrinos within the $\Lambda$CDM model, using joint observations of BNS mergers detected by 3G GW detectors and short $\gamma$-ray burst (GRB) observed by a THESEUS-like mission. They found that the GW data provide a tight upper limit on $m_{\nu,{\rm sterile}}^{\rm eff}$ and show a mild preference for $\Delta N_{\rm eff} > 0$ at the 1.1$\sigma$ level, indicating that GW standard siren observations can significantly tighten the constraints on sterile neutrino parameters.

Taken together, these insights underscore the importance and timeliness of investigating sterile neutrinos within dynamical DE models using future GW observations.
In this paper, we consider three representative dynamical DE models, namely the $w$CDM, holographic DE (HDE), and CPL models. We forecast the prospects for detecting sterile neutrinos in these models using joint GW-GRB observations. The primary aim of this study is to evaluate how future GW standard sirens can improve cosmological constraints on sterile neutrino parameters in dynamical DE models.

This work is organized as follows. In Sec.~\ref{sec2}, we introduce the models, cosmological data, and GW simulation used in this work. In Sec.~\ref{sec3}, we present the constraint results of cosmological parameters from the GW standard siren observations. The conclusion is given in Sec.~\ref{sec4}. 

\section{Methodology}\label{sec2}
In this section, we first introduce the cosmological models used in this study, followed by a brief overview of the cosmological datasets utilized, and finally describe the approach for simulating the GW data.

\begin{table*}[htbp]
\renewcommand\arraystretch{1.7}
\centering
\caption{Cosmological models and parameters used in this work.}
\setlength{\tabcolsep}{14pt}
\resizebox{0.9\textwidth}{!}{%
\begin{tabular}{lcl}
\hline
Model & Parameter \# & Parameters \\ 
\hline
%$\Lambda$CDM+$\nu_s$ & 8 & $\omega_b$, $\omega_c$, $\Theta_{\rm s}$, $\tau$, $n_s$, $\ln (10^{10}A_s)$, $N_{\rm eff}$, $m_{\nu,{\rm{sterile}}}^{\rm{eff}}$ \\ 
$w$CDM+$\nu_s$ & 9 & $\omega_b$, $\omega_c$, $\Theta_{\rm s}$, $\tau$, $\ln (10^{10}A_s)$, $n_s$, $N_{\rm eff}$, $m_{\nu,{\rm{sterile}}}^{\rm{eff}}$, $w$ \\ 
HDE+$\nu_s$ & 9 & $\omega_b$, $\omega_c$, $\Theta_{\rm s}$, $\tau$, $\ln (10^{10}A_s)$, $n_s$, $N_{\rm eff}$, $m_{\nu,{\rm{sterile}}}^{\rm{eff}}$, $c$ \\ 
CPL+$\nu_s$ & 10 & $\omega_b$, $\omega_c$, $\Theta_{\rm s}$, $\tau$, $\ln (10^{10}A_s)$, $n_s$, $N_{\rm eff}$, $m_{\nu,{\rm{sterile}}}^{\rm{eff}}$, $w_0$, $w_a$ \\ 
\hline
\end{tabular}
}
\label{tab:parms}
\end{table*}

\subsection{Models}\label{sec2.1}

The baseline model considered is the $\Lambda$CDM model. The extended model, $\Lambda$CDM with sterile neutrinos ($\Lambda$CDM+$\nu_s$) includes eight base parameters, including the baryon density parameter $\omega_b\equiv \Omega_b h^2$, the cold dark matter density parameter $\omega_c\equiv \Omega_c h^2$, the ratio between the sound horizon and the angular diameter distance at decoupling $\Theta_{\rm{s}}$, the optical depth to the reionization of the universe $\tau$, the amplitude of initial curvature perturbation $A_s$, the power-law spectral index $n_s$, the effective number of relativistic species $N_{\rm eff}$, and the effective sterile neutrino mass $m_{\nu,{\rm sterile}}^{\rm eff}$. 
Note that the true mass of thermally distributed sterile neutrinos is given by $m_{\rm sterile}^{\rm thermal}=(N_{\rm eff}-N_{\rm{SM}})^{-3/4}m_{\nu,{\rm sterile}}^{\rm eff}$, where $N_{\rm{SM}} = 3.044$ is the effective number of active neutrinos in the Standard Model framework \cite{Akita:2020szl, Froustey:2020mcq, Bennett:2020zkv}. 
In addition, the total mass of active neutrinos is fixed at $\sum m_\nu = 0.06$ eV in this study.

In our analysis, we consider three dynamical DE models, with their respective parameters detailed in Table~\ref{tab:parms}, and followed by a brief introduction to each model.

\begin{itemize}

\item The $w$CDM+$\nu_s$ model: Building on the $\Lambda$CDM+$\nu_s$ model, we consider the case where the equation-of-state (EoS) parameter $w$ is treated as a constant free parameter.

\item The HDE+$\nu_s$ model: Extending the $\Lambda$CDM+$\nu_s$ model, we adopt the HDE model~\cite{Li:2004rb}, which introduces an additional parameter $c$ through the energy density definition $\rho_{\rm{de}} = 3 c^2 M^2_{\rm{Pl}} L^{-2}$, where $M_{\rm{Pl}}$ is the reduced Planck mass and $L$ is the future event horizon of the universe. The parameter $c$ is dimensionless and phenomenological, playing an important role in determining the properties of DE in this model. The EoS is given by $w = - \frac{1}{3} - \frac{2}{3} \frac{\sqrt{\Omega_{\rm{de}}}}{c}$. For more details of the HDE model, see, e.g., Refs.~\cite{Zhang:2005hs,Li:2009bn,Li:2013dha,Feng:2016djj}.

\item The CPL+$\nu_s$ model: 
This model further extends $\Lambda$CDM+$\nu_s$ by introducing a time-varying EoS via the CPL parametrization~\cite{Chevallier:2000qy,Linder:2002et}, $w(a) = w_0 + w_a (1 - a)$, where $w_0$ denotes the present-day value of $w$ and $w_a$ characterizes its evolution with the scale factor $a$.

\end{itemize}

\subsection{Cosmological data}\label{sec2.2}
The cosmological observational data used in this study include the CMB, BAO, and SN. For the CMB data, we use the Planck TT, TE, EE spectra at $\ell\geq 30$, the low-$\ell$ temperature commander likelihood, and the low-$\ell$ SimAll EE likelihood from the Planck 2018 release~\cite{Planck:2018vyg}. For the BAO data, we adopt measurements from 6dFGS ~\cite{Beutler:2011hx}, SDSS-MGS~\cite{Ross:2014qpa}, and BOSS-DR12~\cite{BOSS:2016wmc}. For the SN data, we employ the Pantheon sample, which consists of 1048 data points from the Pantheon complation~\cite{Pan-STARRS1:2017jku}.
For simplicity, we use ``CBS'' to denote the joint CMB+BAO+SN dataset.

Note that the primary objective of this work is to investigate the impact of joint GW and GRB observations on the cosmological constraints of sterile neutrino parameters within various dynamical DE models. To facilitate a direct and consistent comparison with previous studies on sterile neutrinos in the $\Lambda$CDM model~\cite{Feng:2024mfx}, we deliberately adopt the same combination of observational datasets (CBS). This approach allows us to isolate and evaluate the specific contribution of GW observations to the improvement in sterile neutrino constraints.

\subsection{Gravitational wave simulation}\label{sec2.1}
In this subsection, we present the method of simulating the joint GW standard sirens and GRB events. We focus on the synergy between the THESEUS-like GRB detector and the 3G GW observations. We utilize the simulation method as prescribed in Refs.~\cite{Han:2023exn,Feng:2024lzh,Feng:2024mfx,Han:2024sxm}.  

Based on the star formation rate~\cite{Vitale:2018yhm,Yang:2021qge,Belgacem:2019tbw}, the merger rate of the BNS density with redshift in the observer frame is
\begin{equation}
R_{\rm m}(z)=\frac{\mathcal{R}_{\rm m}(z)}{1+z} \frac{{\rm d}V(z)}{{\rm d}z},
\label{eq:1}
\end{equation}
where ${\rm d}V(z)/{\rm d}z$ represents the comoving volume element, and $\mathcal{R}_{\rm m}(z)$ is the BNS merger rate in the source frame, defined as
\begin{equation}
	\mathcal{R}_{\rm m}(z)=\int_{t_{\rm min}}^{t_{\rm max}} \mathcal{R}_{\rm f}[t(z)-t_{\rm d}] P(t_{\rm d}){\rm d}t_{\rm d}.
	\label{eq:2}
\end{equation}
Here, $t_{\rm max}=t_{\rm H}$ is the maximum delay time, $t_{\rm min}=20$ Myr is the minimum delay time, $\mathcal{R}_{\rm f}$ is the cosmic star formation rate in the source frame based on the Madau-Dickinson model~\cite{Madau:2014bja}, $t(z)$ is the age of the universe at the time of merger, $t_{\rm d}$ is the delay time between BNS system formation and merger, and $P(t_{\rm d})$ represents the time delay distribution of $t_{\rm d}$, for which we use the exponential time delay model~\cite{Vitale:2018yhm}, expressed as
\begin{equation}
	P(t_{\rm d})=\frac{1}{\tau}{\rm exp}(-t_{\rm d}/\tau),
\end{equation}
with an e-fold time of $\tau=100$ Myr for $t_{\rm d}>t_{\rm min}$.

In our analysis of BNS mergers, we adopt a local comoving merger rate of $\mathcal{R}_{\rm m}(z=0)=920~\rm Gpc^{-3}~yr^{-1}$, which corresponds to the median estimate derived from the O1 LIGO and O2 LIGO/Virgo observation runs~\cite{Eichhorn:2018phj} and remains consistent with the findings from the O3 observation run~\cite{KAGRA:2021duu}.
We simulate a catalog of BNS mergers for a 10-year observation.
For each source, the location $(\theta,\phi)$, the polarization angle $\psi$, the orientation angle $\iota$, and the coalescence phase $\psi_{\rm c}$ are all drawn from uniform distributions. 
%At present, several candidate models exist for the neutron star (NS) mass distribution. However, the impact of different NS mass distributions on cosmological analysis is minimal~\cite{Han:2023exn}. 
To simplify, we adopt a Gaussian mass distribution.
This distribution has a mean NS mass of $1.33~M_{\odot}$ and a standard deviation of $0.09~M_{\odot}$, where $M_{\odot}$ denotes the solar mass~\cite{Ozel:2016oaf,LIGOScientific:2018mvr}.

In the framework of the stationary phase approximation~\cite{Zhang:2017srh}, the Fourier transform of the frequency-domain GW waveform for a detector network (with $N$ detectors) is expressed as~\cite{Wen:2010cr,Zhao:2017cbb}
\begin{equation}
\tilde{\boldsymbol{h}}(f)=e^{-i\boldsymbol\Phi}\boldsymbol h(f),
\end{equation}
where $\boldsymbol\Phi$ is the $N\times N$ diagonal matrix with $\Phi_{ij}=2\pi f\delta_{ij}(\boldsymbol{n\cdot r}_k)$, $\boldsymbol n$ represents the propagation direction of GW, $\boldsymbol r_k$ is the spatial location of the $k$-th detector, and
\begin{equation}
\boldsymbol h(f)=\Big[\frac{h_1(f)}{\sqrt{S_{\rm {n},1}(f)}}, \frac{h_2(f)}{\sqrt{S_{\rm {n},2}(f)}}, \ldots,\frac{h_N(f)}{\sqrt{S_{{\rm n},N}(f)}}\Big ]^{\rm T},
\end{equation}
where $S_{\rm {n},k}(f)$ is the one-side noise power spectral density of the $k$-th detector.

In this study, we consider the waveform in the inspiralling stage for the non-spinning BNS system. We employ the restricted Post-Newtonian approximation up to 3.5 PN order~\cite{Cutler:1992tc,Sathyaprakash:2009xs}. 
The Fourier transform of the GW waveform for the $k$-th detector is  expressed as
\begin{align}
	h_k(f)=&\mathcal A_k f^{-7/6}{\rm exp}
	\{i[2\pi f t_{\rm c}-\pi/4-2\psi_c+2\Psi(f/2)]\nonumber\\ &-\varphi_{k,(2,0)})\},
\end{align}
where the detailed forms of $\Psi(f/2)$ and $\varphi_{k,(2,0)}$ can be found in Refs.~\cite{Cutler:1992tc,Zhao:2017cbb}, and the Fourier amplitude can be written as
\begin{align}
	\mathcal A_k=&\frac{1}{d_{\rm L}}\sqrt{(F_{+,k}(1+\cos^{2}\iota))^{2}+(2F_{\times,k}\cos\iota)^{2}}\nonumber\\ &\times\sqrt{5\pi/96}\pi^{-7/6}\mathcal M^{5/6}_{\rm chirp},
\end{align}
where $d_{\rm L}$ is the luminosity distance to the GW source, $F_{+,k}$ and $F_{\times,k}$ represent the antenna response functions of the $k$-th GW detector, $\mathcal M_{\rm chirp}=(1+z)\eta^{3/5}M$ is the chirp mass of the binary system, $\eta=m_1 m_2/M^2$ is the symmetric mass ratio, and $M=m_1+m_2$ is the total mass of the binary system with $m_1$ and $m_2$ as the masses of the components. We use the GW waveform in the frequency domain, where time $t$ is replaced by $t_{f}=t_{\rm c}-(5 / 256) \mathcal{M}_{\rm chirp}^{-5 / 3}(\pi f)^{-8 / 3}$~\cite{Cutler:1992tc,Zhao:2017cbb}, with $t_{\rm c}$ being the coalescence time.

Following the simulation of the GW catalog, we calculate the signal-to-noise ratio (SNR) for each GW event. Here, we adopt the SNR threshold to be 12 in our simulation. For low-mass systems, the combined SNR for the detection network of $N$ independent detectors can be calculated by
\begin{equation}
\rho=(\tilde{\boldsymbol h}|\tilde{\boldsymbol h})^{1/2}.
\end{equation}
The inner product is defined as
\begin{equation}
(\boldsymbol a|\boldsymbol b)=2\int_{f_{\rm lower}}^{f_{\rm upper}}\{\boldsymbol a(f)\boldsymbol b^*(f)+\boldsymbol a^*(f)\boldsymbol b(f)\}{\rm d}f,
\end{equation}
where $\boldsymbol a$ and $\boldsymbol b$ are column matrices with the same dimension, $*$ denotes conjugate transpose, the lower cutoff frequency is set to $f_{\rm lower}=1$ Hz for ET and $f_{\rm lower}=5$ Hz for CE, and $f_{\rm upper}=2/(6^{3/2}2\pi M_{\rm obs})$ represents the frequency in the last stable orbit, with $M_{\rm obs}=(m_1+m_2)(1+z)$. 

For the short GRB model, we use the Gaussian structured jet profile model, which is based on the analysis of GW170817/GRB170817A~\cite{Howell:2018nhu},
\begin{equation}
L_{\rm iso}(\theta_{\rm v})=L_{\rm on}\exp\left(-\frac{\theta^2_{\rm v}}{2\theta^2_{\rm c}} \right),
\label{eq:jet}
\end{equation}
where $\theta_{\rm v}$ is the viewing angle, $L_{\rm iso}(\theta_{\rm v})$ represents the isotropically equivalent luminosity of short GRB observed at different values of $\theta_{\rm v}$, $L_{\rm on}=L_{\rm iso}(0)$ is the on-axis isotropic luminosity, $\theta_{\rm c}=4.7^{\circ}$ is the characteristic core angle, and the direction of the jet is assumed to be aligned with the binary's orbital angular momentum, namely $\iota=\theta_{\rm v}$.

For the distribution of the short GRB, we assume an empirical broken-power-law luminosity function, expressed as
\begin{equation}
	\Phi(L)\propto
	\begin{cases}
		(L/L_*)^{\alpha}, & L<L_*, \\
		(L/L_*)^{\beta}, & L\ge L_*,
	\end{cases}
	\label{eq:distribution}
\end{equation}
where $L$ is the peak luminosity in the 1--10000 keV energy range in the rest frame assuming isotropic emission, $L_{*}$ is the characteristic parameter separating the two regimes, $\alpha$ and $\beta$ are the characteristic slopes that describe these regimes, respectively.
Following Ref.~\cite{Wanderman:2014eza}, we use $L_{*}=2\times10^{52}$ erg sec$^{-1}$, $\alpha=-1.95$, and $\beta=-3$.
We adopt a standard low end cutoff in luminosity of $L_{\rm min} = 10^{49}$ erg sec$^{-1}$, and we treat the on-axis isotropic luminosity $L_{\rm on}$ as the peak luminosity $L$~\cite{Belgacem:2019tbw,Tan:2020vtc,Yang:2021qge,Han:2023exn}.
For the THESEUS mission~\cite{Stratta:2018ldl}, a GRB detection is recorded if the value of observed flux is larger than the flux threshold $P_{\rm T}=0.2~\rm ph~s^{-1}~cm^{-2}$ in the 50--300 keV band.
For the GRB detection, we take a sky coverage fraction of 0.5 and a duty cycle of 80\% \cite{Stratta:2018ldl}.
Thus from the GW catalog that exceeds the threshold of 12, we can select the GW-GRB events based on the probability distribution $\Phi(L){\rm d}L$.

For a network with $N$ independent interferometers, the Fisher information matrix is defined as
\begin{equation}
F_{ij}=\left(\frac{\partial \tilde{\boldsymbol{h}}}{\partial \theta_i}\Bigg |\frac{\partial \tilde{\boldsymbol{h}}}{\partial \theta_j}\right),
\end{equation}
where $\theta_i$ represents nine GW parameters ($d_{\rm L}$, $\mathcal{M}_{\rm chirp}$, $\eta$, $\theta$, $\phi$, $\iota$, $t_{\rm c}$, $\psi_{\rm c}$, $\psi$) for each GW event. The covariance matrix is equal to the inverse of the Fisher matrix, i.e., ${\rm Cov}_{ij}=(F^{-1})_{ij}$. Therefore, the instrumental error $\sigma_{d_{\rm L}}^{\rm inst}$ of the GW parameter $\theta _i$ is given by $\Delta\theta_i=\sqrt{{\rm Cov}_{ii}}$.

The total uncertainty of the luminosity distance $d_{\rm L}$ can be written as
\begin{align}
\sigma_{d_{\rm L}}&~~=\sqrt{(\sigma_{d_{\rm L}}^{\rm inst})^2+(\sigma_{d_{\rm L}}^{\rm lens})^2+(\sigma_{d_{\rm L}}^{\rm pv})^2},\label{total}
\end{align}
where the instrumental error of $\sigma_{d_{\rm L}}^{\rm inst}$ is estimated by the Fisher information matrix, $\sigma_{d_{\rm L}}^{\rm lens}$ is the weak-lensing error, and $\sigma_{d_{\rm L}}^{\rm pv}$ is the peculiar velocity error.

The $\sigma_{d_{\rm L}}^{\rm lens}$ is adopted from Refs.~\cite{Speri:2020hwc,Hirata:2010ba},
\begin{align}
\sigma_{d_{\rm L}}^{\rm lens}(z)=&\left[1-\frac{0.3}{\pi/2} \arctan(z/0.073)\right]\times d_{\rm L}(z)\nonumber\\ &\times 0.066\left [\frac{1-(1+z)^{-0.25}}{0.25}\right ]^{1.8}.\label{lens}
\end{align}

The $\sigma_{d_{\rm L}}^{\rm pv}$ of the GW source is given by~\cite{Kocsis:2005vv}
\begin{equation}
	\sigma_{d_{\rm L}}^{\rm pv}(z)=d_{\rm L}(z)\times \left [ 1+ \frac{c(1+z)^2}{H(z)d_{\rm L}(z)}\right ]\frac{\sqrt{\langle v^2\rangle}}{c},\label{pv}
\end{equation}
where $c$ is the speed of light in vacuum, $H(z)$ is the Hubble parameter, and $\sqrt{\langle v^2\rangle}$ is the peculiar velocity of the GW source, set to $\sqrt{\langle v^2\rangle}=500\ {\rm km\ s^{-1}}$.

For the GW standard siren observation with $N$ data points, the $\chi^2$ function is given by
\begin{align}
\chi_{\rm GW}^2=\sum\limits_{i=1}^{N}\left[\frac{{d}_L^i-d_{\rm L}({z}_i;\vec{\Omega})}{{\sigma}_{d_{\rm L}}^i}\right]^2,
\label{equa:chi2}
\end{align}
where $\vec{\Omega}$ is the set of cosmological parameters, ${z}_i$, ${d}_L^i$, and ${\sigma}_{d_{\rm L}}^i$ are the $i$-th GW redshift, luminosity distance, and the total error of the luminosity distance, respectively.

In this work, we present the forecast for the search for sterile neutrinos in dynamical DE models using joint GW-GRB observations. We use the public Markov-chain Monte Carlo (MCMC) package \texttt{CosmoMC}~\cite{Lewis:2002ah} to infer the posterior probability distributions of the cosmological parameters.
To demonstrate the impact of simulated GW data on constraining sterile neutrino parameters, we will analyze two scenarios of 3G GW observations: a single ET and the ET-CE-CE network, which consists of one ET detector and two CE-like detectors (one in the United States with a 40 km arm length and another one in Australia with a 20 km arm length, hereafter referred to as ET2CE).
We utilize the ET sensitivity curve provided in Ref. \cite{ETcurve-web} and the CE sensitivity curves from Ref. \cite{CEcurve-web}, as presented in Fig. 1 of Ref. \cite{Feng:2024mfx}.
For the GW detectors, due to the large uncertainty in the duty cycle, we consider only the ideal scenario in which all detectors are assumed to have a 100\% duty cycle, as discussed in Ref. \cite{Zhu:2021ram}.
The geometry of the GW detector, characterized by parameters such as latitude ($\varphi$), longitude ($\lambda$), opening angle ($\zeta$), and arm bisector angle ($\gamma$), is described in detail in Table I of Ref. \cite{Feng:2024mfx}.
To ensure a direct comparison of constraints within the same parameter space, we adopt the best-fit cosmological parameter values from the CBS dataset as the fiducial values for simulating GW data in dynamical DE models.
Thus, in our analysis, we use three data combinations: (1) CBS, (2) CBS+ET, and (3) CBS+ET2CE. We will report the constraint results in the next section.

%%%%%%%%%%%%%%%%%%%%%%%%%%%%%%%%%%%%%%%%%%%%%%%%%%%%%%%%%%%%%%%%
\begin{table*}\small
\setlength\tabcolsep{2.0pt}
\renewcommand{\arraystretch}{2.0}
\caption{\label{tabDE}Fitting results of the $w$CDM+$\nu_s$, HDE+$\nu_s$ and CPL+$\nu_s$ models by using the CBS, CBS+ET, and CBS+ET2CE data combinations. We quote $\pm 1\sigma$ errors for the parameters, but for the parameters $N_{\rm eff}$ and $m_{\nu,{\rm{sterile}}}^{\rm{eff}}$, when central values cannot be determined, we quote the $2\sigma$ upper limits. Here, $H_0$ is in units of ${\rm km}\ {\rm s^{-1}}\ {\rm Mpc^{-1}}$ and $m_{\nu,{\rm{sterile}}}^{\rm{eff}}$ is in units of eV. For a parameter $\xi$, $\sigma(\xi)$ and $\varepsilon(\xi)=\sigma(\xi)/\xi$ represent its absolute and relative errors, respectively.}
\centering
\begin{tabular}{ccccccccccccccccccc}
\hline \multicolumn{1}{c}{ } &&\multicolumn{3}{c}{$w$CDM+$\nu_s$}&&\multicolumn{3}{c}{HDE+$\nu_s$}&&\multicolumn{3}{c}{CPL+$\nu_s$}\\
%\cline{1-1}\cline{3-5}\cline{7-9}
%\cline{3-5}\cline{7-9}\cline{11-13}\cline{15-16}
\cline{3-5}\cline{7-9}\cline{11-13}
 && CBS & CBS+ET & CBS+ET2CE  &&  CBS & CBS+ET & CBS+ET2CE &&  CBS & CBS+ET & CBS+ET2CE\\
\hline
$\sigma(\Omega_m)$&&$0.0078$&$0.0042$&$0.0042$&&$0.0077$&$0.0040$&$0.0038$&&$0.0081$&$0.0049$&$0.0048$\\
$\sigma(H_0)$&&$0.9850$&$0.0655$&$0.0655$&&$1.1300$&$0.0750$&$0.0735$&&$0.9700$&$0.0855$&$0.0880$\\
$\sigma(w/c/w_0)$\tablenote{$w$ is for $w$CDM+$\nu_s$, $c$ is for HDE+$\nu_s$ and $w_0$ is for CPL+$\nu_s$.}&&$0.0360$&$0.0160$&$0.0160$&&$0.031$&$0.015$&$0.015$&&$0.087$&$0.031$&$0.031$\\
$\sigma(w_a)$&&$...$&$...$&$...$&&$...$&$...$&$...$&&$0.39$&$0.18$&$0.18$\\
$\varepsilon(\Omega_m)$&&$2.542\%$&$1.375\%$&$1.375\%$&&$2.516\%$&$1.309\%$&$1.243\%$&&$2.620\%$&$1.589\%$&$1.556\%$\\
$\varepsilon(H_0)$&&$1.432\%$&$0.096\%$&$0.096\%$&&$1.646\%$&$0.109\%$&$0.107\%$&&$1.409\%$&$0.124\%$&$0.128\%$\\
$\varepsilon(w/c/w_0)$&&$3.472\%$&$1.540\%$&$1.540\%$&&$4.851\%$&$2.344\%$&$2.344\%$&&$9.315\%$&$3.326\%$&$3.330\%$\\
$\varepsilon(w_a)$&&$...$&$...$&$...$&&$...$&$...$&$...$&&$78.000\%$&$35.294\%$&$35.294\%$\\
$N_{\rm eff}$&&$<3.340$&$3.142^{+0.032}_{-0.091}$&$3.143^{+0.035}_{-0.087}$&&$<3.562$&$3.217^{+0.068}_{-0.062}$&$3.219\pm0.058$&&$<3.404$&$3.165^{+0.045}_{-0.103}$&$3.162^{+0.044}_{-0.098}$\\
$m_{\nu,{\rm{sterile}}}^{\rm{eff}}$&&$<0.651$&$<0.565$&$<0.564$&&$<0.479$&$<0.121$&$<0.115$&&$<0.707$&$<0.691$&$<0.680$\\
\hline
\end{tabular}
%\end{table}
\end{table*}
%%%%%%%%%%%%%%%%%%%%%%%%%%%%%%%%%%%%%%%%%%%%%%%%%%%%%%%%%%%%%%%%

%%%%%%%%%%%%%%%%%%%%%%%%%%%%%%%%%%%%%%%%%%%%%%%%%%%%%%%%%%%%%%%%
\begin{table}\small
%\begin{table*}\small
\setlength\tabcolsep{2.0pt}
\renewcommand{\arraystretch}{2.0}
\caption{\label{tablcdm} Same as Table~\ref{tabDE}, but in the $\Lambda$CDM+$\nu_s$ model~\cite{Feng:2024mfx}.}
\centering
\begin{tabular}{cccccc}
\hline 
Parameter&& CBS & CBS+ET & CBS+ET2CE\\
\hline
$\sigma(\Omega_m)$&&$0.00640$&$0.00420$&$0.00405$\\
$\sigma(H_0)$&&$0.7500$&$0.0540$&$0.0520$\\
$\varepsilon(\Omega_m)$&&$2.055\%$&$1.357\%$&$1.308\%$\\
$\varepsilon(H_0)$&&$1.101\%$&$0.079\%$&$0.076\%$\\
$N_{\rm eff}$&&$<3.3446$&$3.148^{+0.039}_{-0.094}$&$3.150^{+0.042}_{-0.089}$\\
$m_{\nu,{\rm{sterile}}}^{\rm{eff}}$&&$<0.5789$&$<0.4842$&$<0.4226$\\
\hline
\end{tabular}
\end{table}
%\end{table*}
%%%%%%%%%%%%%%%%%%%%%%%%%%%%%%%%%%%%%%%%%%%%%%%%%%%%%%%%%%%%%%%%

%%%%%%%%%%%%%%%%%%%%%%%%%%%%%%%%%%%%%%%%%%%%%%%%%%%%%%%%%%%%%%%%
\begin{figure*}[!htbp]
\includegraphics[width=0.9\textwidth]{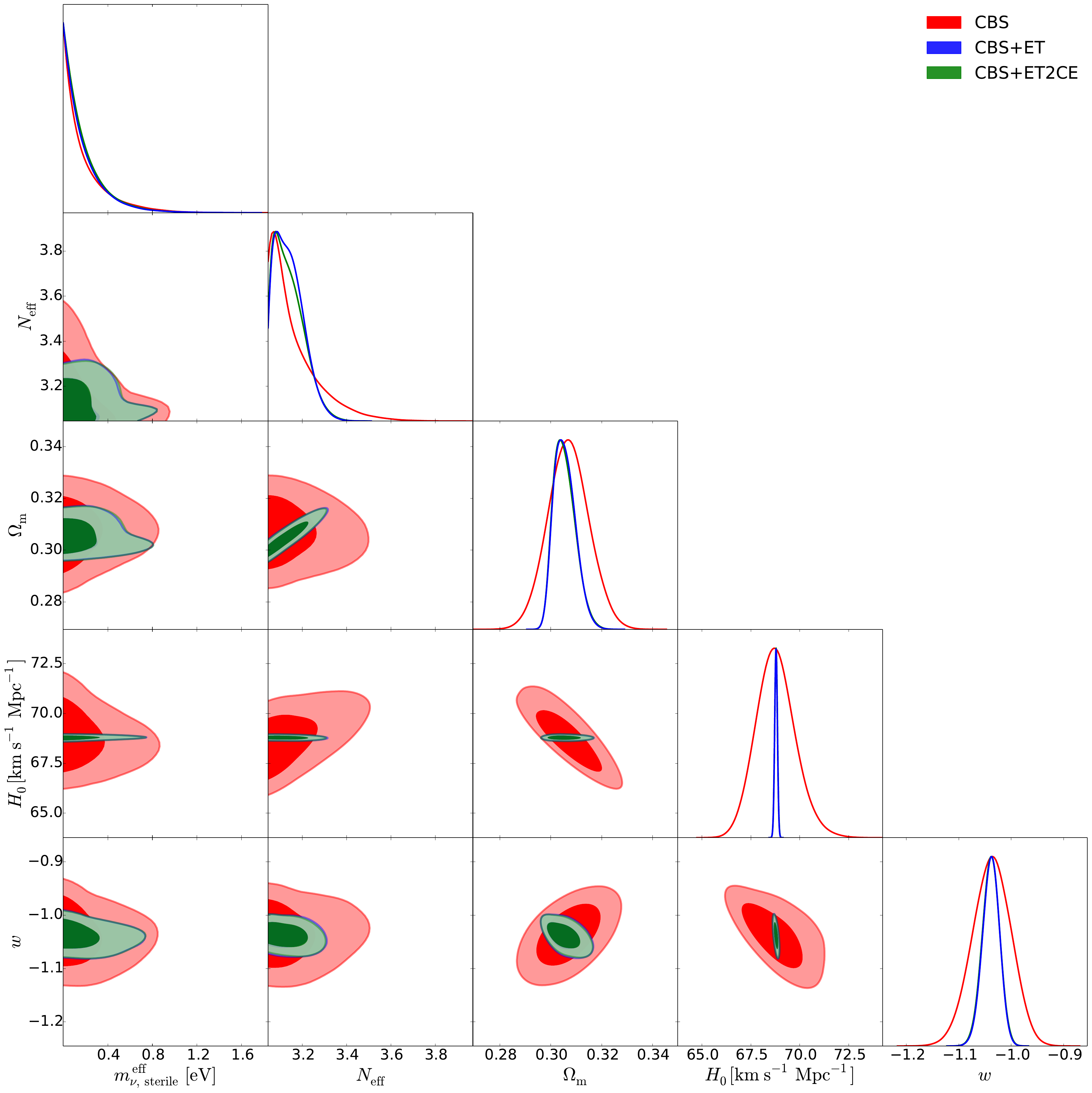}
\centering
\caption{The triangular plots of the marginalized posterior distributions for cosmological parameters in the $w$CDM+$\nu_s$ model, using the CBS, CBS+ET, and CBS+ET2CE data combinations, respectively.}\label{figwcdm}
\end{figure*}
%%%%%%%%%%%%%%%%%%%%%%%%%%%%%%%%%%%%%%%%%%%%%%%%%%%%%%%%%%%%%%%%

%%%%%%%%%%%%%%%%%%%%%%%%%%%%%%%%%%%%%%%%%%%%%%%%%%%%%%%%%%%%%%%%
\begin{figure*}[!htbp]
\includegraphics[width=0.9\textwidth]{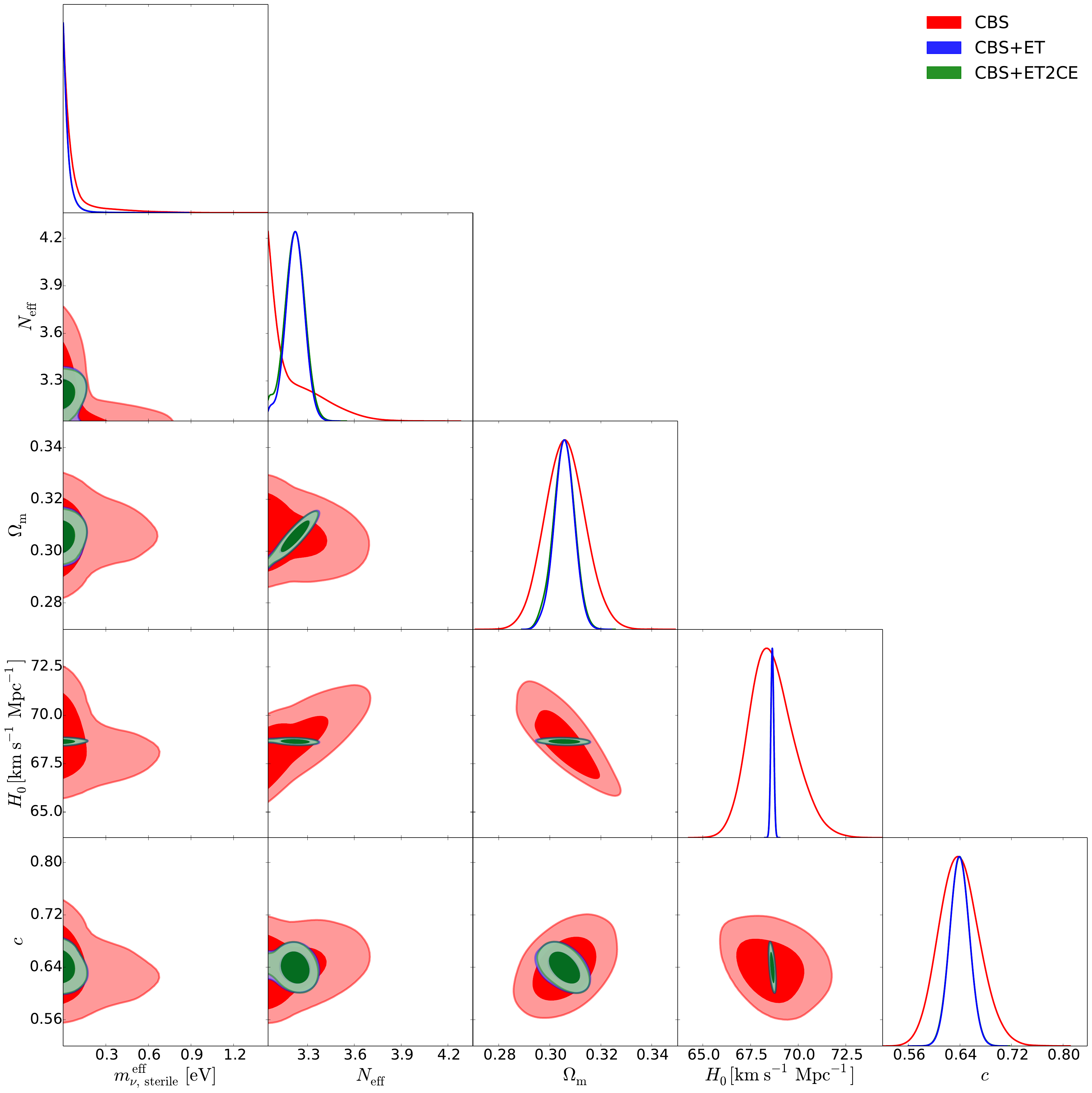}
\centering
\caption{The triangular plots of the marginalized posterior distributions for cosmological parameters in the HDE+$\nu_s$ model, using the CBS, CBS+ET, and CBS+ET2CE data combinations, respectively.}\label{fighde}
\end{figure*}
%%%%%%%%%%%%%%%%%%%%%%%%%%%%%%%%%%%%%%%%%%%%%%%%%%%%%%%%%%%%%%%%

%%%%%%%%%%%%%%%%%%%%%%%%%%%%%%%%%%%%%%%%%%%%%%%%%%%%%%%%%%%%%%%%
\begin{figure*}[!htbp]
\includegraphics[width=0.9\textwidth]{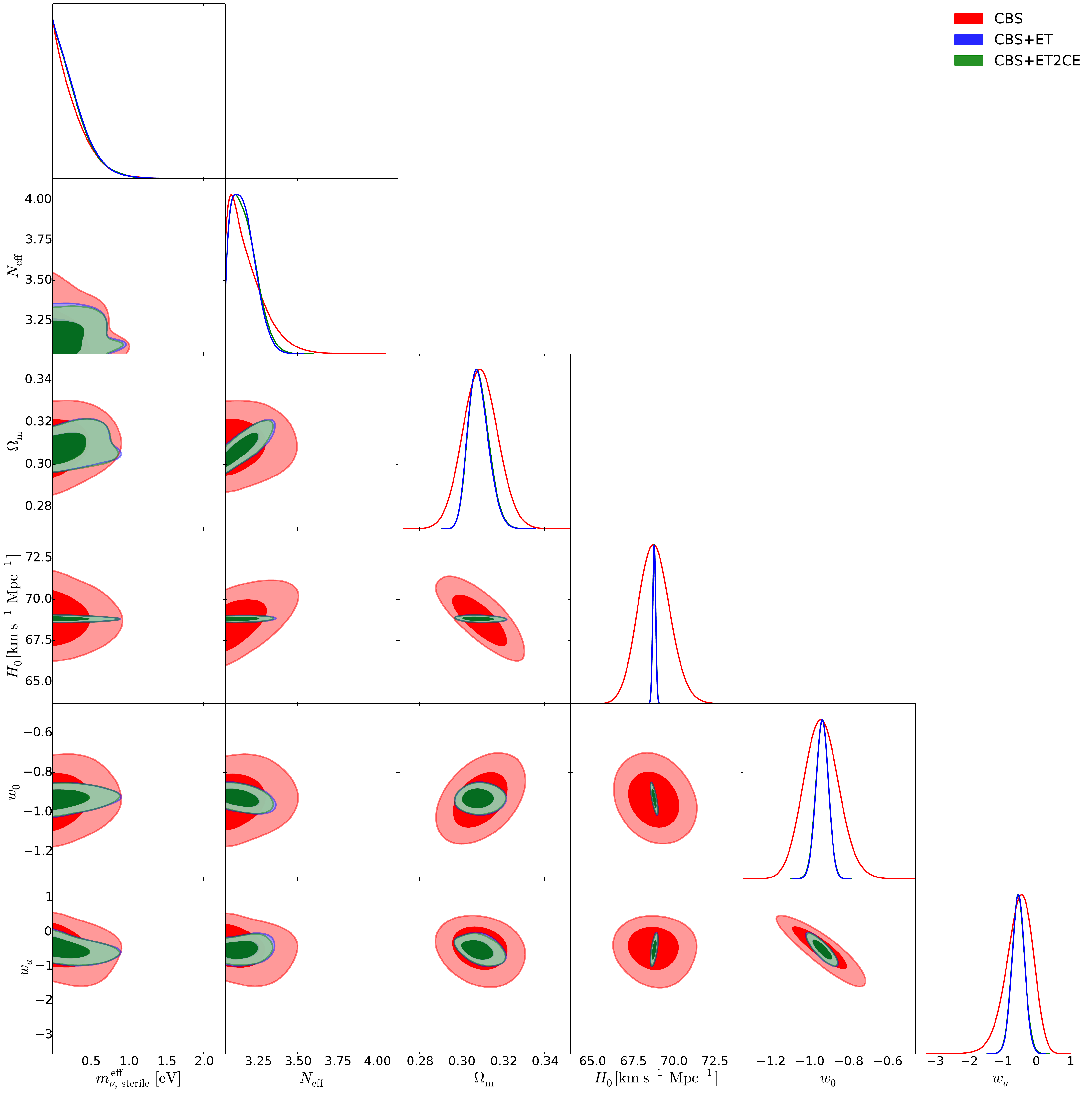}
\centering
\caption{The triangular plots of the marginalized posterior distributions for cosmological parameters in the CPL+$\nu_s$ model, using the CBS, CBS+ET, and CBS+ET2CE data combinations, respectively.}\label{figcpl}
\end{figure*}
%%%%%%%%%%%%%%%%%%%%%%%%%%%%%%%%%%%%%%%%%%%%%%%%%%%%%%%%%%%%%%%%

%%%%%%%%%%%%%%%%%%%%%%%%%%%%%%%%%%%%%%%%%%%%%%%%%%%%%%%%%%%%%%%%
\begin{figure*}[!htbp]
\includegraphics[width=0.3\textwidth]{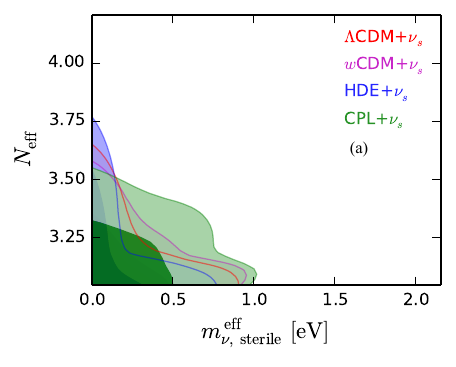}
\includegraphics[width=0.3\textwidth]{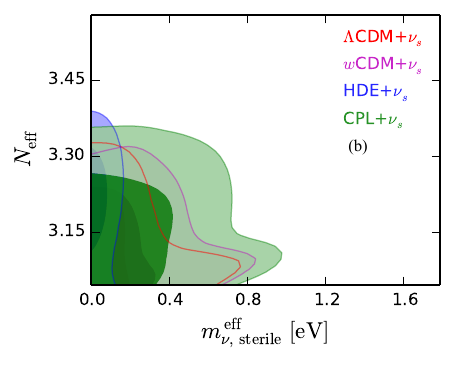}
\includegraphics[width=0.3\textwidth]{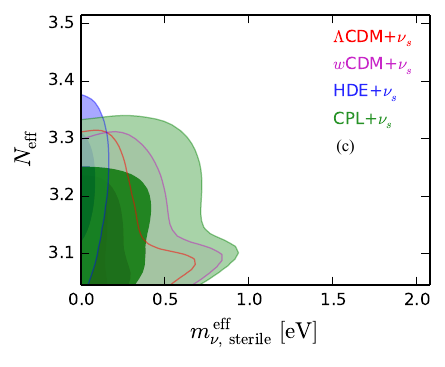}
\centering
\caption{Two-dimensional marginalized posterior contours (1$\sigma$ and 2$\sigma$) in the 
$m_{\nu,{\rm{sterile}}}^{\rm{eff}}$--$N_{\rm eff}$ plane for the $\Lambda$CDM+$\nu_s$, $w$CDM+$\nu_s$, HDE+$\nu_s$, and CPL+$\nu_s$ models using the CBS (a), CBS+ET (b), and CBS+ET2CE (c) data combinations, respectively.}\label{figDE}
\end{figure*}
%%%%%%%%%%%%%%%%%%%%%%%%%%%%%%%%%%%%%%%%%%%%%%%%%%%%%%%%%%%%%%%%

\section{Results and discussion}\label{sec3}
In this section, we present the fitting results for the $w$CDM+$\nu_s$, HDE+$\nu_s$, and CPL+$\nu_s$ models, respectively. We constrain these cosmological models using future GW standard siren observations. It is found that approximately 400 and 640 standard sirens can be detected over 10 years by the ET+THESEUS and ET2CE+THESEUS networks, respectively, with their redshift distributions provided in Ref. \cite{Feng:2024mfx}.
The fitting results are presented in Figs. \ref{figwcdm}--\ref{figDE} as well as Table~\ref{tabDE}. Note that we quote $\pm 1\sigma$ errors for the parameters, but for the parameters $N_{\rm eff}$ and $m_{\nu,{\rm{sterile}}}^{\rm{eff}}$, when central values cannot be determined, we instead quote the $2\sigma$ upper limits.
We use $\sigma(\xi)$ and $\varepsilon(\xi)=\sigma(\xi)/\xi$ to represent the absolute error and the relative error of a parameter $\xi$, respectively. 
For direct comparison, we also reproduce the constraint of the $\Lambda$CDM+$\nu_s$ \cite{Feng:2024mfx} under the CBS, CBS+ET and CBS+ET2CE data combinations in Table~\ref{tablcdm}.

First, we examine how the simulated GW data and the properties of DE affect the constraints on sterile neutrino parameters.

In Table~\ref{tabDE}, using CBS data, we obtain $N_{\rm eff}<3.340$ and $m_{\nu,{\rm{sterile}}}^{\rm{eff}}<0.651$ eV for $w$CDM+$\nu_s$, $N_{\rm eff}<3.562$ and $m_{\nu,{\rm{sterile}}}^{\rm{eff}}<0.479$ eV for HDE+$\nu_s$, $N_{\rm eff}<3.404$ and $m_{\nu,{\rm{sterile}}}^{\rm{eff}}<0.707$ eV for CPL+$\nu_s$.
Using CBS+ET data, we obtain $N_{\rm eff}=3.142^{+0.032}_{-0.091}$ and $m_{\nu,{\rm{sterile}}}^{\rm{eff}}<0.565$ eV for $w$CDM+$\nu_s$, $N_{\rm eff}=3.217^{+0.068}_{-0.062}$ and $m_{\nu,{\rm{sterile}}}^{\rm{eff}}<0.121$ eV for HDE+$\nu_s$, $N_{\rm eff}=3.165^{+0.045}_{-0.103}$ and $m_{\nu,{\rm{sterile}}}^{\rm{eff}}<0.691$ eV for CPL+$\nu_s$.
Using CBS+ET2CE data, we obtain $N_{\rm eff}=3.143^{+0.035}_{-0.087}$ and $m_{\nu,{\rm{sterile}}}^{\rm{eff}}<0.564$ eV for $w$CDM+$\nu_s$, $N_{\rm eff}=3.219\pm0.058$ and $m_{\nu,{\rm{sterile}}}^{\rm{eff}}<0.115$ eV for HDE+$\nu_s$, $N_{\rm eff}=3.162^{+0.044}_{-0.098}$ and $m_{\nu,{\rm{sterile}}}^{\rm{eff}}<0.680$ eV for CPL+$\nu_s$.

We find that the CBS dataset alone does not enable a precise determination of either parameter $N_{\rm eff}$ or $m_{\nu,{\rm{sterile}}}^{\rm{eff}}$, regardless of the dynamical DE model considered; only upper limits can be obtained for both parameters.
However, when GW standard sirens are incorporated into the data combination, the constraints on $N_{\rm eff}$ and $m_{\nu,{\rm{sterile}}}^{\rm{eff}}$ are significantly improved across all dynamical DE models.

The preference for $\Delta N_{\rm eff}>0$ reaches approximately $1.1\sigma$ (CBS+ET and CBS+ET2CE) for the $w$CDM+$\nu_s$ model, $2.8\sigma$ (CBS+ET) and $3.0\sigma$ (CBS+ET2CE) for the HDE+$\nu_s$ model, and $1.2\sigma$ (CBS+ET and CBS+ET2CE) for the CPL+$\nu_s$ model. Moreover, the inclusion of GW data helps reduce the $2\sigma$ upper limit on $m_{\nu,{\rm{sterile}}}^{\rm{eff}}$ by 13.2\%--13.4\%, 74.7\%--76.0\%, and 2.3\%--3.8\% for the $w$CDM+$\nu_s$, HDE+$\nu_s$, and CPL+$\nu_s$ models, respectively. These results clearly demonstrate that GW observations can significantly tighten the constraints on both $N_{\rm eff}$ and $m_{\nu,{\rm{sterile}}}^{\rm{eff}}$ (see also Figs.~\ref{figwcdm}--\ref{figcpl}).
This finding is consistent with previous studies on sterile neutrinos in the $\Lambda$CDM model using GW data~\cite{Feng:2024mfx}.

To explicitly illustrate how the properties of DE influence the constraints on sterile neutrino parameters, Fig.~\ref{figDE} shows the joint constraints on the $\Lambda$CDM+$\nu_s$, $w$CDM+$\nu_s$, HDE+$\nu_s$, and CPL+$\nu_s$ models in the $m_{\nu,{\rm{sterile}}}^{\rm{eff}}$--$N_{\rm eff}$ plane, derived from the CBS, CBS+ET, and CBS+ET2CE data combinations. We find that,  compared to the $\Lambda$CDM+$\nu_s$ model (see also Table~\ref{tablcdm}), in the $w$CDM+$\nu_s$ and CPL+$\nu_s$ models the limits on $m_{\nu,{\rm{sterile}}}^{\rm{eff}}$ become much looser, but in the HDE+$\nu_s$ model the limits become much more stringent. This clearly indicates that the DE properties can evidently affect the upper limits on $m_{\nu,{\rm{sterile}}}^{\rm{eff}}$. 

%This is because the HDE model makes DE more important in the early universe than the $w$CDM and CPL models. This stronger early-time influence enhances the sensitivity of the HDE model to additional radiation components, such as sterile neutrinos. As a result, it gives tighter constraints on $N_{\rm eff}$ and $m_{\nu,{\rm{sterile}}}^{\rm{eff}}$. In contrast, the $w$CDM and CPL models mainly affect the universe at late times, when sterile neutrinos have less impact. So, the constraints on the sterile neutrino mass in these two models become weaker.

Therefore, both GW data and the properties of DE play an important role in constraining the sterile neutrino parameters, with GW observations significantly enhancing the precision of these constraints.

Next, we discuss how the GW data help to improve the constraint accuracies for other cosmological parameters.     

For the $w$CDM+$\nu_s$ model, Fig.~\ref{figwcdm} presents the constraint results from the CBS, CBS+ET, and CBS+ET2CE combinations. It is evident that incorporating GW data significantly improves the constraints on $\Omega_m$, $H_0$, and $w$. Specifically, compared to CBS alone, the inclusion of ET or ET2CE data improves the constraints on $\Omega_m$, $H_0$, and $w$ by 46.2\%, 93.4\%, and 55.6\%, respectively.

For the HDE+$\nu_s$ model, Fig.~\ref{fighde} presents the corresponding constraint results. The inclusion of GW data also leads to notable improvements in the constraints on $\Omega_m$, $H_0$, and the model parameter $c$. Specifically, compared to CBS alone, the inclusion of ET data improves the constraints on $\Omega_m$, $H_0$, and $c$ by 48.1\%, 93.4\% and 51.6\%, respectively. When ET2CE data are included, the corresponding improvements reach 50.6\%, 93.5\% and 51.6\%, respectively.

For the CPL+$\nu_s$ model, Fig.~\ref{figcpl} shows the constraints from the CBS, CBS+ET, and CBS+ET2CE combinations. It is clear that incorporating GW data significantly improves the constraints on $\Omega_m$, $H_0$, $w_0$, and $w_a$.  With the inclusion of ET data, the constraints on $\Omega_m$, $H_0$, $w_0$, and $w_a$ are improved by 39.5\%, 91.2\%, 64.4\%, and 53.8\%, respectively, and by 40.7\%, 90.9\%, 64.4\%, and 53.8\% with the ET2CE data.

These results demonstrate that incorporating GW data significantly enhances the precision of cosmological parameter constraints, including $\Omega_m$, $H_0$, $w$, $c$, $w_0$, and $w_a$.

\section{Conclusion}\label{sec4}
This work aims to forecast the potential for detecting sterile neutrinos in dynamical DE cosmologies using joint GW-GRB observations. We consider three representative models, the $w$CDM, HDE, and CPL models, to investigate how future GW data may improve constraints on sterile neutrino and other cosmological parameters. Two GW observation strategies are explored: the ET alone and the ET2CE network. To evaluate the impact of GW data, we also incorporate existing CMB, BAO, and SN data for comparative and joint analyses.

We find that the properties of DE could significantly influence the constraints on the sterile neutrino parameters. Compared to the $\Lambda$CDM model, in the HDE model the limits on $m_{\nu,{\rm{sterile}}}^{\rm{eff}}$ become much more stringent, but in the $w$CDM model and the CPL model the limits become much looser. 

In addition, we find that the inclusion of GW data significantly improves the constraints on sterile neutrino parameters in dynamical DE models, compared to the current limits derived from CMB+BAO+SN observations. Specifically, the GW data significantly tighten the constraint on $N_{\rm eff}$ and indicate a preference for $\Delta N_{\rm eff}>0$ at approximately 1.1$\sigma$ in the $w$CDM model, 3.0$\sigma$ in the HDE model, and 1.2$\sigma$ in the CPL model. Moreover, the upper limits on $m_{\nu,{\rm{sterile}}}^{\rm{eff}}$ are reduced by 13.2\%--13.4\% for $w$CDM, 74.7\%--76.0\% for HDE, and 2.3\%--3.8\% for CPL, respectively. These results demonstrate that GW observations can substantially enhance the prospects for detecting sterile neutrinos.

Furthermore, we find that the GW data also play a significant role in improving the constraints on other cosmological parameters, including $\Omega_m$, $H_0$, and the EoS parameters of DE, namely, $w$ in the $w$CDM model, $c$ in the HDE model, and $w_0$ and $w_a$ in the CPL model.

These findings highlight that GW observations significantly improve the constraints on most cosmological parameters and enhance the prospects for detecting sterile neutrinos within various DE models. 
With upgrades to GW detectors, a substantial sample of standard sirens is expected to be produced, which will play a crucial role in further advancing sterile neutrino searches.

\begin{acknowledgments}
We thank Tian-Nuo Li and Guo-Hong Du for their helpful discussions. 
This work was supported by the National Natural Science Foundation of China (Grant Nos. 12305069, 11947022,  12533001, 12575049, and 12473001),
the National SKA Program of China (Grants Nos. 2022SKA0110200 and 2022SKA0110203),
the China Manned Space Program (Grant No. CMS-CSST-2025-A02),  the National 111 Project (Grant No. B16009),
and the Program of the Education Department of Liaoning Province (Grant No. JYTMS20231695).

\end{acknowledgments}

\bibliography{DEsterile_gwgrb}

\begin{thebibliography}{181}%
\makeatletter
\providecommand \@ifxundefined [1]{%
 \@ifx{#1\undefined}
}%
\providecommand \@ifnum [1]{%
 \ifnum #1\expandafter \@firstoftwo
 \else \expandafter \@secondoftwo
 \fi
}%
\providecommand \@ifx [1]{%
 \ifx #1\expandafter \@firstoftwo
 \else \expandafter \@secondoftwo
 \fi
}%
\providecommand \natexlab [1]{#1}%
\providecommand \enquote  [1]{``#1''}%
\providecommand \bibnamefont  [1]{#1}%
\providecommand \bibfnamefont [1]{#1}%
\providecommand \citenamefont [1]{#1}%
\providecommand \href@noop [0]{\@secondoftwo}%
\providecommand \href [0]{\begingroup \@sanitize@url \@href}%
\providecommand \@href[1]{\@@startlink{#1}\@@href}%
\providecommand \@@href[1]{\endgroup#1\@@endlink}%
\providecommand \@sanitize@url [0]{\catcode `\\12\catcode `\$12\catcode
  `\&12\catcode `\#12\catcode `\^12\catcode `\_12\catcode `\%12\relax}%
\providecommand \@@startlink[1]{}%
\providecommand \@@endlink[0]{}%
\providecommand \url  [0]{\begingroup\@sanitize@url \@url }%
\providecommand \@url [1]{\endgroup\@href {#1}{\urlprefix }}%
\providecommand \urlprefix  [0]{URL }%
\providecommand \Eprint [0]{\href }%
\providecommand \doibase [0]{http://dx.doi.org/}%
\providecommand \selectlanguage [0]{\@gobble}%
\providecommand \bibinfo  [0]{\@secondoftwo}%
\providecommand \bibfield  [0]{\@secondoftwo}%
\providecommand \translation [1]{[#1]}%
\providecommand \BibitemOpen [0]{}%
\providecommand \bibitemStop [0]{}%
\providecommand \bibitemNoStop [0]{.\EOS\space}%
\providecommand \EOS [0]{\spacefactor3000\relax}%
\providecommand \BibitemShut  [1]{\csname bibitem#1\endcsname}%
\let\auto@bib@innerbib\@empty
%</preamble>
\bibitem [{\citenamefont {Abbott}\ \emph
  {et~al.}(2017{\natexlab{a}})\citenamefont {Abbott} \emph
  {et~al.}}]{LIGOScientific:2017vwq}%
  \BibitemOpen
  \bibfield  {author} {\bibinfo {author} {\bibfnamefont {B.~P.}\ \bibnamefont
  {Abbott}} \emph {et~al.} (\bibinfo {collaboration} {LIGO Scientific,
  Virgo}),\ }\href {\doibase 10.1103/PhysRevLett.119.161101} {\bibfield
  {journal} {\bibinfo  {journal} {Phys. Rev. Lett.}\ }\textbf {\bibinfo
  {volume} {119}},\ \bibinfo {pages} {161101} (\bibinfo {year}
  {2017}{\natexlab{a}})},\ \Eprint {http://arxiv.org/abs/1710.05832}
  {arXiv:1710.05832 [gr-qc]} \BibitemShut {NoStop}%
\bibitem [{\citenamefont {Schutz}(1986)}]{Schutz:1986gp}%
  \BibitemOpen
  \bibfield  {author} {\bibinfo {author} {\bibfnamefont {B.~F.}\ \bibnamefont
  {Schutz}},\ }\href {\doibase 10.1038/323310a0} {\bibfield  {journal}
  {\bibinfo  {journal} {Nature}\ }\textbf {\bibinfo {volume} {323}},\ \bibinfo
  {pages} {310} (\bibinfo {year} {1986})}\BibitemShut {NoStop}%
\bibitem [{\citenamefont {Holz}\ and\ \citenamefont
  {Hughes}(2005)}]{Holz:2005df}%
  \BibitemOpen
  \bibfield  {author} {\bibinfo {author} {\bibfnamefont {D.~E.}\ \bibnamefont
  {Holz}}\ and\ \bibinfo {author} {\bibfnamefont {S.~A.}\ \bibnamefont
  {Hughes}},\ }\href {\doibase 10.1086/431341} {\bibfield  {journal} {\bibinfo
  {journal} {Astrophys. J.}\ }\textbf {\bibinfo {volume} {629}},\ \bibinfo
  {pages} {15} (\bibinfo {year} {2005})},\ \Eprint
  {http://arxiv.org/abs/astro-ph/0504616} {arXiv:astro-ph/0504616} \BibitemShut
  {NoStop}%
\bibitem [{\citenamefont {Dalal}\ \emph {et~al.}(2006)\citenamefont {Dalal},
  \citenamefont {Holz}, \citenamefont {Hughes},\ and\ \citenamefont
  {Jain}}]{Dalal:2006qt}%
  \BibitemOpen
  \bibfield  {author} {\bibinfo {author} {\bibfnamefont {N.}~\bibnamefont
  {Dalal}}, \bibinfo {author} {\bibfnamefont {D.~E.}\ \bibnamefont {Holz}},
  \bibinfo {author} {\bibfnamefont {S.~A.}\ \bibnamefont {Hughes}}, \ and\
  \bibinfo {author} {\bibfnamefont {B.}~\bibnamefont {Jain}},\ }\href {\doibase
  10.1103/PhysRevD.74.063006} {\bibfield  {journal} {\bibinfo  {journal} {Phys.
  Rev. D}\ }\textbf {\bibinfo {volume} {74}},\ \bibinfo {pages} {063006}
  (\bibinfo {year} {2006})},\ \Eprint {http://arxiv.org/abs/astro-ph/0601275}
  {arXiv:astro-ph/0601275} \BibitemShut {NoStop}%
\bibitem [{\citenamefont {Nissanke}\ \emph {et~al.}(2010)\citenamefont
  {Nissanke}, \citenamefont {Holz}, \citenamefont {Hughes}, \citenamefont
  {Dalal},\ and\ \citenamefont {Sievers}}]{Nissanke:2009kt}%
  \BibitemOpen
  \bibfield  {author} {\bibinfo {author} {\bibfnamefont {S.}~\bibnamefont
  {Nissanke}}, \bibinfo {author} {\bibfnamefont {D.~E.}\ \bibnamefont {Holz}},
  \bibinfo {author} {\bibfnamefont {S.~A.}\ \bibnamefont {Hughes}}, \bibinfo
  {author} {\bibfnamefont {N.}~\bibnamefont {Dalal}}, \ and\ \bibinfo {author}
  {\bibfnamefont {J.~L.}\ \bibnamefont {Sievers}},\ }\href {\doibase
  10.1088/0004-637X/725/1/496} {\bibfield  {journal} {\bibinfo  {journal}
  {Astrophys. J.}\ }\textbf {\bibinfo {volume} {725}},\ \bibinfo {pages} {496}
  (\bibinfo {year} {2010})},\ \Eprint {http://arxiv.org/abs/0904.1017}
  {arXiv:0904.1017 [astro-ph.CO]} \BibitemShut {NoStop}%
\bibitem [{\citenamefont {Cutler}\ and\ \citenamefont
  {Holz}(2009)}]{Cutler:2009qv}%
  \BibitemOpen
  \bibfield  {author} {\bibinfo {author} {\bibfnamefont {C.}~\bibnamefont
  {Cutler}}\ and\ \bibinfo {author} {\bibfnamefont {D.~E.}\ \bibnamefont
  {Holz}},\ }\href {\doibase 10.1103/PhysRevD.80.104009} {\bibfield  {journal}
  {\bibinfo  {journal} {Phys. Rev. D}\ }\textbf {\bibinfo {volume} {80}},\
  \bibinfo {pages} {104009} (\bibinfo {year} {2009})},\ \Eprint
  {http://arxiv.org/abs/0906.3752} {arXiv:0906.3752 [astro-ph.CO]} \BibitemShut
  {NoStop}%
\bibitem [{\citenamefont {Camera}\ and\ \citenamefont
  {Nishizawa}(2013)}]{Camera:2013xfa}%
  \BibitemOpen
  \bibfield  {author} {\bibinfo {author} {\bibfnamefont {S.}~\bibnamefont
  {Camera}}\ and\ \bibinfo {author} {\bibfnamefont {A.}~\bibnamefont
  {Nishizawa}},\ }\href {\doibase 10.1103/PhysRevLett.110.151103} {\bibfield
  {journal} {\bibinfo  {journal} {Phys. Rev. Lett.}\ }\textbf {\bibinfo
  {volume} {110}},\ \bibinfo {pages} {151103} (\bibinfo {year} {2013})},\
  \Eprint {http://arxiv.org/abs/1303.5446} {arXiv:1303.5446 [astro-ph.CO]}
  \BibitemShut {NoStop}%
\bibitem [{\citenamefont {Cai}\ and\ \citenamefont {Yang}(2017)}]{Cai:2016sby}%
  \BibitemOpen
  \bibfield  {author} {\bibinfo {author} {\bibfnamefont {R.-G.}\ \bibnamefont
  {Cai}}\ and\ \bibinfo {author} {\bibfnamefont {T.}~\bibnamefont {Yang}},\
  }\href {\doibase 10.1103/PhysRevD.95.044024} {\bibfield  {journal} {\bibinfo
  {journal} {Phys. Rev. D}\ }\textbf {\bibinfo {volume} {95}},\ \bibinfo
  {pages} {044024} (\bibinfo {year} {2017})},\ \Eprint
  {http://arxiv.org/abs/1608.08008} {arXiv:1608.08008 [astro-ph.CO]}
  \BibitemShut {NoStop}%
\bibitem [{\citenamefont {Cai}\ \emph {et~al.}(2017)\citenamefont {Cai},
  \citenamefont {Cao}, \citenamefont {Guo}, \citenamefont {Wang},\ and\
  \citenamefont {Yang}}]{Cai:2017cbj}%
  \BibitemOpen
  \bibfield  {author} {\bibinfo {author} {\bibfnamefont {R.-G.}\ \bibnamefont
  {Cai}}, \bibinfo {author} {\bibfnamefont {Z.}~\bibnamefont {Cao}}, \bibinfo
  {author} {\bibfnamefont {Z.-K.}\ \bibnamefont {Guo}}, \bibinfo {author}
  {\bibfnamefont {S.-J.}\ \bibnamefont {Wang}}, \ and\ \bibinfo {author}
  {\bibfnamefont {T.}~\bibnamefont {Yang}},\ }\href {\doibase
  10.1093/nsr/nwx029} {\bibfield  {journal} {\bibinfo  {journal} {Natl. Sci.
  Rev.}\ }\textbf {\bibinfo {volume} {4}},\ \bibinfo {pages} {687} (\bibinfo
  {year} {2017})},\ \Eprint {http://arxiv.org/abs/1703.00187} {arXiv:1703.00187
  [gr-qc]} \BibitemShut {NoStop}%
\bibitem [{\citenamefont {Cai}\ \emph {et~al.}(2018)\citenamefont {Cai},
  \citenamefont {Liu}, \citenamefont {Liu}, \citenamefont {Wang},\ and\
  \citenamefont {Yang}}]{Cai:2017aea}%
  \BibitemOpen
  \bibfield  {author} {\bibinfo {author} {\bibfnamefont {R.-G.}\ \bibnamefont
  {Cai}}, \bibinfo {author} {\bibfnamefont {T.-B.}\ \bibnamefont {Liu}},
  \bibinfo {author} {\bibfnamefont {X.-W.}\ \bibnamefont {Liu}}, \bibinfo
  {author} {\bibfnamefont {S.-J.}\ \bibnamefont {Wang}}, \ and\ \bibinfo
  {author} {\bibfnamefont {T.}~\bibnamefont {Yang}},\ }\href {\doibase
  10.1103/PhysRevD.97.103005} {\bibfield  {journal} {\bibinfo  {journal} {Phys.
  Rev. D}\ }\textbf {\bibinfo {volume} {97}},\ \bibinfo {pages} {103005}
  (\bibinfo {year} {2018})},\ \Eprint {http://arxiv.org/abs/1712.00952}
  {arXiv:1712.00952 [astro-ph.CO]} \BibitemShut {NoStop}%
\bibitem [{\citenamefont {Chen}\ \emph {et~al.}(2018)\citenamefont {Chen},
  \citenamefont {Fishbach},\ and\ \citenamefont {Holz}}]{Chen:2017rfc}%
  \BibitemOpen
  \bibfield  {author} {\bibinfo {author} {\bibfnamefont {H.-Y.}\ \bibnamefont
  {Chen}}, \bibinfo {author} {\bibfnamefont {M.}~\bibnamefont {Fishbach}}, \
  and\ \bibinfo {author} {\bibfnamefont {D.~E.}\ \bibnamefont {Holz}},\ }\href
  {\doibase 10.1038/s41586-018-0606-0} {\bibfield  {journal} {\bibinfo
  {journal} {Nature}\ }\textbf {\bibinfo {volume} {562}},\ \bibinfo {pages}
  {545} (\bibinfo {year} {2018})},\ \Eprint {http://arxiv.org/abs/1712.06531}
  {arXiv:1712.06531 [astro-ph.CO]} \BibitemShut {NoStop}%
\bibitem [{\citenamefont {Wang}\ \emph
  {et~al.}(2018{\natexlab{a}})\citenamefont {Wang}, \citenamefont {Zhang},
  \citenamefont {Zhang},\ and\ \citenamefont {Zhang}}]{Wang:2018lun}%
  \BibitemOpen
  \bibfield  {author} {\bibinfo {author} {\bibfnamefont {L.-F.}\ \bibnamefont
  {Wang}}, \bibinfo {author} {\bibfnamefont {X.-N.}\ \bibnamefont {Zhang}},
  \bibinfo {author} {\bibfnamefont {J.-F.}\ \bibnamefont {Zhang}}, \ and\
  \bibinfo {author} {\bibfnamefont {X.}~\bibnamefont {Zhang}},\ }\href
  {\doibase 10.1016/j.physletb.2018.05.027} {\bibfield  {journal} {\bibinfo
  {journal} {Phys. Lett. B}\ }\textbf {\bibinfo {volume} {782}},\ \bibinfo
  {pages} {87} (\bibinfo {year} {2018}{\natexlab{a}})},\ \Eprint
  {http://arxiv.org/abs/1802.04720} {arXiv:1802.04720 [astro-ph.CO]}
  \BibitemShut {NoStop}%
\bibitem [{\citenamefont {Vitale}\ and\ \citenamefont
  {Chen}(2018)}]{Vitale:2018wlg}%
  \BibitemOpen
  \bibfield  {author} {\bibinfo {author} {\bibfnamefont {S.}~\bibnamefont
  {Vitale}}\ and\ \bibinfo {author} {\bibfnamefont {H.-Y.}\ \bibnamefont
  {Chen}},\ }\href {\doibase 10.1103/PhysRevLett.121.021303} {\bibfield
  {journal} {\bibinfo  {journal} {Phys. Rev. Lett.}\ }\textbf {\bibinfo
  {volume} {121}},\ \bibinfo {pages} {021303} (\bibinfo {year} {2018})},\
  \Eprint {http://arxiv.org/abs/1804.07337} {arXiv:1804.07337 [astro-ph.CO]}
  \BibitemShut {NoStop}%
\bibitem [{\citenamefont {Belgacem}\ \emph {et~al.}(2019)\citenamefont
  {Belgacem}, \citenamefont {Dirian}, \citenamefont {Foffa}, \citenamefont
  {Howell}, \citenamefont {Maggiore},\ and\ \citenamefont
  {Regimbau}}]{Belgacem:2019tbw}%
  \BibitemOpen
  \bibfield  {author} {\bibinfo {author} {\bibfnamefont {E.}~\bibnamefont
  {Belgacem}}, \bibinfo {author} {\bibfnamefont {Y.}~\bibnamefont {Dirian}},
  \bibinfo {author} {\bibfnamefont {S.}~\bibnamefont {Foffa}}, \bibinfo
  {author} {\bibfnamefont {E.~J.}\ \bibnamefont {Howell}}, \bibinfo {author}
  {\bibfnamefont {M.}~\bibnamefont {Maggiore}}, \ and\ \bibinfo {author}
  {\bibfnamefont {T.}~\bibnamefont {Regimbau}},\ }\href {\doibase
  10.1088/1475-7516/2019/08/015} {\bibfield  {journal} {\bibinfo  {journal}
  {JCAP}\ }\textbf {\bibinfo {volume} {08}},\ \bibinfo {pages} {015} (\bibinfo
  {year} {2019})},\ \Eprint {http://arxiv.org/abs/1907.01487} {arXiv:1907.01487
  [astro-ph.CO]} \BibitemShut {NoStop}%
\bibitem [{\citenamefont {D'Agostino}\ and\ \citenamefont
  {Nunes}(2019)}]{DAgostino:2019hvh}%
  \BibitemOpen
  \bibfield  {author} {\bibinfo {author} {\bibfnamefont {R.}~\bibnamefont
  {D'Agostino}}\ and\ \bibinfo {author} {\bibfnamefont {R.~C.}\ \bibnamefont
  {Nunes}},\ }\href {\doibase 10.1103/PhysRevD.100.044041} {\bibfield
  {journal} {\bibinfo  {journal} {Phys. Rev. D}\ }\textbf {\bibinfo {volume}
  {100}},\ \bibinfo {pages} {044041} (\bibinfo {year} {2019})},\ \Eprint
  {http://arxiv.org/abs/1907.05516} {arXiv:1907.05516 [gr-qc]} \BibitemShut
  {NoStop}%
\bibitem [{\citenamefont {Howlett}\ and\ \citenamefont
  {Davis}(2020)}]{Howlett:2019mdh}%
  \BibitemOpen
  \bibfield  {author} {\bibinfo {author} {\bibfnamefont {C.}~\bibnamefont
  {Howlett}}\ and\ \bibinfo {author} {\bibfnamefont {T.~M.}\ \bibnamefont
  {Davis}},\ }\href {\doibase 10.1093/mnras/staa049} {\bibfield  {journal}
  {\bibinfo  {journal} {Mon. Not. Roy. Astron. Soc.}\ }\textbf {\bibinfo
  {volume} {492}},\ \bibinfo {pages} {3803} (\bibinfo {year} {2020})},\ \Eprint
  {http://arxiv.org/abs/1909.00587} {arXiv:1909.00587 [astro-ph.CO]}
  \BibitemShut {NoStop}%
\bibitem [{\citenamefont {Zhao}\ \emph {et~al.}(2020)\citenamefont {Zhao},
  \citenamefont {Wang}, \citenamefont {Zhang},\ and\ \citenamefont
  {Zhang}}]{Zhao:2019gyk}%
  \BibitemOpen
  \bibfield  {author} {\bibinfo {author} {\bibfnamefont {Z.-W.}\ \bibnamefont
  {Zhao}}, \bibinfo {author} {\bibfnamefont {L.-F.}\ \bibnamefont {Wang}},
  \bibinfo {author} {\bibfnamefont {J.-F.}\ \bibnamefont {Zhang}}, \ and\
  \bibinfo {author} {\bibfnamefont {X.}~\bibnamefont {Zhang}},\ }\href
  {\doibase 10.1016/j.scib.2020.04.032} {\bibfield  {journal} {\bibinfo
  {journal} {Sci. Bull.}\ }\textbf {\bibinfo {volume} {65}},\ \bibinfo {pages}
  {1340} (\bibinfo {year} {2020})},\ \Eprint {http://arxiv.org/abs/1912.11629}
  {arXiv:1912.11629 [astro-ph.CO]} \BibitemShut {NoStop}%
\bibitem [{\citenamefont {Borhanian}\ \emph {et~al.}(2020)\citenamefont
  {Borhanian}, \citenamefont {Dhani}, \citenamefont {Gupta}, \citenamefont
  {Arun},\ and\ \citenamefont {Sathyaprakash}}]{Borhanian:2020vyr}%
  \BibitemOpen
  \bibfield  {author} {\bibinfo {author} {\bibfnamefont {S.}~\bibnamefont
  {Borhanian}}, \bibinfo {author} {\bibfnamefont {A.}~\bibnamefont {Dhani}},
  \bibinfo {author} {\bibfnamefont {A.}~\bibnamefont {Gupta}}, \bibinfo
  {author} {\bibfnamefont {K.~G.}\ \bibnamefont {Arun}}, \ and\ \bibinfo
  {author} {\bibfnamefont {B.~S.}\ \bibnamefont {Sathyaprakash}},\ }\href
  {\doibase 10.3847/2041-8213/abcaf5} {\bibfield  {journal} {\bibinfo
  {journal} {Astrophys. J. Lett.}\ }\textbf {\bibinfo {volume} {905}},\
  \bibinfo {pages} {L28} (\bibinfo {year} {2020})},\ \Eprint
  {http://arxiv.org/abs/2007.02883} {arXiv:2007.02883 [astro-ph.CO]}
  \BibitemShut {NoStop}%
\bibitem [{\citenamefont {Qi}\ \emph {et~al.}(2021)\citenamefont {Qi},
  \citenamefont {Jin}, \citenamefont {Fan}, \citenamefont {Zhang},\ and\
  \citenamefont {Zhang}}]{Qi:2021iic}%
  \BibitemOpen
  \bibfield  {author} {\bibinfo {author} {\bibfnamefont {J.-Z.}\ \bibnamefont
  {Qi}}, \bibinfo {author} {\bibfnamefont {S.-J.}\ \bibnamefont {Jin}},
  \bibinfo {author} {\bibfnamefont {X.-L.}\ \bibnamefont {Fan}}, \bibinfo
  {author} {\bibfnamefont {J.-F.}\ \bibnamefont {Zhang}}, \ and\ \bibinfo
  {author} {\bibfnamefont {X.}~\bibnamefont {Zhang}},\ }\href {\doibase
  10.1088/1475-7516/2021/12/042} {\bibfield  {journal} {\bibinfo  {journal}
  {JCAP}\ }\textbf {\bibinfo {volume} {12}},\ \bibinfo {pages} {042} (\bibinfo
  {year} {2021})},\ \Eprint {http://arxiv.org/abs/2102.01292} {arXiv:2102.01292
  [astro-ph.CO]} \BibitemShut {NoStop}%
\bibitem [{\citenamefont {Yu}\ \emph {et~al.}(2021)\citenamefont {Yu},
  \citenamefont {Song}, \citenamefont {Ai}, \citenamefont {Gao}, \citenamefont
  {Wang}, \citenamefont {Wang}, \citenamefont {Lu}, \citenamefont {Fang},\ and\
  \citenamefont {Zhao}}]{Yu:2021nvx}%
  \BibitemOpen
  \bibfield  {author} {\bibinfo {author} {\bibfnamefont {J.}~\bibnamefont
  {Yu}}, \bibinfo {author} {\bibfnamefont {H.}~\bibnamefont {Song}}, \bibinfo
  {author} {\bibfnamefont {S.}~\bibnamefont {Ai}}, \bibinfo {author}
  {\bibfnamefont {H.}~\bibnamefont {Gao}}, \bibinfo {author} {\bibfnamefont
  {F.}~\bibnamefont {Wang}}, \bibinfo {author} {\bibfnamefont {Y.}~\bibnamefont
  {Wang}}, \bibinfo {author} {\bibfnamefont {Y.}~\bibnamefont {Lu}}, \bibinfo
  {author} {\bibfnamefont {W.}~\bibnamefont {Fang}}, \ and\ \bibinfo {author}
  {\bibfnamefont {W.}~\bibnamefont {Zhao}},\ }\href {\doibase
  10.3847/1538-4357/ac0628} {\bibfield  {journal} {\bibinfo  {journal}
  {Astrophys. J.}\ }\textbf {\bibinfo {volume} {916}},\ \bibinfo {pages} {54}
  (\bibinfo {year} {2021})},\ \Eprint {http://arxiv.org/abs/2104.12374}
  {arXiv:2104.12374 [astro-ph.HE]} \BibitemShut {NoStop}%
\bibitem [{\citenamefont {Zhu}\ \emph {et~al.}(2022)\citenamefont {Zhu},
  \citenamefont {Xie}, \citenamefont {Hu}, \citenamefont {Liu}, \citenamefont
  {Li}, \citenamefont {Napolitano}, \citenamefont {Tang}, \citenamefont
  {Zhang},\ and\ \citenamefont {Mei}}]{Zhu:2021bpp}%
  \BibitemOpen
  \bibfield  {author} {\bibinfo {author} {\bibfnamefont {L.-G.}\ \bibnamefont
  {Zhu}}, \bibinfo {author} {\bibfnamefont {L.-H.}\ \bibnamefont {Xie}},
  \bibinfo {author} {\bibfnamefont {Y.-M.}\ \bibnamefont {Hu}}, \bibinfo
  {author} {\bibfnamefont {S.}~\bibnamefont {Liu}}, \bibinfo {author}
  {\bibfnamefont {E.-K.}\ \bibnamefont {Li}}, \bibinfo {author} {\bibfnamefont
  {N.~R.}\ \bibnamefont {Napolitano}}, \bibinfo {author} {\bibfnamefont
  {B.-T.}\ \bibnamefont {Tang}}, \bibinfo {author} {\bibfnamefont {J.-d.}\
  \bibnamefont {Zhang}}, \ and\ \bibinfo {author} {\bibfnamefont
  {J.}~\bibnamefont {Mei}},\ }\href {\doibase 10.1007/s11433-021-1859-9}
  {\bibfield  {journal} {\bibinfo  {journal} {Sci. China Phys. Mech. Astron.}\
  }\textbf {\bibinfo {volume} {65}},\ \bibinfo {pages} {259811} (\bibinfo
  {year} {2022})},\ \Eprint {http://arxiv.org/abs/2110.05224} {arXiv:2110.05224
  [astro-ph.CO]} \BibitemShut {NoStop}%
\bibitem [{\citenamefont {Zheng}\ \emph {et~al.}(2022)\citenamefont {Zheng},
  \citenamefont {Chen}, \citenamefont {Xu},\ and\ \citenamefont
  {Zhu}}]{Zheng:2022gfi}%
  \BibitemOpen
  \bibfield  {author} {\bibinfo {author} {\bibfnamefont {J.}~\bibnamefont
  {Zheng}}, \bibinfo {author} {\bibfnamefont {Y.}~\bibnamefont {Chen}},
  \bibinfo {author} {\bibfnamefont {T.}~\bibnamefont {Xu}}, \ and\ \bibinfo
  {author} {\bibfnamefont {Z.-H.}\ \bibnamefont {Zhu}},\ }\href {\doibase
  10.1140/epjp/s13360-022-02718-3} {\bibfield  {journal} {\bibinfo  {journal}
  {Eur. Phys. J. Plus}\ }\textbf {\bibinfo {volume} {137}},\ \bibinfo {pages}
  {509} (\bibinfo {year} {2022})},\ \Eprint {http://arxiv.org/abs/2201.07011}
  {arXiv:2201.07011 [astro-ph.CO]} \BibitemShut {NoStop}%
\bibitem [{\citenamefont {Ezquiaga}\ and\ \citenamefont
  {Holz}(2022)}]{Ezquiaga:2022zkx}%
  \BibitemOpen
  \bibfield  {author} {\bibinfo {author} {\bibfnamefont {J.~M.}\ \bibnamefont
  {Ezquiaga}}\ and\ \bibinfo {author} {\bibfnamefont {D.~E.}\ \bibnamefont
  {Holz}},\ }\href {\doibase 10.1103/PhysRevLett.129.061102} {\bibfield
  {journal} {\bibinfo  {journal} {Phys. Rev. Lett.}\ }\textbf {\bibinfo
  {volume} {129}},\ \bibinfo {pages} {061102} (\bibinfo {year} {2022})},\
  \Eprint {http://arxiv.org/abs/2202.08240} {arXiv:2202.08240 [astro-ph.CO]}
  \BibitemShut {NoStop}%
\bibitem [{\citenamefont {Jin}\ \emph {et~al.}(2022)\citenamefont {Jin},
  \citenamefont {Zhu}, \citenamefont {Wang}, \citenamefont {Li}, \citenamefont
  {Zhang},\ and\ \citenamefont {Zhang}}]{Jin:2022tdf}%
  \BibitemOpen
  \bibfield  {author} {\bibinfo {author} {\bibfnamefont {S.-J.}\ \bibnamefont
  {Jin}}, \bibinfo {author} {\bibfnamefont {R.-Q.}\ \bibnamefont {Zhu}},
  \bibinfo {author} {\bibfnamefont {L.-F.}\ \bibnamefont {Wang}}, \bibinfo
  {author} {\bibfnamefont {H.-L.}\ \bibnamefont {Li}}, \bibinfo {author}
  {\bibfnamefont {J.-F.}\ \bibnamefont {Zhang}}, \ and\ \bibinfo {author}
  {\bibfnamefont {X.}~\bibnamefont {Zhang}},\ }\href {\doibase
  10.1088/1572-9494/ac7b76} {\bibfield  {journal} {\bibinfo  {journal} {Commun.
  Theor. Phys.}\ }\textbf {\bibinfo {volume} {74}},\ \bibinfo {pages} {105404}
  (\bibinfo {year} {2022})},\ \Eprint {http://arxiv.org/abs/2204.04689}
  {arXiv:2204.04689 [astro-ph.CO]} \BibitemShut {NoStop}%
\bibitem [{\citenamefont {Auclair}\ \emph {et~al.}(2023)\citenamefont {Auclair}
  \emph {et~al.}}]{LISACosmologyWorkingGroup:2022jok}%
  \BibitemOpen
  \bibfield  {author} {\bibinfo {author} {\bibfnamefont {P.}~\bibnamefont
  {Auclair}} \emph {et~al.} (\bibinfo {collaboration} {LISA Cosmology Working
  Group}),\ }\href {\doibase 10.1007/s41114-023-00045-2} {\bibfield  {journal}
  {\bibinfo  {journal} {Living Rev. Rel.}\ }\textbf {\bibinfo {volume} {26}},\
  \bibinfo {pages} {5} (\bibinfo {year} {2023})},\ \Eprint
  {http://arxiv.org/abs/2204.05434} {arXiv:2204.05434 [astro-ph.CO]}
  \BibitemShut {NoStop}%
\bibitem [{\citenamefont {Hou}\ \emph {et~al.}(2023)\citenamefont {Hou},
  \citenamefont {Qi}, \citenamefont {Han}, \citenamefont {Zhang}, \citenamefont
  {Cao},\ and\ \citenamefont {Zhang}}]{Hou:2022rvk}%
  \BibitemOpen
  \bibfield  {author} {\bibinfo {author} {\bibfnamefont {W.-T.}\ \bibnamefont
  {Hou}}, \bibinfo {author} {\bibfnamefont {J.-Z.}\ \bibnamefont {Qi}},
  \bibinfo {author} {\bibfnamefont {T.}~\bibnamefont {Han}}, \bibinfo {author}
  {\bibfnamefont {J.-F.}\ \bibnamefont {Zhang}}, \bibinfo {author}
  {\bibfnamefont {S.}~\bibnamefont {Cao}}, \ and\ \bibinfo {author}
  {\bibfnamefont {X.}~\bibnamefont {Zhang}},\ }\href {\doibase
  10.1088/1475-7516/2023/05/017} {\bibfield  {journal} {\bibinfo  {journal}
  {JCAP}\ }\textbf {\bibinfo {volume} {05}},\ \bibinfo {pages} {017} (\bibinfo
  {year} {2023})},\ \Eprint {http://arxiv.org/abs/2211.10087} {arXiv:2211.10087
  [astro-ph.CO]} \BibitemShut {NoStop}%
\bibitem [{\citenamefont {Song}\ \emph {et~al.}(2024)\citenamefont {Song},
  \citenamefont {Wang}, \citenamefont {Li}, \citenamefont {Zhao}, \citenamefont
  {Zhang}, \citenamefont {Zhao},\ and\ \citenamefont {Zhang}}]{Song:2022siz}%
  \BibitemOpen
  \bibfield  {author} {\bibinfo {author} {\bibfnamefont {J.-Y.}\ \bibnamefont
  {Song}}, \bibinfo {author} {\bibfnamefont {L.-F.}\ \bibnamefont {Wang}},
  \bibinfo {author} {\bibfnamefont {Y.}~\bibnamefont {Li}}, \bibinfo {author}
  {\bibfnamefont {Z.-W.}\ \bibnamefont {Zhao}}, \bibinfo {author}
  {\bibfnamefont {J.-F.}\ \bibnamefont {Zhang}}, \bibinfo {author}
  {\bibfnamefont {W.}~\bibnamefont {Zhao}}, \ and\ \bibinfo {author}
  {\bibfnamefont {X.}~\bibnamefont {Zhang}},\ }\href {\doibase
  10.1007/s11433-023-2260-2} {\bibfield  {journal} {\bibinfo  {journal} {Sci.
  China Phys. Mech. Astron.}\ }\textbf {\bibinfo {volume} {67}},\ \bibinfo
  {pages} {230411} (\bibinfo {year} {2024})},\ \Eprint
  {http://arxiv.org/abs/2212.00531} {arXiv:2212.00531 [astro-ph.CO]}
  \BibitemShut {NoStop}%
\bibitem [{\citenamefont {Zhu}\ and\ \citenamefont {Chen}(2023)}]{Zhu:2023jti}%
  \BibitemOpen
  \bibfield  {author} {\bibinfo {author} {\bibfnamefont {L.-G.}\ \bibnamefont
  {Zhu}}\ and\ \bibinfo {author} {\bibfnamefont {X.}~\bibnamefont {Chen}},\
  }\href {\doibase 10.3847/1538-4357/acc24b} {\bibfield  {journal} {\bibinfo
  {journal} {Astrophys. J.}\ }\textbf {\bibinfo {volume} {948}},\ \bibinfo
  {pages} {26} (\bibinfo {year} {2023})},\ \Eprint
  {http://arxiv.org/abs/2302.10621} {arXiv:2302.10621 [astro-ph.CO]}
  \BibitemShut {NoStop}%
\bibitem [{\citenamefont {Muttoni}\ \emph {et~al.}(2023)\citenamefont
  {Muttoni}, \citenamefont {Laghi}, \citenamefont {Tamanini}, \citenamefont
  {Marsat},\ and\ \citenamefont {Izquierdo-Villalba}}]{Muttoni:2023prw}%
  \BibitemOpen
  \bibfield  {author} {\bibinfo {author} {\bibfnamefont {N.}~\bibnamefont
  {Muttoni}}, \bibinfo {author} {\bibfnamefont {D.}~\bibnamefont {Laghi}},
  \bibinfo {author} {\bibfnamefont {N.}~\bibnamefont {Tamanini}}, \bibinfo
  {author} {\bibfnamefont {S.}~\bibnamefont {Marsat}}, \ and\ \bibinfo {author}
  {\bibfnamefont {D.}~\bibnamefont {Izquierdo-Villalba}},\ }\href {\doibase
  10.1103/PhysRevD.108.043543} {\bibfield  {journal} {\bibinfo  {journal}
  {Phys. Rev. D}\ }\textbf {\bibinfo {volume} {108}},\ \bibinfo {pages}
  {043543} (\bibinfo {year} {2023})},\ \Eprint
  {http://arxiv.org/abs/2303.10693} {arXiv:2303.10693 [astro-ph.CO]}
  \BibitemShut {NoStop}%
\bibitem [{\citenamefont {Branchesi}\ \emph {et~al.}(2023)\citenamefont
  {Branchesi} \emph {et~al.}}]{Branchesi:2023mws}%
  \BibitemOpen
  \bibfield  {author} {\bibinfo {author} {\bibfnamefont {M.}~\bibnamefont
  {Branchesi}} \emph {et~al.},\ }\href {\doibase 10.1088/1475-7516/2023/07/068}
  {\bibfield  {journal} {\bibinfo  {journal} {JCAP}\ }\textbf {\bibinfo
  {volume} {07}},\ \bibinfo {pages} {068} (\bibinfo {year} {2023})},\ \Eprint
  {http://arxiv.org/abs/2303.15923} {arXiv:2303.15923 [gr-qc]} \BibitemShut
  {NoStop}%
\bibitem [{\citenamefont {Chen}\ \emph {et~al.}(2024)\citenamefont {Chen},
  \citenamefont {Talbot},\ and\ \citenamefont {Chase}}]{Chen:2023dgw}%
  \BibitemOpen
  \bibfield  {author} {\bibinfo {author} {\bibfnamefont {H.-Y.}\ \bibnamefont
  {Chen}}, \bibinfo {author} {\bibfnamefont {C.}~\bibnamefont {Talbot}}, \ and\
  \bibinfo {author} {\bibfnamefont {E.~A.}\ \bibnamefont {Chase}},\ }\href
  {\doibase 10.1103/PhysRevLett.132.191003} {\bibfield  {journal} {\bibinfo
  {journal} {Phys. Rev. Lett.}\ }\textbf {\bibinfo {volume} {132}},\ \bibinfo
  {pages} {191003} (\bibinfo {year} {2024})},\ \Eprint
  {http://arxiv.org/abs/2307.10402} {arXiv:2307.10402 [astro-ph.CO]}
  \BibitemShut {NoStop}%
\bibitem [{\citenamefont {Han}\ \emph {et~al.}(2024)\citenamefont {Han},
  \citenamefont {Jin}, \citenamefont {Zhang},\ and\ \citenamefont
  {Zhang}}]{Han:2023exn}%
  \BibitemOpen
  \bibfield  {author} {\bibinfo {author} {\bibfnamefont {T.}~\bibnamefont
  {Han}}, \bibinfo {author} {\bibfnamefont {S.-J.}\ \bibnamefont {Jin}},
  \bibinfo {author} {\bibfnamefont {J.-F.}\ \bibnamefont {Zhang}}, \ and\
  \bibinfo {author} {\bibfnamefont {X.}~\bibnamefont {Zhang}},\ }\href
  {\doibase 10.1140/epjc/s10052-024-12999-w} {\bibfield  {journal} {\bibinfo
  {journal} {Eur. Phys. J. C}\ }\textbf {\bibinfo {volume} {84}},\ \bibinfo
  {pages} {663} (\bibinfo {year} {2024})},\ \Eprint
  {http://arxiv.org/abs/2309.14965} {arXiv:2309.14965 [astro-ph.CO]}
  \BibitemShut {NoStop}%
\bibitem [{\citenamefont {Li}\ \emph {et~al.}(2024{\natexlab{a}})\citenamefont
  {Li}, \citenamefont {Jin}, \citenamefont {Li}, \citenamefont {Zhang},\ and\
  \citenamefont {Zhang}}]{Li:2023gtu}%
  \BibitemOpen
  \bibfield  {author} {\bibinfo {author} {\bibfnamefont {T.-N.}\ \bibnamefont
  {Li}}, \bibinfo {author} {\bibfnamefont {S.-J.}\ \bibnamefont {Jin}},
  \bibinfo {author} {\bibfnamefont {H.-L.}\ \bibnamefont {Li}}, \bibinfo
  {author} {\bibfnamefont {J.-F.}\ \bibnamefont {Zhang}}, \ and\ \bibinfo
  {author} {\bibfnamefont {X.}~\bibnamefont {Zhang}},\ }\href {\doibase
  10.3847/1538-4357/ad1bc9} {\bibfield  {journal} {\bibinfo  {journal}
  {Astrophys. J.}\ }\textbf {\bibinfo {volume} {963}},\ \bibinfo {pages} {52}
  (\bibinfo {year} {2024}{\natexlab{a}})},\ \Eprint
  {http://arxiv.org/abs/2310.15879} {arXiv:2310.15879 [astro-ph.CO]}
  \BibitemShut {NoStop}%
\bibitem [{\citenamefont {Yu}\ \emph {et~al.}(2024)\citenamefont {Yu},
  \citenamefont {Liu}, \citenamefont {Yang}, \citenamefont {Wang},
  \citenamefont {Zhang}, \citenamefont {Zhang},\ and\ \citenamefont
  {Zhao}}]{Yu:2023ico}%
  \BibitemOpen
  \bibfield  {author} {\bibinfo {author} {\bibfnamefont {J.}~\bibnamefont
  {Yu}}, \bibinfo {author} {\bibfnamefont {Z.}~\bibnamefont {Liu}}, \bibinfo
  {author} {\bibfnamefont {X.}~\bibnamefont {Yang}}, \bibinfo {author}
  {\bibfnamefont {Y.}~\bibnamefont {Wang}}, \bibinfo {author} {\bibfnamefont
  {P.}~\bibnamefont {Zhang}}, \bibinfo {author} {\bibfnamefont
  {X.}~\bibnamefont {Zhang}}, \ and\ \bibinfo {author} {\bibfnamefont
  {W.}~\bibnamefont {Zhao}},\ }\href {\doibase 10.3847/1538-4365/ad0ece}
  {\bibfield  {journal} {\bibinfo  {journal} {Astrophys. J. Suppl.}\ }\textbf
  {\bibinfo {volume} {270}},\ \bibinfo {pages} {24} (\bibinfo {year} {2024})},\
  \Eprint {http://arxiv.org/abs/2311.11588} {arXiv:2311.11588 [astro-ph.HE]}
  \BibitemShut {NoStop}%
\bibitem [{\citenamefont {Dong}\ \emph {et~al.}(2025)\citenamefont {Dong},
  \citenamefont {Song}, \citenamefont {Jin}, \citenamefont {Zhang},\ and\
  \citenamefont {Zhang}}]{Dong:2024bvw}%
  \BibitemOpen
  \bibfield  {author} {\bibinfo {author} {\bibfnamefont {Y.-Y.}\ \bibnamefont
  {Dong}}, \bibinfo {author} {\bibfnamefont {J.-Y.}\ \bibnamefont {Song}},
  \bibinfo {author} {\bibfnamefont {S.-J.}\ \bibnamefont {Jin}}, \bibinfo
  {author} {\bibfnamefont {J.-F.}\ \bibnamefont {Zhang}}, \ and\ \bibinfo
  {author} {\bibfnamefont {X.}~\bibnamefont {Zhang}},\ }\href {\doibase
  10.1088/1475-7516/2025/05/046} {\bibfield  {journal} {\bibinfo  {journal}
  {JCAP}\ }\textbf {\bibinfo {volume} {05}},\ \bibinfo {pages} {046} (\bibinfo
  {year} {2025})},\ \Eprint {http://arxiv.org/abs/2404.18188} {arXiv:2404.18188
  [astro-ph.CO]} \BibitemShut {NoStop}%
\bibitem [{\citenamefont {Feng}\ \emph {et~al.}(2024)\citenamefont {Feng},
  \citenamefont {Han}, \citenamefont {Zhang},\ and\ \citenamefont
  {Zhang}}]{Feng:2024lzh}%
  \BibitemOpen
  \bibfield  {author} {\bibinfo {author} {\bibfnamefont {L.}~\bibnamefont
  {Feng}}, \bibinfo {author} {\bibfnamefont {T.}~\bibnamefont {Han}}, \bibinfo
  {author} {\bibfnamefont {J.-F.}\ \bibnamefont {Zhang}}, \ and\ \bibinfo
  {author} {\bibfnamefont {X.}~\bibnamefont {Zhang}},\ }\href {\doibase
  10.1088/1674-1137/ad5ae4} {\bibfield  {journal} {\bibinfo  {journal} {Chin.
  Phys. C}\ }\textbf {\bibinfo {volume} {48}},\ \bibinfo {pages} {095104}
  (\bibinfo {year} {2024})},\ \Eprint {http://arxiv.org/abs/2404.19530}
  {arXiv:2404.19530 [astro-ph.CO]} \BibitemShut {NoStop}%
\bibitem [{\citenamefont {Zheng}\ \emph {et~al.}(2024)\citenamefont {Zheng},
  \citenamefont {Liu},\ and\ \citenamefont {Qi}}]{Zheng:2024mbo}%
  \BibitemOpen
  \bibfield  {author} {\bibinfo {author} {\bibfnamefont {J.}~\bibnamefont
  {Zheng}}, \bibinfo {author} {\bibfnamefont {X.-H.}\ \bibnamefont {Liu}}, \
  and\ \bibinfo {author} {\bibfnamefont {J.-Z.}\ \bibnamefont {Qi}},\ }\href
  {\doibase 10.3847/1538-4357/ad7bb5} {\bibfield  {journal} {\bibinfo
  {journal} {Astrophys. J.}\ }\textbf {\bibinfo {volume} {975}},\ \bibinfo
  {pages} {215} (\bibinfo {year} {2024})},\ \Eprint
  {http://arxiv.org/abs/2407.05686} {arXiv:2407.05686 [astro-ph.CO]}
  \BibitemShut {NoStop}%
\bibitem [{\citenamefont {Xiao}\ \emph {et~al.}(2025)\citenamefont {Xiao},
  \citenamefont {Shao}, \citenamefont {Wang}, \citenamefont {Song},
  \citenamefont {Feng}, \citenamefont {Zhang},\ and\ \citenamefont
  {Zhang}}]{Xiao:2024nmi}%
  \BibitemOpen
  \bibfield  {author} {\bibinfo {author} {\bibfnamefont {S.-R.}\ \bibnamefont
  {Xiao}}, \bibinfo {author} {\bibfnamefont {Y.}~\bibnamefont {Shao}}, \bibinfo
  {author} {\bibfnamefont {L.-F.}\ \bibnamefont {Wang}}, \bibinfo {author}
  {\bibfnamefont {J.-Y.}\ \bibnamefont {Song}}, \bibinfo {author}
  {\bibfnamefont {L.}~\bibnamefont {Feng}}, \bibinfo {author} {\bibfnamefont
  {J.-F.}\ \bibnamefont {Zhang}}, \ and\ \bibinfo {author} {\bibfnamefont
  {X.}~\bibnamefont {Zhang}},\ }\href {\doibase 10.1088/1475-7516/2025/04/060}
  {\bibfield  {journal} {\bibinfo  {journal} {JCAP}\ }\textbf {\bibinfo
  {volume} {04}},\ \bibinfo {pages} {060} (\bibinfo {year} {2025})},\ \Eprint
  {http://arxiv.org/abs/2408.00609} {arXiv:2408.00609 [astro-ph.CO]}
  \BibitemShut {NoStop}%
\bibitem [{\citenamefont {Feng}\ \emph
  {et~al.}(2025{\natexlab{a}})\citenamefont {Feng}, \citenamefont {Han},
  \citenamefont {Zhang},\ and\ \citenamefont {Zhang}}]{Feng:2024mfx}%
  \BibitemOpen
  \bibfield  {author} {\bibinfo {author} {\bibfnamefont {L.}~\bibnamefont
  {Feng}}, \bibinfo {author} {\bibfnamefont {T.}~\bibnamefont {Han}}, \bibinfo
  {author} {\bibfnamefont {J.-F.}\ \bibnamefont {Zhang}}, \ and\ \bibinfo
  {author} {\bibfnamefont {X.}~\bibnamefont {Zhang}},\ }\href {\doibase
  10.1088/1572-9494/ad9c3e} {\bibfield  {journal} {\bibinfo  {journal} {Commun.
  Theor. Phys.}\ }\textbf {\bibinfo {volume} {77}},\ \bibinfo {pages} {065403}
  (\bibinfo {year} {2025}{\natexlab{a}})},\ \Eprint
  {http://arxiv.org/abs/2409.04453} {arXiv:2409.04453 [astro-ph.HE]}
  \BibitemShut {NoStop}%
\bibitem [{\citenamefont {Han}\ \emph {et~al.}(2025{\natexlab{a}})\citenamefont
  {Han}, \citenamefont {Li}, \citenamefont {Zhang},\ and\ \citenamefont
  {Zhang}}]{Han:2024sxm}%
  \BibitemOpen
  \bibfield  {author} {\bibinfo {author} {\bibfnamefont {T.}~\bibnamefont
  {Han}}, \bibinfo {author} {\bibfnamefont {Z.}~\bibnamefont {Li}}, \bibinfo
  {author} {\bibfnamefont {J.-F.}\ \bibnamefont {Zhang}}, \ and\ \bibinfo
  {author} {\bibfnamefont {X.}~\bibnamefont {Zhang}},\ }\href {\doibase
  10.3390/universe11030085} {\bibfield  {journal} {\bibinfo  {journal}
  {Universe}\ }\textbf {\bibinfo {volume} {11}},\ \bibinfo {pages} {85}
  (\bibinfo {year} {2025}{\natexlab{a}})},\ \Eprint
  {http://arxiv.org/abs/2412.06873} {arXiv:2412.06873 [astro-ph.CO]}
  \BibitemShut {NoStop}%
\bibitem [{\citenamefont {Han}\ \emph {et~al.}(2025{\natexlab{b}})\citenamefont
  {Han}, \citenamefont {Zhang},\ and\ \citenamefont {Zhang}}]{Han:2025fii}%
  \BibitemOpen
  \bibfield  {author} {\bibinfo {author} {\bibfnamefont {T.}~\bibnamefont
  {Han}}, \bibinfo {author} {\bibfnamefont {J.-F.}\ \bibnamefont {Zhang}}, \
  and\ \bibinfo {author} {\bibfnamefont {X.}~\bibnamefont {Zhang}},\
  }\href@noop {} {\  (\bibinfo {year} {2025}{\natexlab{b}})},\ \Eprint
  {http://arxiv.org/abs/2504.17741} {arXiv:2504.17741 [astro-ph.CO]}
  \BibitemShut {NoStop}%
\bibitem [{\citenamefont {Jin}\ \emph {et~al.}(2025)\citenamefont {Jin},
  \citenamefont {Song}, \citenamefont {Sun}, \citenamefont {Xiao},
  \citenamefont {Wang}, \citenamefont {Wang}, \citenamefont {Zhang},\ and\
  \citenamefont {Zhang}}]{Jin:2025dvf}%
  \BibitemOpen
  \bibfield  {author} {\bibinfo {author} {\bibfnamefont {S.-J.}\ \bibnamefont
  {Jin}}, \bibinfo {author} {\bibfnamefont {J.-Y.}\ \bibnamefont {Song}},
  \bibinfo {author} {\bibfnamefont {T.-Y.}\ \bibnamefont {Sun}}, \bibinfo
  {author} {\bibfnamefont {S.-R.}\ \bibnamefont {Xiao}}, \bibinfo {author}
  {\bibfnamefont {H.}~\bibnamefont {Wang}}, \bibinfo {author} {\bibfnamefont
  {L.-F.}\ \bibnamefont {Wang}}, \bibinfo {author} {\bibfnamefont {J.-F.}\
  \bibnamefont {Zhang}}, \ and\ \bibinfo {author} {\bibfnamefont
  {X.}~\bibnamefont {Zhang}},\ }\href@noop {} {\  (\bibinfo {year} {2025})},\
  \Eprint {http://arxiv.org/abs/2507.12965} {arXiv:2507.12965 [astro-ph.CO]}
  \BibitemShut {NoStop}%
\bibitem [{ET-()}]{ET-web}%
  \BibitemOpen
  \href@noop {} {\enquote {\bibinfo {title} {{\rm ET}},}\ }\bibinfo
  {howpublished} {\url{https://www.et-gw.eu/}}\BibitemShut {NoStop}%
\bibitem [{\citenamefont {Punturo}\ \emph {et~al.}(2010)\citenamefont {Punturo}
  \emph {et~al.}}]{Punturo:2010zz}%
  \BibitemOpen
  \bibfield  {author} {\bibinfo {author} {\bibfnamefont {M.}~\bibnamefont
  {Punturo}} \emph {et~al.},\ }\href {\doibase 10.1088/0264-9381/27/19/194002}
  {\bibfield  {journal} {\bibinfo  {journal} {Class. Quant. Grav.}\ }\textbf
  {\bibinfo {volume} {27}},\ \bibinfo {pages} {194002} (\bibinfo {year}
  {2010})}\BibitemShut {NoStop}%
\bibitem [{CE-()}]{CE-web}%
  \BibitemOpen
  \href@noop {} {\enquote {\bibinfo {title} {{\rm CE}},}\ }\bibinfo
  {howpublished} {\url{https://cosmicexplorer.org/}}\BibitemShut {NoStop}%
\bibitem [{\citenamefont {Abbott}\ \emph
  {et~al.}(2017{\natexlab{b}})\citenamefont {Abbott} \emph
  {et~al.}}]{Evans:2016mbw}%
  \BibitemOpen
  \bibfield  {author} {\bibinfo {author} {\bibfnamefont {B.~P.}\ \bibnamefont
  {Abbott}} \emph {et~al.} (\bibinfo {collaboration} {LIGO Scientific}),\
  }\href {\doibase 10.1088/1361-6382/aa51f4} {\bibfield  {journal} {\bibinfo
  {journal} {Class. Quant. Grav.}\ }\textbf {\bibinfo {volume} {34}},\ \bibinfo
  {pages} {044001} (\bibinfo {year} {2017}{\natexlab{b}})},\ \Eprint
  {http://arxiv.org/abs/1607.08697} {arXiv:1607.08697 [astro-ph.IM]}
  \BibitemShut {NoStop}%
\bibitem [{\citenamefont {Lesgourgues}\ and\ \citenamefont
  {Pastor}(2006)}]{Lesgourgues:2006nd}%
  \BibitemOpen
  \bibfield  {author} {\bibinfo {author} {\bibfnamefont {J.}~\bibnamefont
  {Lesgourgues}}\ and\ \bibinfo {author} {\bibfnamefont {S.}~\bibnamefont
  {Pastor}},\ }\href {\doibase 10.1016/j.physrep.2006.04.001} {\bibfield
  {journal} {\bibinfo  {journal} {Phys. Rept.}\ }\textbf {\bibinfo {volume}
  {429}},\ \bibinfo {pages} {307} (\bibinfo {year} {2006})},\ \Eprint
  {http://arxiv.org/abs/astro-ph/0603494} {arXiv:astro-ph/0603494} \BibitemShut
  {NoStop}%
\bibitem [{\citenamefont {Aguilar}\ \emph {et~al.}(2001)\citenamefont {Aguilar}
  \emph {et~al.}}]{LSND:2001aii}%
  \BibitemOpen
  \bibfield  {author} {\bibinfo {author} {\bibfnamefont {A.}~\bibnamefont
  {Aguilar}} \emph {et~al.} (\bibinfo {collaboration} {LSND}),\ }\href
  {\doibase 10.1103/PhysRevD.64.112007} {\bibfield  {journal} {\bibinfo
  {journal} {Phys. Rev. D}\ }\textbf {\bibinfo {volume} {64}},\ \bibinfo
  {pages} {112007} (\bibinfo {year} {2001})},\ \Eprint
  {http://arxiv.org/abs/hep-ex/0104049} {arXiv:hep-ex/0104049} \BibitemShut
  {NoStop}%
\bibitem [{\citenamefont {Mention}\ \emph {et~al.}(2011)\citenamefont
  {Mention}, \citenamefont {Fechner}, \citenamefont {Lasserre}, \citenamefont
  {Mueller}, \citenamefont {Lhuillier}, \citenamefont {Cribier},\ and\
  \citenamefont {Letourneau}}]{Mention:2011rk}%
  \BibitemOpen
  \bibfield  {author} {\bibinfo {author} {\bibfnamefont {G.}~\bibnamefont
  {Mention}}, \bibinfo {author} {\bibfnamefont {M.}~\bibnamefont {Fechner}},
  \bibinfo {author} {\bibfnamefont {T.}~\bibnamefont {Lasserre}}, \bibinfo
  {author} {\bibfnamefont {T.~A.}\ \bibnamefont {Mueller}}, \bibinfo {author}
  {\bibfnamefont {D.}~\bibnamefont {Lhuillier}}, \bibinfo {author}
  {\bibfnamefont {M.}~\bibnamefont {Cribier}}, \ and\ \bibinfo {author}
  {\bibfnamefont {A.}~\bibnamefont {Letourneau}},\ }\href {\doibase
  10.1103/PhysRevD.83.073006} {\bibfield  {journal} {\bibinfo  {journal} {Phys.
  Rev. D}\ }\textbf {\bibinfo {volume} {83}},\ \bibinfo {pages} {073006}
  (\bibinfo {year} {2011})},\ \Eprint {http://arxiv.org/abs/1101.2755}
  {arXiv:1101.2755 [hep-ex]} \BibitemShut {NoStop}%
\bibitem [{\citenamefont {Conrad}\ \emph {et~al.}(2013)\citenamefont {Conrad},
  \citenamefont {Ignarra}, \citenamefont {Karagiorgi}, \citenamefont
  {Shaevitz},\ and\ \citenamefont {Spitz}}]{Conrad:2012qt}%
  \BibitemOpen
  \bibfield  {author} {\bibinfo {author} {\bibfnamefont {J.~M.}\ \bibnamefont
  {Conrad}}, \bibinfo {author} {\bibfnamefont {C.~M.}\ \bibnamefont {Ignarra}},
  \bibinfo {author} {\bibfnamefont {G.}~\bibnamefont {Karagiorgi}}, \bibinfo
  {author} {\bibfnamefont {M.~H.}\ \bibnamefont {Shaevitz}}, \ and\ \bibinfo
  {author} {\bibfnamefont {J.}~\bibnamefont {Spitz}},\ }\href {\doibase
  10.1155/2013/163897} {\bibfield  {journal} {\bibinfo  {journal} {Adv. High
  Energy Phys.}\ }\textbf {\bibinfo {volume} {2013}},\ \bibinfo {pages}
  {163897} (\bibinfo {year} {2013})},\ \Eprint {http://arxiv.org/abs/1207.4765}
  {arXiv:1207.4765 [hep-ex]} \BibitemShut {NoStop}%
\bibitem [{\citenamefont {Giunti}\ \emph {et~al.}(2012)\citenamefont {Giunti},
  \citenamefont {Laveder}, \citenamefont {Li}, \citenamefont {Liu},\ and\
  \citenamefont {Long}}]{Giunti:2012tn}%
  \BibitemOpen
  \bibfield  {author} {\bibinfo {author} {\bibfnamefont {C.}~\bibnamefont
  {Giunti}}, \bibinfo {author} {\bibfnamefont {M.}~\bibnamefont {Laveder}},
  \bibinfo {author} {\bibfnamefont {Y.~F.}\ \bibnamefont {Li}}, \bibinfo
  {author} {\bibfnamefont {Q.~Y.}\ \bibnamefont {Liu}}, \ and\ \bibinfo
  {author} {\bibfnamefont {H.~W.}\ \bibnamefont {Long}},\ }\href {\doibase
  10.1103/PhysRevD.86.113014} {\bibfield  {journal} {\bibinfo  {journal} {Phys.
  Rev. D}\ }\textbf {\bibinfo {volume} {86}},\ \bibinfo {pages} {113014}
  (\bibinfo {year} {2012})},\ \Eprint {http://arxiv.org/abs/1210.5715}
  {arXiv:1210.5715 [hep-ph]} \BibitemShut {NoStop}%
\bibitem [{\citenamefont {Kopp}\ \emph {et~al.}(2013)\citenamefont {Kopp},
  \citenamefont {Machado}, \citenamefont {Maltoni},\ and\ \citenamefont
  {Schwetz}}]{Kopp:2013vaa}%
  \BibitemOpen
  \bibfield  {author} {\bibinfo {author} {\bibfnamefont {J.}~\bibnamefont
  {Kopp}}, \bibinfo {author} {\bibfnamefont {P.~A.~N.}\ \bibnamefont
  {Machado}}, \bibinfo {author} {\bibfnamefont {M.}~\bibnamefont {Maltoni}}, \
  and\ \bibinfo {author} {\bibfnamefont {T.}~\bibnamefont {Schwetz}},\ }\href
  {\doibase 10.1007/JHEP05(2013)050} {\bibfield  {journal} {\bibinfo  {journal}
  {JHEP}\ }\textbf {\bibinfo {volume} {05}},\ \bibinfo {pages} {050} (\bibinfo
  {year} {2013})},\ \Eprint {http://arxiv.org/abs/1303.3011} {arXiv:1303.3011
  [hep-ph]} \BibitemShut {NoStop}%
\bibitem [{\citenamefont {Giunti}\ \emph {et~al.}(2013)\citenamefont {Giunti},
  \citenamefont {Laveder}, \citenamefont {Li},\ and\ \citenamefont
  {Long}}]{Giunti:2013aea}%
  \BibitemOpen
  \bibfield  {author} {\bibinfo {author} {\bibfnamefont {C.}~\bibnamefont
  {Giunti}}, \bibinfo {author} {\bibfnamefont {M.}~\bibnamefont {Laveder}},
  \bibinfo {author} {\bibfnamefont {Y.~F.}\ \bibnamefont {Li}}, \ and\ \bibinfo
  {author} {\bibfnamefont {H.~W.}\ \bibnamefont {Long}},\ }\href {\doibase
  10.1103/PhysRevD.88.073008} {\bibfield  {journal} {\bibinfo  {journal} {Phys.
  Rev. D}\ }\textbf {\bibinfo {volume} {88}},\ \bibinfo {pages} {073008}
  (\bibinfo {year} {2013})},\ \Eprint {http://arxiv.org/abs/1308.5288}
  {arXiv:1308.5288 [hep-ph]} \BibitemShut {NoStop}%
\bibitem [{\citenamefont {Gariazzo}\ \emph {et~al.}(2013)\citenamefont
  {Gariazzo}, \citenamefont {Giunti},\ and\ \citenamefont
  {Laveder}}]{Gariazzo:2013gua}%
  \BibitemOpen
  \bibfield  {author} {\bibinfo {author} {\bibfnamefont {S.}~\bibnamefont
  {Gariazzo}}, \bibinfo {author} {\bibfnamefont {C.}~\bibnamefont {Giunti}}, \
  and\ \bibinfo {author} {\bibfnamefont {M.}~\bibnamefont {Laveder}},\ }\href
  {\doibase 10.1007/JHEP11(2013)211} {\bibfield  {journal} {\bibinfo  {journal}
  {JHEP}\ }\textbf {\bibinfo {volume} {11}},\ \bibinfo {pages} {211} (\bibinfo
  {year} {2013})},\ \Eprint {http://arxiv.org/abs/1309.3192} {arXiv:1309.3192
  [hep-ph]} \BibitemShut {NoStop}%
\bibitem [{\citenamefont {Aguilar-Arevalo}\ \emph {et~al.}(2018)\citenamefont
  {Aguilar-Arevalo} \emph {et~al.}}]{MiniBooNE:2018esg}%
  \BibitemOpen
  \bibfield  {author} {\bibinfo {author} {\bibfnamefont {A.~A.}\ \bibnamefont
  {Aguilar-Arevalo}} \emph {et~al.} (\bibinfo {collaboration} {MiniBooNE}),\
  }\href {\doibase 10.1103/PhysRevLett.121.221801} {\bibfield  {journal}
  {\bibinfo  {journal} {Phys. Rev. Lett.}\ }\textbf {\bibinfo {volume} {121}},\
  \bibinfo {pages} {221801} (\bibinfo {year} {2018})},\ \Eprint
  {http://arxiv.org/abs/1805.12028} {arXiv:1805.12028 [hep-ex]} \BibitemShut
  {NoStop}%
\bibitem [{\citenamefont {Adamson}\ \emph {et~al.}(2016)\citenamefont {Adamson}
  \emph {et~al.}}]{DayaBay:2016lkk}%
  \BibitemOpen
  \bibfield  {author} {\bibinfo {author} {\bibfnamefont {P.}~\bibnamefont
  {Adamson}} \emph {et~al.} (\bibinfo {collaboration} {Daya Bay, MINOS}),\
  }\href {\doibase 10.1103/PhysRevLett.117.151801} {\bibfield  {journal}
  {\bibinfo  {journal} {Phys. Rev. Lett.}\ }\textbf {\bibinfo {volume} {117}},\
  \bibinfo {pages} {151801} (\bibinfo {year} {2016})},\ \bibinfo {note}
  {[Addendum: Phys.Rev.Lett. 117, 209901 (2016)]},\ \Eprint
  {http://arxiv.org/abs/1607.01177} {arXiv:1607.01177 [hep-ex]} \BibitemShut
  {NoStop}%
\bibitem [{\citenamefont {An}\ \emph {et~al.}(2024)\citenamefont {An} \emph
  {et~al.}}]{DayaBay:2024nip}%
  \BibitemOpen
  \bibfield  {author} {\bibinfo {author} {\bibfnamefont {F.~P.}\ \bibnamefont
  {An}} \emph {et~al.} (\bibinfo {collaboration} {Daya Bay}),\ }\href {\doibase
  10.1103/PhysRevLett.133.051801} {\bibfield  {journal} {\bibinfo  {journal}
  {Phys. Rev. Lett.}\ }\textbf {\bibinfo {volume} {133}},\ \bibinfo {pages}
  {051801} (\bibinfo {year} {2024})},\ \Eprint
  {http://arxiv.org/abs/2404.01687} {arXiv:2404.01687 [hep-ex]} \BibitemShut
  {NoStop}%
\bibitem [{\citenamefont {Aartsen}\ \emph {et~al.}(2016)\citenamefont {Aartsen}
  \emph {et~al.}}]{IceCube:2016rnb}%
  \BibitemOpen
  \bibfield  {author} {\bibinfo {author} {\bibfnamefont {M.~G.}\ \bibnamefont
  {Aartsen}} \emph {et~al.} (\bibinfo {collaboration} {IceCube}),\ }\href
  {\doibase 10.1103/PhysRevLett.117.071801} {\bibfield  {journal} {\bibinfo
  {journal} {Phys. Rev. Lett.}\ }\textbf {\bibinfo {volume} {117}},\ \bibinfo
  {pages} {071801} (\bibinfo {year} {2016})},\ \Eprint
  {http://arxiv.org/abs/1605.01990} {arXiv:1605.01990 [hep-ex]} \BibitemShut
  {NoStop}%
\bibitem [{\citenamefont {Abbasi}\ \emph {et~al.}(2024)\citenamefont {Abbasi}
  \emph {et~al.}}]{IceCubeCollaboration:2024nle}%
  \BibitemOpen
  \bibfield  {author} {\bibinfo {author} {\bibfnamefont {R.}~\bibnamefont
  {Abbasi}} \emph {et~al.} (\bibinfo {collaboration} {(IceCube
  Collaboration){\ensuremath{\parallel}}, IceCube}),\ }\href {\doibase
  10.1103/PhysRevLett.133.201804} {\bibfield  {journal} {\bibinfo  {journal}
  {Phys. Rev. Lett.}\ }\textbf {\bibinfo {volume} {133}},\ \bibinfo {pages}
  {201804} (\bibinfo {year} {2024})},\ \Eprint
  {http://arxiv.org/abs/2405.08070} {arXiv:2405.08070 [hep-ex]} \BibitemShut
  {NoStop}%
\bibitem [{\citenamefont {An}\ \emph {et~al.}(2016)\citenamefont {An} \emph
  {et~al.}}]{JUNO:2015zny}%
  \BibitemOpen
  \bibfield  {author} {\bibinfo {author} {\bibfnamefont {F.}~\bibnamefont {An}}
  \emph {et~al.} (\bibinfo {collaboration} {JUNO}),\ }\href {\doibase
  10.1088/0954-3899/43/3/030401} {\bibfield  {journal} {\bibinfo  {journal} {J.
  Phys. G}\ }\textbf {\bibinfo {volume} {43}},\ \bibinfo {pages} {030401}
  (\bibinfo {year} {2016})},\ \Eprint {http://arxiv.org/abs/1507.05613}
  {arXiv:1507.05613 [physics.ins-det]} \BibitemShut {NoStop}%
\bibitem [{\citenamefont {Forero}\ \emph {et~al.}(2021)\citenamefont {Forero},
  \citenamefont {Parke}, \citenamefont {Ternes},\ and\ \citenamefont
  {Funchal}}]{Forero:2021lax}%
  \BibitemOpen
  \bibfield  {author} {\bibinfo {author} {\bibfnamefont {D.~V.}\ \bibnamefont
  {Forero}}, \bibinfo {author} {\bibfnamefont {S.~J.}\ \bibnamefont {Parke}},
  \bibinfo {author} {\bibfnamefont {C.~A.}\ \bibnamefont {Ternes}}, \ and\
  \bibinfo {author} {\bibfnamefont {R.~Z.}\ \bibnamefont {Funchal}},\ }\href
  {\doibase 10.1103/PhysRevD.104.113004} {\bibfield  {journal} {\bibinfo
  {journal} {Phys. Rev. D}\ }\textbf {\bibinfo {volume} {104}},\ \bibinfo
  {pages} {113004} (\bibinfo {year} {2021})},\ \Eprint
  {http://arxiv.org/abs/2107.12410} {arXiv:2107.12410 [hep-ph]} \BibitemShut
  {NoStop}%
\bibitem [{\citenamefont {Abusleme}\ \emph {et~al.}(2025)\citenamefont
  {Abusleme} \emph {et~al.}}]{JUNO:2024jaw}%
  \BibitemOpen
  \bibfield  {author} {\bibinfo {author} {\bibfnamefont {A.}~\bibnamefont
  {Abusleme}} \emph {et~al.} (\bibinfo {collaboration} {JUNO}),\ }\href
  {\doibase 10.1088/1674-1137/ad7f3e} {\bibfield  {journal} {\bibinfo
  {journal} {Chin. Phys. C}\ }\textbf {\bibinfo {volume} {49}},\ \bibinfo
  {pages} {033104} (\bibinfo {year} {2025})},\ \Eprint
  {http://arxiv.org/abs/2405.18008} {arXiv:2405.18008 [hep-ex]} \BibitemShut
  {NoStop}%
\bibitem [{\citenamefont {Adamson}\ \emph {et~al.}(2013)\citenamefont {Adamson}
  \emph {et~al.}}]{MINOS:2013utc}%
  \BibitemOpen
  \bibfield  {author} {\bibinfo {author} {\bibfnamefont {P.}~\bibnamefont
  {Adamson}} \emph {et~al.} (\bibinfo {collaboration} {MINOS}),\ }\href
  {\doibase 10.1103/PhysRevLett.110.251801} {\bibfield  {journal} {\bibinfo
  {journal} {Phys. Rev. Lett.}\ }\textbf {\bibinfo {volume} {110}},\ \bibinfo
  {pages} {251801} (\bibinfo {year} {2013})},\ \Eprint
  {http://arxiv.org/abs/1304.6335} {arXiv:1304.6335 [hep-ex]} \BibitemShut
  {NoStop}%
\bibitem [{\citenamefont {Aartsen}\ \emph {et~al.}(2020)\citenamefont {Aartsen}
  \emph {et~al.}}]{IceCube-Gen2:2019fet}%
  \BibitemOpen
  \bibfield  {author} {\bibinfo {author} {\bibfnamefont {M.~G.}\ \bibnamefont
  {Aartsen}} \emph {et~al.} (\bibinfo {collaboration} {IceCube-Gen2}),\ }\href
  {\doibase 10.1103/PhysRevD.101.032006} {\bibfield  {journal} {\bibinfo
  {journal} {Phys. Rev. D}\ }\textbf {\bibinfo {volume} {101}},\ \bibinfo
  {pages} {032006} (\bibinfo {year} {2020})},\ \Eprint
  {http://arxiv.org/abs/1911.06745} {arXiv:1911.06745 [hep-ex]} \BibitemShut
  {NoStop}%
\bibitem [{\citenamefont {Hu}\ \emph {et~al.}(1998)\citenamefont {Hu},
  \citenamefont {Eisenstein},\ and\ \citenamefont {Tegmark}}]{Hu:1997mj}%
  \BibitemOpen
  \bibfield  {author} {\bibinfo {author} {\bibfnamefont {W.}~\bibnamefont
  {Hu}}, \bibinfo {author} {\bibfnamefont {D.~J.}\ \bibnamefont {Eisenstein}},
  \ and\ \bibinfo {author} {\bibfnamefont {M.}~\bibnamefont {Tegmark}},\ }\href
  {\doibase 10.1103/PhysRevLett.80.5255} {\bibfield  {journal} {\bibinfo
  {journal} {Phys. Rev. Lett.}\ }\textbf {\bibinfo {volume} {80}},\ \bibinfo
  {pages} {5255} (\bibinfo {year} {1998})},\ \Eprint
  {http://arxiv.org/abs/astro-ph/9712057} {arXiv:astro-ph/9712057} \BibitemShut
  {NoStop}%
\bibitem [{\citenamefont {de~Holanda}\ and\ \citenamefont
  {Smirnov}(2011)}]{deHolanda:2010am}%
  \BibitemOpen
  \bibfield  {author} {\bibinfo {author} {\bibfnamefont {P.~C.}\ \bibnamefont
  {de~Holanda}}\ and\ \bibinfo {author} {\bibfnamefont {A.~Y.}\ \bibnamefont
  {Smirnov}},\ }\href {\doibase 10.1103/PhysRevD.83.113011} {\bibfield
  {journal} {\bibinfo  {journal} {Phys. Rev. D}\ }\textbf {\bibinfo {volume}
  {83}},\ \bibinfo {pages} {113011} (\bibinfo {year} {2011})},\ \Eprint
  {http://arxiv.org/abs/1012.5627} {arXiv:1012.5627 [hep-ph]} \BibitemShut
  {NoStop}%
\bibitem [{\citenamefont {Li}\ \emph {et~al.}(2013{\natexlab{a}})\citenamefont
  {Li}, \citenamefont {Wang}, \citenamefont {Li},\ and\ \citenamefont
  {Zhang}}]{Li:2012spm}%
  \BibitemOpen
  \bibfield  {author} {\bibinfo {author} {\bibfnamefont {Y.-H.}\ \bibnamefont
  {Li}}, \bibinfo {author} {\bibfnamefont {S.}~\bibnamefont {Wang}}, \bibinfo
  {author} {\bibfnamefont {X.-D.}\ \bibnamefont {Li}}, \ and\ \bibinfo {author}
  {\bibfnamefont {X.}~\bibnamefont {Zhang}},\ }\href {\doibase
  10.1088/1475-7516/2013/02/033} {\bibfield  {journal} {\bibinfo  {journal}
  {JCAP}\ }\textbf {\bibinfo {volume} {02}},\ \bibinfo {pages} {033} (\bibinfo
  {year} {2013}{\natexlab{a}})},\ \Eprint {http://arxiv.org/abs/1207.6679}
  {arXiv:1207.6679 [astro-ph.CO]} \BibitemShut {NoStop}%
\bibitem [{\citenamefont {Wang}\ \emph {et~al.}(2012)\citenamefont {Wang},
  \citenamefont {Meng}, \citenamefont {Zhang}, \citenamefont {Shan},
  \citenamefont {Gong}, \citenamefont {Tao}, \citenamefont {Chen},\ and\
  \citenamefont {Huang}}]{Wang:2012vh}%
  \BibitemOpen
  \bibfield  {author} {\bibinfo {author} {\bibfnamefont {X.}~\bibnamefont
  {Wang}}, \bibinfo {author} {\bibfnamefont {X.-L.}\ \bibnamefont {Meng}},
  \bibinfo {author} {\bibfnamefont {T.-J.}\ \bibnamefont {Zhang}}, \bibinfo
  {author} {\bibfnamefont {H.}~\bibnamefont {Shan}}, \bibinfo {author}
  {\bibfnamefont {Y.}~\bibnamefont {Gong}}, \bibinfo {author} {\bibfnamefont
  {C.}~\bibnamefont {Tao}}, \bibinfo {author} {\bibfnamefont {X.}~\bibnamefont
  {Chen}}, \ and\ \bibinfo {author} {\bibfnamefont {Y.~F.}\ \bibnamefont
  {Huang}},\ }\href {\doibase 10.1088/1475-7516/2012/11/018} {\bibfield
  {journal} {\bibinfo  {journal} {JCAP}\ }\textbf {\bibinfo {volume} {11}},\
  \bibinfo {pages} {018} (\bibinfo {year} {2012})},\ \Eprint
  {http://arxiv.org/abs/1210.2136} {arXiv:1210.2136 [astro-ph.CO]} \BibitemShut
  {NoStop}%
\bibitem [{\citenamefont {Palazzo}(2013)}]{Palazzo:2013me}%
  \BibitemOpen
  \bibfield  {author} {\bibinfo {author} {\bibfnamefont {A.}~\bibnamefont
  {Palazzo}},\ }\href {\doibase 10.1142/S0217732313300048} {\bibfield
  {journal} {\bibinfo  {journal} {Mod. Phys. Lett. A}\ }\textbf {\bibinfo
  {volume} {28}},\ \bibinfo {pages} {1330004} (\bibinfo {year} {2013})},\
  \Eprint {http://arxiv.org/abs/1302.1102} {arXiv:1302.1102 [hep-ph]}
  \BibitemShut {NoStop}%
\bibitem [{\citenamefont {Wyman}\ \emph {et~al.}(2014)\citenamefont {Wyman},
  \citenamefont {Rudd}, \citenamefont {Vanderveld},\ and\ \citenamefont
  {Hu}}]{Wyman:2013lza}%
  \BibitemOpen
  \bibfield  {author} {\bibinfo {author} {\bibfnamefont {M.}~\bibnamefont
  {Wyman}}, \bibinfo {author} {\bibfnamefont {D.~H.}\ \bibnamefont {Rudd}},
  \bibinfo {author} {\bibfnamefont {R.~A.}\ \bibnamefont {Vanderveld}}, \ and\
  \bibinfo {author} {\bibfnamefont {W.}~\bibnamefont {Hu}},\ }\href {\doibase
  10.1103/PhysRevLett.112.051302} {\bibfield  {journal} {\bibinfo  {journal}
  {Phys. Rev. Lett.}\ }\textbf {\bibinfo {volume} {112}},\ \bibinfo {pages}
  {051302} (\bibinfo {year} {2014})},\ \Eprint {http://arxiv.org/abs/1307.7715}
  {arXiv:1307.7715 [astro-ph.CO]} \BibitemShut {NoStop}%
\bibitem [{\citenamefont {Battye}\ and\ \citenamefont
  {Moss}(2014)}]{Battye:2013xqa}%
  \BibitemOpen
  \bibfield  {author} {\bibinfo {author} {\bibfnamefont {R.~A.}\ \bibnamefont
  {Battye}}\ and\ \bibinfo {author} {\bibfnamefont {A.}~\bibnamefont {Moss}},\
  }\href {\doibase 10.1103/PhysRevLett.112.051303} {\bibfield  {journal}
  {\bibinfo  {journal} {Phys. Rev. Lett.}\ }\textbf {\bibinfo {volume} {112}},\
  \bibinfo {pages} {051303} (\bibinfo {year} {2014})},\ \Eprint
  {http://arxiv.org/abs/1308.5870} {arXiv:1308.5870 [astro-ph.CO]} \BibitemShut
  {NoStop}%
\bibitem [{\citenamefont {Zhang}\ \emph
  {et~al.}(2015{\natexlab{a}})\citenamefont {Zhang}, \citenamefont {Li},\ and\
  \citenamefont {Zhang}}]{Zhang:2014dxk}%
  \BibitemOpen
  \bibfield  {author} {\bibinfo {author} {\bibfnamefont {J.-F.}\ \bibnamefont
  {Zhang}}, \bibinfo {author} {\bibfnamefont {Y.-H.}\ \bibnamefont {Li}}, \
  and\ \bibinfo {author} {\bibfnamefont {X.}~\bibnamefont {Zhang}},\ }\href
  {\doibase 10.1016/j.physletb.2014.12.012} {\bibfield  {journal} {\bibinfo
  {journal} {Phys. Lett. B}\ }\textbf {\bibinfo {volume} {740}},\ \bibinfo
  {pages} {359} (\bibinfo {year} {2015}{\natexlab{a}})},\ \Eprint
  {http://arxiv.org/abs/1403.7028} {arXiv:1403.7028 [astro-ph.CO]} \BibitemShut
  {NoStop}%
\bibitem [{\citenamefont {Dvorkin}\ \emph {et~al.}(2014)\citenamefont
  {Dvorkin}, \citenamefont {Wyman}, \citenamefont {Rudd},\ and\ \citenamefont
  {Hu}}]{Dvorkin:2014lea}%
  \BibitemOpen
  \bibfield  {author} {\bibinfo {author} {\bibfnamefont {C.}~\bibnamefont
  {Dvorkin}}, \bibinfo {author} {\bibfnamefont {M.}~\bibnamefont {Wyman}},
  \bibinfo {author} {\bibfnamefont {D.~H.}\ \bibnamefont {Rudd}}, \ and\
  \bibinfo {author} {\bibfnamefont {W.}~\bibnamefont {Hu}},\ }\href {\doibase
  10.1103/PhysRevD.90.083503} {\bibfield  {journal} {\bibinfo  {journal} {Phys.
  Rev. D}\ }\textbf {\bibinfo {volume} {90}},\ \bibinfo {pages} {083503}
  (\bibinfo {year} {2014})},\ \Eprint {http://arxiv.org/abs/1403.8049}
  {arXiv:1403.8049 [astro-ph.CO]} \BibitemShut {NoStop}%
\bibitem [{\citenamefont {Archidiacono}\ \emph {et~al.}(2014)\citenamefont
  {Archidiacono}, \citenamefont {Fornengo}, \citenamefont {Gariazzo},
  \citenamefont {Giunti}, \citenamefont {Hannestad},\ and\ \citenamefont
  {Laveder}}]{Archidiacono:2014apa}%
  \BibitemOpen
  \bibfield  {author} {\bibinfo {author} {\bibfnamefont {M.}~\bibnamefont
  {Archidiacono}}, \bibinfo {author} {\bibfnamefont {N.}~\bibnamefont
  {Fornengo}}, \bibinfo {author} {\bibfnamefont {S.}~\bibnamefont {Gariazzo}},
  \bibinfo {author} {\bibfnamefont {C.}~\bibnamefont {Giunti}}, \bibinfo
  {author} {\bibfnamefont {S.}~\bibnamefont {Hannestad}}, \ and\ \bibinfo
  {author} {\bibfnamefont {M.}~\bibnamefont {Laveder}},\ }\href {\doibase
  10.1088/1475-7516/2014/06/031} {\bibfield  {journal} {\bibinfo  {journal}
  {JCAP}\ }\textbf {\bibinfo {volume} {06}},\ \bibinfo {pages} {031} (\bibinfo
  {year} {2014})},\ \Eprint {http://arxiv.org/abs/1404.1794} {arXiv:1404.1794
  [astro-ph.CO]} \BibitemShut {NoStop}%
\bibitem [{\citenamefont {Zhang}\ \emph
  {et~al.}(2014{\natexlab{a}})\citenamefont {Zhang}, \citenamefont {Li},\ and\
  \citenamefont {Zhang}}]{Zhang:2014nta}%
  \BibitemOpen
  \bibfield  {author} {\bibinfo {author} {\bibfnamefont {J.-F.}\ \bibnamefont
  {Zhang}}, \bibinfo {author} {\bibfnamefont {Y.-H.}\ \bibnamefont {Li}}, \
  and\ \bibinfo {author} {\bibfnamefont {X.}~\bibnamefont {Zhang}},\ }\href
  {\doibase 10.1140/epjc/s10052-014-2954-8} {\bibfield  {journal} {\bibinfo
  {journal} {Eur. Phys. J. C}\ }\textbf {\bibinfo {volume} {74}},\ \bibinfo
  {pages} {2954} (\bibinfo {year} {2014}{\natexlab{a}})},\ \Eprint
  {http://arxiv.org/abs/1404.3598} {arXiv:1404.3598 [astro-ph.CO]} \BibitemShut
  {NoStop}%
\bibitem [{\citenamefont {An}\ \emph {et~al.}(2014)\citenamefont {An} \emph
  {et~al.}}]{DayaBay:2014fct}%
  \BibitemOpen
  \bibfield  {author} {\bibinfo {author} {\bibfnamefont {F.~P.}\ \bibnamefont
  {An}} \emph {et~al.} (\bibinfo {collaboration} {Daya Bay}),\ }\href {\doibase
  10.1103/PhysRevLett.113.141802} {\bibfield  {journal} {\bibinfo  {journal}
  {Phys. Rev. Lett.}\ }\textbf {\bibinfo {volume} {113}},\ \bibinfo {pages}
  {141802} (\bibinfo {year} {2014})},\ \Eprint {http://arxiv.org/abs/1407.7259}
  {arXiv:1407.7259 [hep-ex]} \BibitemShut {NoStop}%
\bibitem [{\citenamefont {Zhang}\ \emph
  {et~al.}(2014{\natexlab{b}})\citenamefont {Zhang}, \citenamefont {Geng},\
  and\ \citenamefont {Zhang}}]{Zhang:2014ifa}%
  \BibitemOpen
  \bibfield  {author} {\bibinfo {author} {\bibfnamefont {J.-F.}\ \bibnamefont
  {Zhang}}, \bibinfo {author} {\bibfnamefont {J.-J.}\ \bibnamefont {Geng}}, \
  and\ \bibinfo {author} {\bibfnamefont {X.}~\bibnamefont {Zhang}},\ }\href
  {\doibase 10.1088/1475-7516/2014/10/044} {\bibfield  {journal} {\bibinfo
  {journal} {JCAP}\ }\textbf {\bibinfo {volume} {10}},\ \bibinfo {pages} {044}
  (\bibinfo {year} {2014}{\natexlab{b}})},\ \Eprint
  {http://arxiv.org/abs/1408.0481} {arXiv:1408.0481 [astro-ph.CO]} \BibitemShut
  {NoStop}%
\bibitem [{\citenamefont {Zhang}\ \emph
  {et~al.}(2015{\natexlab{b}})\citenamefont {Zhang}, \citenamefont {Zhao},
  \citenamefont {Li},\ and\ \citenamefont {Zhang}}]{Zhang:2015rha}%
  \BibitemOpen
  \bibfield  {author} {\bibinfo {author} {\bibfnamefont {J.-F.}\ \bibnamefont
  {Zhang}}, \bibinfo {author} {\bibfnamefont {M.-M.}\ \bibnamefont {Zhao}},
  \bibinfo {author} {\bibfnamefont {Y.-H.}\ \bibnamefont {Li}}, \ and\ \bibinfo
  {author} {\bibfnamefont {X.}~\bibnamefont {Zhang}},\ }\href {\doibase
  10.1088/1475-7516/2015/04/038} {\bibfield  {journal} {\bibinfo  {journal}
  {JCAP}\ }\textbf {\bibinfo {volume} {04}},\ \bibinfo {pages} {038} (\bibinfo
  {year} {2015}{\natexlab{b}})},\ \Eprint {http://arxiv.org/abs/1502.04028}
  {arXiv:1502.04028 [astro-ph.CO]} \BibitemShut {NoStop}%
\bibitem [{\citenamefont {Geng}\ \emph {et~al.}(2016)\citenamefont {Geng},
  \citenamefont {Lee}, \citenamefont {Myrzakulov}, \citenamefont {Sami},\ and\
  \citenamefont {Saridakis}}]{Geng:2015haa}%
  \BibitemOpen
  \bibfield  {author} {\bibinfo {author} {\bibfnamefont {C.-Q.}\ \bibnamefont
  {Geng}}, \bibinfo {author} {\bibfnamefont {C.-C.}\ \bibnamefont {Lee}},
  \bibinfo {author} {\bibfnamefont {R.}~\bibnamefont {Myrzakulov}}, \bibinfo
  {author} {\bibfnamefont {M.}~\bibnamefont {Sami}}, \ and\ \bibinfo {author}
  {\bibfnamefont {E.~N.}\ \bibnamefont {Saridakis}},\ }\href {\doibase
  10.1088/1475-7516/2016/01/049} {\bibfield  {journal} {\bibinfo  {journal}
  {JCAP}\ }\textbf {\bibinfo {volume} {01}},\ \bibinfo {pages} {049} (\bibinfo
  {year} {2016})},\ \Eprint {http://arxiv.org/abs/1504.08141} {arXiv:1504.08141
  [astro-ph.CO]} \BibitemShut {NoStop}%
\bibitem [{\citenamefont {Zhang}(2016)}]{Zhang:2015uhk}%
  \BibitemOpen
  \bibfield  {author} {\bibinfo {author} {\bibfnamefont {X.}~\bibnamefont
  {Zhang}},\ }\href {\doibase 10.1103/PhysRevD.93.083011} {\bibfield  {journal}
  {\bibinfo  {journal} {Phys. Rev. D}\ }\textbf {\bibinfo {volume} {93}},\
  \bibinfo {pages} {083011} (\bibinfo {year} {2016})},\ \Eprint
  {http://arxiv.org/abs/1511.02651} {arXiv:1511.02651 [astro-ph.CO]}
  \BibitemShut {NoStop}%
\bibitem [{\citenamefont {Huang}\ \emph {et~al.}(2016)\citenamefont {Huang},
  \citenamefont {Wang},\ and\ \citenamefont {Wang}}]{Huang:2015wrx}%
  \BibitemOpen
  \bibfield  {author} {\bibinfo {author} {\bibfnamefont {Q.-G.}\ \bibnamefont
  {Huang}}, \bibinfo {author} {\bibfnamefont {K.}~\bibnamefont {Wang}}, \ and\
  \bibinfo {author} {\bibfnamefont {S.}~\bibnamefont {Wang}},\ }\href {\doibase
  10.1140/epjc/s10052-016-4334-z} {\bibfield  {journal} {\bibinfo  {journal}
  {Eur. Phys. J. C}\ }\textbf {\bibinfo {volume} {76}},\ \bibinfo {pages} {489}
  (\bibinfo {year} {2016})},\ \Eprint {http://arxiv.org/abs/1512.05899}
  {arXiv:1512.05899 [astro-ph.CO]} \BibitemShut {NoStop}%
\bibitem [{\citenamefont {Chen}\ \emph {et~al.}(2016)\citenamefont {Chen},
  \citenamefont {Ratra}, \citenamefont {Biesiada}, \citenamefont {Li},\ and\
  \citenamefont {Zhu}}]{Chen:2016eyp}%
  \BibitemOpen
  \bibfield  {author} {\bibinfo {author} {\bibfnamefont {Y.}~\bibnamefont
  {Chen}}, \bibinfo {author} {\bibfnamefont {B.}~\bibnamefont {Ratra}},
  \bibinfo {author} {\bibfnamefont {M.}~\bibnamefont {Biesiada}}, \bibinfo
  {author} {\bibfnamefont {S.}~\bibnamefont {Li}}, \ and\ \bibinfo {author}
  {\bibfnamefont {Z.-H.}\ \bibnamefont {Zhu}},\ }\href {\doibase
  10.3847/0004-637X/829/2/61} {\bibfield  {journal} {\bibinfo  {journal}
  {Astrophys. J.}\ }\textbf {\bibinfo {volume} {829}},\ \bibinfo {pages} {61}
  (\bibinfo {year} {2016})},\ \Eprint {http://arxiv.org/abs/1603.07115}
  {arXiv:1603.07115 [astro-ph.CO]} \BibitemShut {NoStop}%
\bibitem [{\citenamefont {Giusarma}\ \emph {et~al.}(2016)\citenamefont
  {Giusarma}, \citenamefont {Gerbino}, \citenamefont {Mena}, \citenamefont
  {Vagnozzi}, \citenamefont {Ho},\ and\ \citenamefont
  {Freese}}]{Giusarma:2016phn}%
  \BibitemOpen
  \bibfield  {author} {\bibinfo {author} {\bibfnamefont {E.}~\bibnamefont
  {Giusarma}}, \bibinfo {author} {\bibfnamefont {M.}~\bibnamefont {Gerbino}},
  \bibinfo {author} {\bibfnamefont {O.}~\bibnamefont {Mena}}, \bibinfo {author}
  {\bibfnamefont {S.}~\bibnamefont {Vagnozzi}}, \bibinfo {author}
  {\bibfnamefont {S.}~\bibnamefont {Ho}}, \ and\ \bibinfo {author}
  {\bibfnamefont {K.}~\bibnamefont {Freese}},\ }\href {\doibase
  10.1103/PhysRevD.94.083522} {\bibfield  {journal} {\bibinfo  {journal} {Phys.
  Rev. D}\ }\textbf {\bibinfo {volume} {94}},\ \bibinfo {pages} {083522}
  (\bibinfo {year} {2016})},\ \Eprint {http://arxiv.org/abs/1605.04320}
  {arXiv:1605.04320 [astro-ph.CO]} \BibitemShut {NoStop}%
\bibitem [{\citenamefont {Wang}\ \emph {et~al.}(2016)\citenamefont {Wang},
  \citenamefont {Wang}, \citenamefont {Xia},\ and\ \citenamefont
  {Zhang}}]{Wang:2016tsz}%
  \BibitemOpen
  \bibfield  {author} {\bibinfo {author} {\bibfnamefont {S.}~\bibnamefont
  {Wang}}, \bibinfo {author} {\bibfnamefont {Y.-F.}\ \bibnamefont {Wang}},
  \bibinfo {author} {\bibfnamefont {D.-M.}\ \bibnamefont {Xia}}, \ and\
  \bibinfo {author} {\bibfnamefont {X.}~\bibnamefont {Zhang}},\ }\href
  {\doibase 10.1103/PhysRevD.94.083519} {\bibfield  {journal} {\bibinfo
  {journal} {Phys. Rev. D}\ }\textbf {\bibinfo {volume} {94}},\ \bibinfo
  {pages} {083519} (\bibinfo {year} {2016})},\ \Eprint
  {http://arxiv.org/abs/1608.00672} {arXiv:1608.00672 [astro-ph.CO]}
  \BibitemShut {NoStop}%
\bibitem [{\citenamefont {Xu}\ and\ \citenamefont {Huang}(2018)}]{Xu:2016ddc}%
  \BibitemOpen
  \bibfield  {author} {\bibinfo {author} {\bibfnamefont {L.}~\bibnamefont
  {Xu}}\ and\ \bibinfo {author} {\bibfnamefont {Q.-G.}\ \bibnamefont {Huang}},\
  }\href {\doibase 10.1007/s11433-017-9125-0} {\bibfield  {journal} {\bibinfo
  {journal} {Sci. China Phys. Mech. Astron.}\ }\textbf {\bibinfo {volume}
  {61}},\ \bibinfo {pages} {039521} (\bibinfo {year} {2018})},\ \Eprint
  {http://arxiv.org/abs/1611.05178} {arXiv:1611.05178 [astro-ph.CO]}
  \BibitemShut {NoStop}%
\bibitem [{\citenamefont {Vagnozzi}\ \emph {et~al.}(2017)\citenamefont
  {Vagnozzi}, \citenamefont {Giusarma}, \citenamefont {Mena}, \citenamefont
  {Freese}, \citenamefont {Gerbino}, \citenamefont {Ho},\ and\ \citenamefont
  {Lattanzi}}]{Vagnozzi:2017ovm}%
  \BibitemOpen
  \bibfield  {author} {\bibinfo {author} {\bibfnamefont {S.}~\bibnamefont
  {Vagnozzi}}, \bibinfo {author} {\bibfnamefont {E.}~\bibnamefont {Giusarma}},
  \bibinfo {author} {\bibfnamefont {O.}~\bibnamefont {Mena}}, \bibinfo {author}
  {\bibfnamefont {K.}~\bibnamefont {Freese}}, \bibinfo {author} {\bibfnamefont
  {M.}~\bibnamefont {Gerbino}}, \bibinfo {author} {\bibfnamefont
  {S.}~\bibnamefont {Ho}}, \ and\ \bibinfo {author} {\bibfnamefont
  {M.}~\bibnamefont {Lattanzi}},\ }\href {\doibase 10.1103/PhysRevD.96.123503}
  {\bibfield  {journal} {\bibinfo  {journal} {Phys. Rev. D}\ }\textbf {\bibinfo
  {volume} {96}},\ \bibinfo {pages} {123503} (\bibinfo {year} {2017})},\
  \Eprint {http://arxiv.org/abs/1701.08172} {arXiv:1701.08172 [astro-ph.CO]}
  \BibitemShut {NoStop}%
\bibitem [{\citenamefont {Li}\ \emph {et~al.}(2018)\citenamefont {Li},
  \citenamefont {Zhang}, \citenamefont {Du}, \citenamefont {Zhou},\ and\
  \citenamefont {Xu}}]{Li:2017iur}%
  \BibitemOpen
  \bibfield  {author} {\bibinfo {author} {\bibfnamefont {E.-K.}\ \bibnamefont
  {Li}}, \bibinfo {author} {\bibfnamefont {H.}~\bibnamefont {Zhang}}, \bibinfo
  {author} {\bibfnamefont {M.}~\bibnamefont {Du}}, \bibinfo {author}
  {\bibfnamefont {Z.-H.}\ \bibnamefont {Zhou}}, \ and\ \bibinfo {author}
  {\bibfnamefont {L.}~\bibnamefont {Xu}},\ }\href {\doibase
  10.1088/1475-7516/2018/08/042} {\bibfield  {journal} {\bibinfo  {journal}
  {JCAP}\ }\textbf {\bibinfo {volume} {08}},\ \bibinfo {pages} {042} (\bibinfo
  {year} {2018})},\ \Eprint {http://arxiv.org/abs/1703.01554} {arXiv:1703.01554
  [astro-ph.CO]} \BibitemShut {NoStop}%
\bibitem [{\citenamefont {Yang}\ \emph {et~al.}(2017)\citenamefont {Yang},
  \citenamefont {Nunes}, \citenamefont {Pan},\ and\ \citenamefont
  {Mota}}]{Yang:2017amu}%
  \BibitemOpen
  \bibfield  {author} {\bibinfo {author} {\bibfnamefont {W.}~\bibnamefont
  {Yang}}, \bibinfo {author} {\bibfnamefont {R.~C.}\ \bibnamefont {Nunes}},
  \bibinfo {author} {\bibfnamefont {S.}~\bibnamefont {Pan}}, \ and\ \bibinfo
  {author} {\bibfnamefont {D.~F.}\ \bibnamefont {Mota}},\ }\href {\doibase
  10.1103/PhysRevD.95.103522} {\bibfield  {journal} {\bibinfo  {journal} {Phys.
  Rev. D}\ }\textbf {\bibinfo {volume} {95}},\ \bibinfo {pages} {103522}
  (\bibinfo {year} {2017})},\ \Eprint {http://arxiv.org/abs/1703.02556}
  {arXiv:1703.02556 [astro-ph.CO]} \BibitemShut {NoStop}%
\bibitem [{\citenamefont {Feng}\ \emph {et~al.}(2017)\citenamefont {Feng},
  \citenamefont {Zhang},\ and\ \citenamefont {Zhang}}]{Feng:2017nss}%
  \BibitemOpen
  \bibfield  {author} {\bibinfo {author} {\bibfnamefont {L.}~\bibnamefont
  {Feng}}, \bibinfo {author} {\bibfnamefont {J.-F.}\ \bibnamefont {Zhang}}, \
  and\ \bibinfo {author} {\bibfnamefont {X.}~\bibnamefont {Zhang}},\ }\href
  {\doibase 10.1140/epjc/s10052-017-4986-3} {\bibfield  {journal} {\bibinfo
  {journal} {Eur. Phys. J. C}\ }\textbf {\bibinfo {volume} {77}},\ \bibinfo
  {pages} {418} (\bibinfo {year} {2017})},\ \Eprint
  {http://arxiv.org/abs/1703.04884} {arXiv:1703.04884 [astro-ph.CO]}
  \BibitemShut {NoStop}%
\bibitem [{\citenamefont {Feng}\ \emph {et~al.}(2018)\citenamefont {Feng},
  \citenamefont {Zhang},\ and\ \citenamefont {Zhang}}]{Feng:2017mfs}%
  \BibitemOpen
  \bibfield  {author} {\bibinfo {author} {\bibfnamefont {L.}~\bibnamefont
  {Feng}}, \bibinfo {author} {\bibfnamefont {J.-F.}\ \bibnamefont {Zhang}}, \
  and\ \bibinfo {author} {\bibfnamefont {X.}~\bibnamefont {Zhang}},\ }\href
  {\doibase 10.1007/s11433-017-9150-3} {\bibfield  {journal} {\bibinfo
  {journal} {Sci. China Phys. Mech. Astron.}\ }\textbf {\bibinfo {volume}
  {61}},\ \bibinfo {pages} {050411} (\bibinfo {year} {2018})},\ \Eprint
  {http://arxiv.org/abs/1706.06913} {arXiv:1706.06913 [astro-ph.CO]}
  \BibitemShut {NoStop}%
\bibitem [{\citenamefont {Wang}\ \emph
  {et~al.}(2018{\natexlab{b}})\citenamefont {Wang}, \citenamefont {Wang},\ and\
  \citenamefont {Xia}}]{Wang:2017htc}%
  \BibitemOpen
  \bibfield  {author} {\bibinfo {author} {\bibfnamefont {S.}~\bibnamefont
  {Wang}}, \bibinfo {author} {\bibfnamefont {Y.-F.}\ \bibnamefont {Wang}}, \
  and\ \bibinfo {author} {\bibfnamefont {D.-M.}\ \bibnamefont {Xia}},\ }\href
  {\doibase 10.1088/1674-1137/42/6/065103} {\bibfield  {journal} {\bibinfo
  {journal} {Chin. Phys. C}\ }\textbf {\bibinfo {volume} {42}},\ \bibinfo
  {pages} {065103} (\bibinfo {year} {2018}{\natexlab{b}})},\ \Eprint
  {http://arxiv.org/abs/1707.00588} {arXiv:1707.00588 [astro-ph.CO]}
  \BibitemShut {NoStop}%
\bibitem [{\citenamefont {Feng}\ \emph {et~al.}(2019)\citenamefont {Feng},
  \citenamefont {Zhang},\ and\ \citenamefont {Zhang}}]{Feng:2017usu}%
  \BibitemOpen
  \bibfield  {author} {\bibinfo {author} {\bibfnamefont {L.}~\bibnamefont
  {Feng}}, \bibinfo {author} {\bibfnamefont {J.-F.}\ \bibnamefont {Zhang}}, \
  and\ \bibinfo {author} {\bibfnamefont {X.}~\bibnamefont {Zhang}},\ }\href
  {\doibase 10.1016/j.dark.2018.100261} {\bibfield  {journal} {\bibinfo
  {journal} {Phys. Dark Univ.}\ }\textbf {\bibinfo {volume} {23}},\ \bibinfo
  {pages} {100261} (\bibinfo {year} {2019})},\ \Eprint
  {http://arxiv.org/abs/1712.03148} {arXiv:1712.03148 [astro-ph.CO]}
  \BibitemShut {NoStop}%
\bibitem [{\citenamefont {Vagnozzi}\ \emph {et~al.}(2018)\citenamefont
  {Vagnozzi}, \citenamefont {Dhawan}, \citenamefont {Gerbino}, \citenamefont
  {Freese}, \citenamefont {Goobar},\ and\ \citenamefont
  {Mena}}]{Vagnozzi:2018jhn}%
  \BibitemOpen
  \bibfield  {author} {\bibinfo {author} {\bibfnamefont {S.}~\bibnamefont
  {Vagnozzi}}, \bibinfo {author} {\bibfnamefont {S.}~\bibnamefont {Dhawan}},
  \bibinfo {author} {\bibfnamefont {M.}~\bibnamefont {Gerbino}}, \bibinfo
  {author} {\bibfnamefont {K.}~\bibnamefont {Freese}}, \bibinfo {author}
  {\bibfnamefont {A.}~\bibnamefont {Goobar}}, \ and\ \bibinfo {author}
  {\bibfnamefont {O.}~\bibnamefont {Mena}},\ }\href {\doibase
  10.1103/PhysRevD.98.083501} {\bibfield  {journal} {\bibinfo  {journal} {Phys.
  Rev. D}\ }\textbf {\bibinfo {volume} {98}},\ \bibinfo {pages} {083501}
  (\bibinfo {year} {2018})},\ \Eprint {http://arxiv.org/abs/1801.08553}
  {arXiv:1801.08553 [astro-ph.CO]} \BibitemShut {NoStop}%
\bibitem [{\citenamefont {Knee}\ \emph {et~al.}(2019)\citenamefont {Knee},
  \citenamefont {Contreras},\ and\ \citenamefont {Scott}}]{Knee:2018rvj}%
  \BibitemOpen
  \bibfield  {author} {\bibinfo {author} {\bibfnamefont {A.~M.}\ \bibnamefont
  {Knee}}, \bibinfo {author} {\bibfnamefont {D.}~\bibnamefont {Contreras}}, \
  and\ \bibinfo {author} {\bibfnamefont {D.}~\bibnamefont {Scott}},\ }\href
  {\doibase 10.1088/1475-7516/2019/07/039} {\bibfield  {journal} {\bibinfo
  {journal} {JCAP}\ }\textbf {\bibinfo {volume} {07}},\ \bibinfo {pages} {039}
  (\bibinfo {year} {2019})},\ \Eprint {http://arxiv.org/abs/1812.02102}
  {arXiv:1812.02102 [astro-ph.CO]} \BibitemShut {NoStop}%
\bibitem [{\citenamefont {Feng}\ \emph
  {et~al.}(2020{\natexlab{a}})\citenamefont {Feng}, \citenamefont {Li},
  \citenamefont {Zhang},\ and\ \citenamefont {Zhang}}]{Feng:2019mym}%
  \BibitemOpen
  \bibfield  {author} {\bibinfo {author} {\bibfnamefont {L.}~\bibnamefont
  {Feng}}, \bibinfo {author} {\bibfnamefont {H.-L.}\ \bibnamefont {Li}},
  \bibinfo {author} {\bibfnamefont {J.-F.}\ \bibnamefont {Zhang}}, \ and\
  \bibinfo {author} {\bibfnamefont {X.}~\bibnamefont {Zhang}},\ }\href
  {\doibase 10.1007/s11433-019-9431-9} {\bibfield  {journal} {\bibinfo
  {journal} {Sci. China Phys. Mech. Astron.}\ }\textbf {\bibinfo {volume}
  {63}},\ \bibinfo {pages} {220401} (\bibinfo {year} {2020}{\natexlab{a}})},\
  \Eprint {http://arxiv.org/abs/1903.08848} {arXiv:1903.08848 [astro-ph.CO]}
  \BibitemShut {NoStop}%
\bibitem [{\citenamefont {Feng}\ \emph
  {et~al.}(2020{\natexlab{b}})\citenamefont {Feng}, \citenamefont {He},
  \citenamefont {Li}, \citenamefont {Zhang},\ and\ \citenamefont
  {Zhang}}]{Feng:2019jqa}%
  \BibitemOpen
  \bibfield  {author} {\bibinfo {author} {\bibfnamefont {L.}~\bibnamefont
  {Feng}}, \bibinfo {author} {\bibfnamefont {D.-Z.}\ \bibnamefont {He}},
  \bibinfo {author} {\bibfnamefont {H.-L.}\ \bibnamefont {Li}}, \bibinfo
  {author} {\bibfnamefont {J.-F.}\ \bibnamefont {Zhang}}, \ and\ \bibinfo
  {author} {\bibfnamefont {X.}~\bibnamefont {Zhang}},\ }\href {\doibase
  10.1007/s11433-019-1511-8} {\bibfield  {journal} {\bibinfo  {journal} {Sci.
  China Phys. Mech. Astron.}\ }\textbf {\bibinfo {volume} {63}},\ \bibinfo
  {pages} {290404} (\bibinfo {year} {2020}{\natexlab{b}})},\ \Eprint
  {http://arxiv.org/abs/1910.03872} {arXiv:1910.03872 [astro-ph.CO]}
  \BibitemShut {NoStop}%
\bibitem [{\citenamefont {Liu}\ and\ \citenamefont {Miao}(2020)}]{Liu:2020vgn}%
  \BibitemOpen
  \bibfield  {author} {\bibinfo {author} {\bibfnamefont {Z.}~\bibnamefont
  {Liu}}\ and\ \bibinfo {author} {\bibfnamefont {H.}~\bibnamefont {Miao}},\
  }\href {\doibase 10.1142/S0218271820500881} {\bibfield  {journal} {\bibinfo
  {journal} {Int. J. Mod. Phys. D}\ }\textbf {\bibinfo {volume} {29}},\
  \bibinfo {pages} {2050088} (\bibinfo {year} {2020})},\ \Eprint
  {http://arxiv.org/abs/2002.05563} {arXiv:2002.05563 [astro-ph.CO]}
  \BibitemShut {NoStop}%
\bibitem [{\citenamefont {Yang}\ \emph {et~al.}(2020)\citenamefont {Yang},
  \citenamefont {Di~Valentino}, \citenamefont {Mena},\ and\ \citenamefont
  {Pan}}]{Yang:2020tax}%
  \BibitemOpen
  \bibfield  {author} {\bibinfo {author} {\bibfnamefont {W.}~\bibnamefont
  {Yang}}, \bibinfo {author} {\bibfnamefont {E.}~\bibnamefont {Di~Valentino}},
  \bibinfo {author} {\bibfnamefont {O.}~\bibnamefont {Mena}}, \ and\ \bibinfo
  {author} {\bibfnamefont {S.}~\bibnamefont {Pan}},\ }\href {\doibase
  10.1103/PhysRevD.102.023535} {\bibfield  {journal} {\bibinfo  {journal}
  {Phys. Rev. D}\ }\textbf {\bibinfo {volume} {102}},\ \bibinfo {pages}
  {023535} (\bibinfo {year} {2020})},\ \Eprint
  {http://arxiv.org/abs/2003.12552} {arXiv:2003.12552 [astro-ph.CO]}
  \BibitemShut {NoStop}%
\bibitem [{\citenamefont {Zhang}\ \emph {et~al.}(2020)\citenamefont {Zhang},
  \citenamefont {Zhang},\ and\ \citenamefont {Zhang}}]{Zhang:2020mox}%
  \BibitemOpen
  \bibfield  {author} {\bibinfo {author} {\bibfnamefont {M.}~\bibnamefont
  {Zhang}}, \bibinfo {author} {\bibfnamefont {J.-F.}\ \bibnamefont {Zhang}}, \
  and\ \bibinfo {author} {\bibfnamefont {X.}~\bibnamefont {Zhang}},\ }\href
  {\doibase 10.1088/1572-9494/abbb84} {\bibfield  {journal} {\bibinfo
  {journal} {Commun. Theor. Phys.}\ }\textbf {\bibinfo {volume} {72}},\
  \bibinfo {pages} {125402} (\bibinfo {year} {2020})},\ \Eprint
  {http://arxiv.org/abs/2005.04647} {arXiv:2005.04647 [astro-ph.CO]}
  \BibitemShut {NoStop}%
\bibitem [{\citenamefont {Yang}\ \emph {et~al.}(2021)\citenamefont {Yang},
  \citenamefont {Di~Valentino}, \citenamefont {Pan},\ and\ \citenamefont
  {Mena}}]{Yang:2020ope}%
  \BibitemOpen
  \bibfield  {author} {\bibinfo {author} {\bibfnamefont {W.}~\bibnamefont
  {Yang}}, \bibinfo {author} {\bibfnamefont {E.}~\bibnamefont {Di~Valentino}},
  \bibinfo {author} {\bibfnamefont {S.}~\bibnamefont {Pan}}, \ and\ \bibinfo
  {author} {\bibfnamefont {O.}~\bibnamefont {Mena}},\ }\href {\doibase
  10.1016/j.dark.2020.100762} {\bibfield  {journal} {\bibinfo  {journal} {Phys.
  Dark Univ.}\ }\textbf {\bibinfo {volume} {31}},\ \bibinfo {pages} {100762}
  (\bibinfo {year} {2021})},\ \Eprint {http://arxiv.org/abs/2007.02927}
  {arXiv:2007.02927 [astro-ph.CO]} \BibitemShut {NoStop}%
\bibitem [{\citenamefont {Feng}\ \emph {et~al.}(2022)\citenamefont {Feng},
  \citenamefont {Guo}, \citenamefont {Zhang},\ and\ \citenamefont
  {Zhang}}]{Feng:2021ipq}%
  \BibitemOpen
  \bibfield  {author} {\bibinfo {author} {\bibfnamefont {L.}~\bibnamefont
  {Feng}}, \bibinfo {author} {\bibfnamefont {R.-Y.}\ \bibnamefont {Guo}},
  \bibinfo {author} {\bibfnamefont {J.-F.}\ \bibnamefont {Zhang}}, \ and\
  \bibinfo {author} {\bibfnamefont {X.}~\bibnamefont {Zhang}},\ }\href
  {\doibase 10.1016/j.physletb.2022.136940} {\bibfield  {journal} {\bibinfo
  {journal} {Phys. Lett. B}\ }\textbf {\bibinfo {volume} {827}},\ \bibinfo
  {pages} {136940} (\bibinfo {year} {2022})},\ \Eprint
  {http://arxiv.org/abs/2109.06111} {arXiv:2109.06111 [astro-ph.CO]}
  \BibitemShut {NoStop}%
\bibitem [{\citenamefont {Di~Valentino}\ \emph {et~al.}(2022)\citenamefont
  {Di~Valentino}, \citenamefont {Gariazzo}, \citenamefont {Giunti},
  \citenamefont {Mena}, \citenamefont {Pan},\ and\ \citenamefont
  {Yang}}]{DiValentino:2021rjj}%
  \BibitemOpen
  \bibfield  {author} {\bibinfo {author} {\bibfnamefont {E.}~\bibnamefont
  {Di~Valentino}}, \bibinfo {author} {\bibfnamefont {S.}~\bibnamefont
  {Gariazzo}}, \bibinfo {author} {\bibfnamefont {C.}~\bibnamefont {Giunti}},
  \bibinfo {author} {\bibfnamefont {O.}~\bibnamefont {Mena}}, \bibinfo {author}
  {\bibfnamefont {S.}~\bibnamefont {Pan}}, \ and\ \bibinfo {author}
  {\bibfnamefont {W.}~\bibnamefont {Yang}},\ }\href {\doibase
  10.1103/PhysRevD.105.103511} {\bibfield  {journal} {\bibinfo  {journal}
  {Phys. Rev. D}\ }\textbf {\bibinfo {volume} {105}},\ \bibinfo {pages}
  {103511} (\bibinfo {year} {2022})},\ \Eprint
  {http://arxiv.org/abs/2110.03990} {arXiv:2110.03990 [astro-ph.CO]}
  \BibitemShut {NoStop}%
\bibitem [{\citenamefont {Tanseri}\ \emph {et~al.}(2022)\citenamefont
  {Tanseri}, \citenamefont {Hagstotz}, \citenamefont {Vagnozzi}, \citenamefont
  {Giusarma},\ and\ \citenamefont {Freese}}]{Tanseri:2022zfe}%
  \BibitemOpen
  \bibfield  {author} {\bibinfo {author} {\bibfnamefont {I.}~\bibnamefont
  {Tanseri}}, \bibinfo {author} {\bibfnamefont {S.}~\bibnamefont {Hagstotz}},
  \bibinfo {author} {\bibfnamefont {S.}~\bibnamefont {Vagnozzi}}, \bibinfo
  {author} {\bibfnamefont {E.}~\bibnamefont {Giusarma}}, \ and\ \bibinfo
  {author} {\bibfnamefont {K.}~\bibnamefont {Freese}},\ }\href {\doibase
  10.1016/j.jheap.2022.07.002} {\bibfield  {journal} {\bibinfo  {journal}
  {JHEAp}\ }\textbf {\bibinfo {volume} {36}},\ \bibinfo {pages} {1} (\bibinfo
  {year} {2022})},\ \Eprint {http://arxiv.org/abs/2207.01913} {arXiv:2207.01913
  [astro-ph.CO]} \BibitemShut {NoStop}%
\bibitem [{\citenamefont {Chernikov}\ and\ \citenamefont
  {Ivanchik}(2022)}]{Chernikov:2022mdn}%
  \BibitemOpen
  \bibfield  {author} {\bibinfo {author} {\bibfnamefont {P.~A.}\ \bibnamefont
  {Chernikov}}\ and\ \bibinfo {author} {\bibfnamefont {A.~V.}\ \bibnamefont
  {Ivanchik}},\ }\href {\doibase 10.1134/S1063773722110056} {\bibfield
  {journal} {\bibinfo  {journal} {Astron. Lett.}\ }\textbf {\bibinfo {volume}
  {48}},\ \bibinfo {pages} {689} (\bibinfo {year} {2022})},\ \Eprint
  {http://arxiv.org/abs/2302.05251} {arXiv:2302.05251 [astro-ph.CO]}
  \BibitemShut {NoStop}%
\bibitem [{\citenamefont {Pang}\ \emph {et~al.}(2024)\citenamefont {Pang},
  \citenamefont {Zhang},\ and\ \citenamefont {Huang}}]{Pang:2023joc}%
  \BibitemOpen
  \bibfield  {author} {\bibinfo {author} {\bibfnamefont {Y.-H.}\ \bibnamefont
  {Pang}}, \bibinfo {author} {\bibfnamefont {X.}~\bibnamefont {Zhang}}, \ and\
  \bibinfo {author} {\bibfnamefont {Q.-G.}\ \bibnamefont {Huang}},\ }\href
  {\doibase 10.1088/1674-1137/ad34c0} {\bibfield  {journal} {\bibinfo
  {journal} {Chin. Phys. C}\ }\textbf {\bibinfo {volume} {48}},\ \bibinfo
  {pages} {065102} (\bibinfo {year} {2024})},\ \Eprint
  {http://arxiv.org/abs/2312.07188} {arXiv:2312.07188 [astro-ph.CO]}
  \BibitemShut {NoStop}%
\bibitem [{\citenamefont {Pan}\ \emph {et~al.}(2024)\citenamefont {Pan},
  \citenamefont {Seto}, \citenamefont {Takahashi},\ and\ \citenamefont
  {Toda}}]{Pan:2023frx}%
  \BibitemOpen
  \bibfield  {author} {\bibinfo {author} {\bibfnamefont {S.}~\bibnamefont
  {Pan}}, \bibinfo {author} {\bibfnamefont {O.}~\bibnamefont {Seto}}, \bibinfo
  {author} {\bibfnamefont {T.}~\bibnamefont {Takahashi}}, \ and\ \bibinfo
  {author} {\bibfnamefont {Y.}~\bibnamefont {Toda}},\ }\href {\doibase
  10.1103/PhysRevD.110.083524} {\bibfield  {journal} {\bibinfo  {journal}
  {Phys. Rev. D}\ }\textbf {\bibinfo {volume} {110}},\ \bibinfo {pages}
  {083524} (\bibinfo {year} {2024})},\ \Eprint
  {http://arxiv.org/abs/2312.15435} {arXiv:2312.15435 [astro-ph.CO]}
  \BibitemShut {NoStop}%
\bibitem [{\citenamefont {Abdul~Karim}\ \emph {et~al.}(2025)\citenamefont
  {Abdul~Karim} \emph {et~al.}}]{DESI:2025zgx}%
  \BibitemOpen
  \bibfield  {author} {\bibinfo {author} {\bibfnamefont {M.}~\bibnamefont
  {Abdul~Karim}} \emph {et~al.} (\bibinfo {collaboration} {DESI}),\ }\href@noop
  {} {\  (\bibinfo {year} {2025})},\ \Eprint {http://arxiv.org/abs/2503.14738}
  {arXiv:2503.14738 [astro-ph.CO]} \BibitemShut {NoStop}%
\bibitem [{\citenamefont {Adame}\ \emph {et~al.}(2025)\citenamefont {Adame}
  \emph {et~al.}}]{DESI:2024mwx}%
  \BibitemOpen
  \bibfield  {author} {\bibinfo {author} {\bibfnamefont {A.~G.}\ \bibnamefont
  {Adame}} \emph {et~al.} (\bibinfo {collaboration} {DESI}),\ }\href {\doibase
  10.1088/1475-7516/2025/02/021} {\bibfield  {journal} {\bibinfo  {journal}
  {JCAP}\ }\textbf {\bibinfo {volume} {02}},\ \bibinfo {pages} {021} (\bibinfo
  {year} {2025})},\ \Eprint {http://arxiv.org/abs/2404.03002} {arXiv:2404.03002
  [astro-ph.CO]} \BibitemShut {NoStop}%
\bibitem [{\citenamefont {Giar{\`e}}\ \emph
  {et~al.}(2024{\natexlab{a}})\citenamefont {Giar{\`e}}, \citenamefont
  {Sabogal}, \citenamefont {Nunes},\ and\ \citenamefont
  {Di~Valentino}}]{Giare:2024smz}%
  \BibitemOpen
  \bibfield  {author} {\bibinfo {author} {\bibfnamefont {W.}~\bibnamefont
  {Giar{\`e}}}, \bibinfo {author} {\bibfnamefont {M.~A.}\ \bibnamefont
  {Sabogal}}, \bibinfo {author} {\bibfnamefont {R.~C.}\ \bibnamefont {Nunes}},
  \ and\ \bibinfo {author} {\bibfnamefont {E.}~\bibnamefont {Di~Valentino}},\
  }\href {\doibase 10.1103/PhysRevLett.133.251003} {\bibfield  {journal}
  {\bibinfo  {journal} {Phys. Rev. Lett.}\ }\textbf {\bibinfo {volume} {133}},\
  \bibinfo {pages} {251003} (\bibinfo {year} {2024}{\natexlab{a}})},\ \Eprint
  {http://arxiv.org/abs/2404.15232} {arXiv:2404.15232 [astro-ph.CO]}
  \BibitemShut {NoStop}%
\bibitem [{\citenamefont {Wang}\ and\ \citenamefont
  {Piao}(2024)}]{Wang:2024dka}%
  \BibitemOpen
  \bibfield  {author} {\bibinfo {author} {\bibfnamefont {H.}~\bibnamefont
  {Wang}}\ and\ \bibinfo {author} {\bibfnamefont {Y.-S.}\ \bibnamefont
  {Piao}},\ }\href@noop {} {\  (\bibinfo {year} {2024})},\ \Eprint
  {http://arxiv.org/abs/2404.18579} {arXiv:2404.18579 [astro-ph.CO]}
  \BibitemShut {NoStop}%
\bibitem [{\citenamefont {Li}\ \emph {et~al.}(2024{\natexlab{b}})\citenamefont
  {Li}, \citenamefont {Wu}, \citenamefont {Du}, \citenamefont {Jin},
  \citenamefont {Li}, \citenamefont {Zhang},\ and\ \citenamefont
  {Zhang}}]{Li:2024qso}%
  \BibitemOpen
  \bibfield  {author} {\bibinfo {author} {\bibfnamefont {T.-N.}\ \bibnamefont
  {Li}}, \bibinfo {author} {\bibfnamefont {P.-J.}\ \bibnamefont {Wu}}, \bibinfo
  {author} {\bibfnamefont {G.-H.}\ \bibnamefont {Du}}, \bibinfo {author}
  {\bibfnamefont {S.-J.}\ \bibnamefont {Jin}}, \bibinfo {author} {\bibfnamefont
  {H.-L.}\ \bibnamefont {Li}}, \bibinfo {author} {\bibfnamefont {J.-F.}\
  \bibnamefont {Zhang}}, \ and\ \bibinfo {author} {\bibfnamefont
  {X.}~\bibnamefont {Zhang}},\ }\href {\doibase 10.3847/1538-4357/ad87f0}
  {\bibfield  {journal} {\bibinfo  {journal} {Astrophys. J.}\ }\textbf
  {\bibinfo {volume} {976}},\ \bibinfo {pages} {1} (\bibinfo {year}
  {2024}{\natexlab{b}})},\ \Eprint {http://arxiv.org/abs/2407.14934}
  {arXiv:2407.14934 [astro-ph.CO]} \BibitemShut {NoStop}%
\bibitem [{\citenamefont {Giar{\`e}}\ \emph
  {et~al.}(2024{\natexlab{b}})\citenamefont {Giar{\`e}}, \citenamefont
  {Najafi}, \citenamefont {Pan}, \citenamefont {Di~Valentino},\ and\
  \citenamefont {Firouzjaee}}]{Giare:2024gpk}%
  \BibitemOpen
  \bibfield  {author} {\bibinfo {author} {\bibfnamefont {W.}~\bibnamefont
  {Giar{\`e}}}, \bibinfo {author} {\bibfnamefont {M.}~\bibnamefont {Najafi}},
  \bibinfo {author} {\bibfnamefont {S.}~\bibnamefont {Pan}}, \bibinfo {author}
  {\bibfnamefont {E.}~\bibnamefont {Di~Valentino}}, \ and\ \bibinfo {author}
  {\bibfnamefont {J.~T.}\ \bibnamefont {Firouzjaee}},\ }\href {\doibase
  10.1088/1475-7516/2024/10/035} {\bibfield  {journal} {\bibinfo  {journal}
  {JCAP}\ }\textbf {\bibinfo {volume} {10}},\ \bibinfo {pages} {035} (\bibinfo
  {year} {2024}{\natexlab{b}})},\ \Eprint {http://arxiv.org/abs/2407.16689}
  {arXiv:2407.16689 [astro-ph.CO]} \BibitemShut {NoStop}%
\bibitem [{\citenamefont {Dinda}\ and\ \citenamefont
  {Maartens}(2025)}]{Dinda:2024ktd}%
  \BibitemOpen
  \bibfield  {author} {\bibinfo {author} {\bibfnamefont {B.~R.}\ \bibnamefont
  {Dinda}}\ and\ \bibinfo {author} {\bibfnamefont {R.}~\bibnamefont
  {Maartens}},\ }\href {\doibase 10.1088/1475-7516/2025/01/120} {\bibfield
  {journal} {\bibinfo  {journal} {JCAP}\ }\textbf {\bibinfo {volume} {01}},\
  \bibinfo {pages} {120} (\bibinfo {year} {2025})},\ \Eprint
  {http://arxiv.org/abs/2407.17252} {arXiv:2407.17252 [astro-ph.CO]}
  \BibitemShut {NoStop}%
\bibitem [{\citenamefont {Sabogal}\ \emph {et~al.}(2024)\citenamefont
  {Sabogal}, \citenamefont {Silva}, \citenamefont {Nunes}, \citenamefont
  {Kumar}, \citenamefont {Di~Valentino},\ and\ \citenamefont
  {Giar{\`e}}}]{Sabogal:2024yha}%
  \BibitemOpen
  \bibfield  {author} {\bibinfo {author} {\bibfnamefont {M.~A.}\ \bibnamefont
  {Sabogal}}, \bibinfo {author} {\bibfnamefont {E.}~\bibnamefont {Silva}},
  \bibinfo {author} {\bibfnamefont {R.~C.}\ \bibnamefont {Nunes}}, \bibinfo
  {author} {\bibfnamefont {S.}~\bibnamefont {Kumar}}, \bibinfo {author}
  {\bibfnamefont {E.}~\bibnamefont {Di~Valentino}}, \ and\ \bibinfo {author}
  {\bibfnamefont {W.}~\bibnamefont {Giar{\`e}}},\ }\href {\doibase
  10.1103/PhysRevD.110.123508} {\bibfield  {journal} {\bibinfo  {journal}
  {Phys. Rev. D}\ }\textbf {\bibinfo {volume} {110}},\ \bibinfo {pages}
  {123508} (\bibinfo {year} {2024})},\ \Eprint
  {http://arxiv.org/abs/2408.12403} {arXiv:2408.12403 [astro-ph.CO]}
  \BibitemShut {NoStop}%
\bibitem [{\citenamefont {Escamilla}\ \emph {et~al.}(2024)\citenamefont
  {Escamilla}, \citenamefont {{\"O}z{\"u}lker}, \citenamefont {Akarsu},
  \citenamefont {Di~Valentino},\ and\ \citenamefont
  {V{\'a}zquez}}]{Escamilla:2024ahl}%
  \BibitemOpen
  \bibfield  {author} {\bibinfo {author} {\bibfnamefont {L.~A.}\ \bibnamefont
  {Escamilla}}, \bibinfo {author} {\bibfnamefont {E.}~\bibnamefont
  {{\"O}z{\"u}lker}}, \bibinfo {author} {\bibfnamefont {{\"O}.}~\bibnamefont
  {Akarsu}}, \bibinfo {author} {\bibfnamefont {E.}~\bibnamefont
  {Di~Valentino}}, \ and\ \bibinfo {author} {\bibfnamefont {J.~A.}\
  \bibnamefont {V{\'a}zquez}},\ }\href@noop {} {\  (\bibinfo {year} {2024})},\
  \Eprint {http://arxiv.org/abs/2408.12516} {arXiv:2408.12516 [astro-ph.CO]}
  \BibitemShut {NoStop}%
\bibitem [{\citenamefont {Li}\ \emph {et~al.}(2025{\natexlab{a}})\citenamefont
  {Li}, \citenamefont {Li}, \citenamefont {Du}, \citenamefont {Wu},
  \citenamefont {Feng}, \citenamefont {Zhang},\ and\ \citenamefont
  {Zhang}}]{Li:2024qus}%
  \BibitemOpen
  \bibfield  {author} {\bibinfo {author} {\bibfnamefont {T.-N.}\ \bibnamefont
  {Li}}, \bibinfo {author} {\bibfnamefont {Y.-H.}\ \bibnamefont {Li}}, \bibinfo
  {author} {\bibfnamefont {G.-H.}\ \bibnamefont {Du}}, \bibinfo {author}
  {\bibfnamefont {P.-J.}\ \bibnamefont {Wu}}, \bibinfo {author} {\bibfnamefont
  {L.}~\bibnamefont {Feng}}, \bibinfo {author} {\bibfnamefont {J.-F.}\
  \bibnamefont {Zhang}}, \ and\ \bibinfo {author} {\bibfnamefont
  {X.}~\bibnamefont {Zhang}},\ }\href {\doibase
  10.1140/epjc/s10052-025-14279-7} {\bibfield  {journal} {\bibinfo  {journal}
  {Eur. Phys. J. C}\ }\textbf {\bibinfo {volume} {85}},\ \bibinfo {pages} {608}
  (\bibinfo {year} {2025}{\natexlab{a}})},\ \Eprint
  {http://arxiv.org/abs/2411.08639} {arXiv:2411.08639 [astro-ph.CO]}
  \BibitemShut {NoStop}%
\bibitem [{\citenamefont {Li}\ and\ \citenamefont
  {Wang}(2025{\natexlab{a}})}]{Li:2024bwr}%
  \BibitemOpen
  \bibfield  {author} {\bibinfo {author} {\bibfnamefont {J.-X.}\ \bibnamefont
  {Li}}\ and\ \bibinfo {author} {\bibfnamefont {S.}~\bibnamefont {Wang}},\
  }\href {\doibase 10.1088/1475-7516/2025/07/047} {\bibfield  {journal}
  {\bibinfo  {journal} {JCAP}\ }\textbf {\bibinfo {volume} {07}},\ \bibinfo
  {pages} {047} (\bibinfo {year} {2025}{\natexlab{a}})},\ \Eprint
  {http://arxiv.org/abs/2412.09064} {arXiv:2412.09064 [astro-ph.CO]}
  \BibitemShut {NoStop}%
\bibitem [{\citenamefont {Li}\ \emph {et~al.}(2025{\natexlab{b}})\citenamefont
  {Li}, \citenamefont {Du}, \citenamefont {Li}, \citenamefont {Wu},
  \citenamefont {Jin}, \citenamefont {Zhang},\ and\ \citenamefont
  {Zhang}}]{Li:2025owk}%
  \BibitemOpen
  \bibfield  {author} {\bibinfo {author} {\bibfnamefont {T.-N.}\ \bibnamefont
  {Li}}, \bibinfo {author} {\bibfnamefont {G.-H.}\ \bibnamefont {Du}}, \bibinfo
  {author} {\bibfnamefont {Y.-H.}\ \bibnamefont {Li}}, \bibinfo {author}
  {\bibfnamefont {P.-J.}\ \bibnamefont {Wu}}, \bibinfo {author} {\bibfnamefont
  {S.-J.}\ \bibnamefont {Jin}}, \bibinfo {author} {\bibfnamefont {J.-F.}\
  \bibnamefont {Zhang}}, \ and\ \bibinfo {author} {\bibfnamefont
  {X.}~\bibnamefont {Zhang}},\ }\href@noop {} {\  (\bibinfo {year}
  {2025}{\natexlab{b}})},\ \Eprint {http://arxiv.org/abs/2501.07361}
  {arXiv:2501.07361 [astro-ph.CO]} \BibitemShut {NoStop}%
\bibitem [{\citenamefont {Huang}\ \emph {et~al.}(2025)\citenamefont {Huang},
  \citenamefont {Cai},\ and\ \citenamefont {Wang}}]{Huang:2025som}%
  \BibitemOpen
  \bibfield  {author} {\bibinfo {author} {\bibfnamefont {L.}~\bibnamefont
  {Huang}}, \bibinfo {author} {\bibfnamefont {R.-G.}\ \bibnamefont {Cai}}, \
  and\ \bibinfo {author} {\bibfnamefont {S.-J.}\ \bibnamefont {Wang}},\ }\href
  {\doibase 10.1007/s11433-025-2754-5} {\bibfield  {journal} {\bibinfo
  {journal} {Sci. China Phys. Mech. Astron.}\ }\textbf {\bibinfo {volume}
  {68}},\ \bibinfo {pages} {100413} (\bibinfo {year} {2025})},\ \Eprint
  {http://arxiv.org/abs/2502.04212} {arXiv:2502.04212 [astro-ph.CO]}
  \BibitemShut {NoStop}%
\bibitem [{\citenamefont {Pang}\ \emph {et~al.}(2025)\citenamefont {Pang},
  \citenamefont {Zhang},\ and\ \citenamefont {Huang}}]{Pang:2025lvh}%
  \BibitemOpen
  \bibfield  {author} {\bibinfo {author} {\bibfnamefont {Y.-H.}\ \bibnamefont
  {Pang}}, \bibinfo {author} {\bibfnamefont {X.}~\bibnamefont {Zhang}}, \ and\
  \bibinfo {author} {\bibfnamefont {Q.-G.}\ \bibnamefont {Huang}},\ }\href
  {\doibase 10.1007/s11433-025-2713-8} {\bibfield  {journal} {\bibinfo
  {journal} {Sci. China Phys. Mech. Astron.}\ }\textbf {\bibinfo {volume}
  {68}},\ \bibinfo {pages} {280410} (\bibinfo {year} {2025})},\ \Eprint
  {http://arxiv.org/abs/2503.21600} {arXiv:2503.21600 [astro-ph.CO]}
  \BibitemShut {NoStop}%
\bibitem [{\citenamefont {You}\ \emph {et~al.}(2025)\citenamefont {You},
  \citenamefont {Wang},\ and\ \citenamefont {Yang}}]{You:2025uon}%
  \BibitemOpen
  \bibfield  {author} {\bibinfo {author} {\bibfnamefont {C.}~\bibnamefont
  {You}}, \bibinfo {author} {\bibfnamefont {D.}~\bibnamefont {Wang}}, \ and\
  \bibinfo {author} {\bibfnamefont {T.}~\bibnamefont {Yang}},\ }\href {\doibase
  10.1103/f6v7-n9fr} {\bibfield  {journal} {\bibinfo  {journal} {Phys. Rev. D}\
  }\textbf {\bibinfo {volume} {112}},\ \bibinfo {pages} {043503} (\bibinfo
  {year} {2025})},\ \Eprint {http://arxiv.org/abs/2504.00985} {arXiv:2504.00985
  [astro-ph.CO]} \BibitemShut {NoStop}%
\bibitem [{\citenamefont {Pan}\ \emph {et~al.}(2025)\citenamefont {Pan},
  \citenamefont {Paul}, \citenamefont {Saridakis},\ and\ \citenamefont
  {Yang}}]{Pan:2025qwy}%
  \BibitemOpen
  \bibfield  {author} {\bibinfo {author} {\bibfnamefont {S.}~\bibnamefont
  {Pan}}, \bibinfo {author} {\bibfnamefont {S.}~\bibnamefont {Paul}}, \bibinfo
  {author} {\bibfnamefont {E.~N.}\ \bibnamefont {Saridakis}}, \ and\ \bibinfo
  {author} {\bibfnamefont {W.}~\bibnamefont {Yang}},\ }\href@noop {} {\
  (\bibinfo {year} {2025})},\ \Eprint {http://arxiv.org/abs/2504.00994}
  {arXiv:2504.00994 [astro-ph.CO]} \BibitemShut {NoStop}%
\bibitem [{\citenamefont {Wu}(2025)}]{Wu:2025wyk}%
  \BibitemOpen
  \bibfield  {author} {\bibinfo {author} {\bibfnamefont {P.-J.}\ \bibnamefont
  {Wu}},\ }\href@noop {} {\  (\bibinfo {year} {2025})},\ \Eprint
  {http://arxiv.org/abs/2504.09054} {arXiv:2504.09054 [astro-ph.CO]}
  \BibitemShut {NoStop}%
\bibitem [{\citenamefont {Cheng}\ \emph {et~al.}(2025)\citenamefont {Cheng},
  \citenamefont {Di~Valentino}, \citenamefont {Escamilla}, \citenamefont
  {Sen},\ and\ \citenamefont {Visinelli}}]{Cheng:2025lod}%
  \BibitemOpen
  \bibfield  {author} {\bibinfo {author} {\bibfnamefont {H.}~\bibnamefont
  {Cheng}}, \bibinfo {author} {\bibfnamefont {E.}~\bibnamefont {Di~Valentino}},
  \bibinfo {author} {\bibfnamefont {L.~A.}\ \bibnamefont {Escamilla}}, \bibinfo
  {author} {\bibfnamefont {A.~A.}\ \bibnamefont {Sen}}, \ and\ \bibinfo
  {author} {\bibfnamefont {L.}~\bibnamefont {Visinelli}},\ }\href@noop {} {\
  (\bibinfo {year} {2025})},\ \Eprint {http://arxiv.org/abs/2505.02932}
  {arXiv:2505.02932 [astro-ph.CO]} \BibitemShut {NoStop}%
\bibitem [{\citenamefont {Yang}\ \emph {et~al.}(2025)\citenamefont {Yang},
  \citenamefont {Dai},\ and\ \citenamefont {Wang}}]{Yang:2025boq}%
  \BibitemOpen
  \bibfield  {author} {\bibinfo {author} {\bibfnamefont {Y.}~\bibnamefont
  {Yang}}, \bibinfo {author} {\bibfnamefont {X.}~\bibnamefont {Dai}}, \ and\
  \bibinfo {author} {\bibfnamefont {Y.}~\bibnamefont {Wang}},\ }\href {\doibase
  10.1103/8ync-vrtz} {\bibfield  {journal} {\bibinfo  {journal} {Phys. Rev. D}\
  }\textbf {\bibinfo {volume} {111}},\ \bibinfo {pages} {103534} (\bibinfo
  {year} {2025})},\ \Eprint {http://arxiv.org/abs/2505.09879} {arXiv:2505.09879
  [astro-ph.CO]} \BibitemShut {NoStop}%
\bibitem [{\citenamefont {Ling}\ \emph {et~al.}(2025)\citenamefont {Ling},
  \citenamefont {Du}, \citenamefont {Li}, \citenamefont {Zhang}, \citenamefont
  {Wang},\ and\ \citenamefont {Zhang}}]{Ling:2025lmw}%
  \BibitemOpen
  \bibfield  {author} {\bibinfo {author} {\bibfnamefont {J.-L.}\ \bibnamefont
  {Ling}}, \bibinfo {author} {\bibfnamefont {G.-H.}\ \bibnamefont {Du}},
  \bibinfo {author} {\bibfnamefont {T.-N.}\ \bibnamefont {Li}}, \bibinfo
  {author} {\bibfnamefont {J.-F.}\ \bibnamefont {Zhang}}, \bibinfo {author}
  {\bibfnamefont {S.-J.}\ \bibnamefont {Wang}}, \ and\ \bibinfo {author}
  {\bibfnamefont {X.}~\bibnamefont {Zhang}},\ }\href@noop {} {\  (\bibinfo
  {year} {2025})},\ \Eprint {http://arxiv.org/abs/2505.22369} {arXiv:2505.22369
  [astro-ph.CO]} \BibitemShut {NoStop}%
\bibitem [{\citenamefont {Barua}\ and\ \citenamefont
  {Desai}(2025)}]{Barua:2025ypw}%
  \BibitemOpen
  \bibfield  {author} {\bibinfo {author} {\bibfnamefont {S.}~\bibnamefont
  {Barua}}\ and\ \bibinfo {author} {\bibfnamefont {S.}~\bibnamefont {Desai}},\
  }\href {\doibase 10.1016/j.dark.2025.101995} {\bibfield  {journal} {\bibinfo
  {journal} {Phys. Dark Univ.}\ }\textbf {\bibinfo {volume} {49}},\ \bibinfo
  {pages} {101995} (\bibinfo {year} {2025})},\ \Eprint
  {http://arxiv.org/abs/2506.12709} {arXiv:2506.12709 [astro-ph.CO]}
  \BibitemShut {NoStop}%
\bibitem [{\citenamefont {Li}\ and\ \citenamefont {Zhang}(2025)}]{Li:2025ula}%
  \BibitemOpen
  \bibfield  {author} {\bibinfo {author} {\bibfnamefont {Y.-H.}\ \bibnamefont
  {Li}}\ and\ \bibinfo {author} {\bibfnamefont {X.}~\bibnamefont {Zhang}},\
  }\href@noop {} {\  (\bibinfo {year} {2025})},\ \Eprint
  {http://arxiv.org/abs/2506.18477} {arXiv:2506.18477 [astro-ph.CO]}
  \BibitemShut {NoStop}%
\bibitem [{\citenamefont {{\"O}z{\"u}lker}\ \emph {et~al.}(2025)\citenamefont
  {{\"O}z{\"u}lker}, \citenamefont {Di~Valentino},\ and\ \citenamefont
  {Giar{\`e}}}]{Ozulker:2025ehg}%
  \BibitemOpen
  \bibfield  {author} {\bibinfo {author} {\bibfnamefont {E.}~\bibnamefont
  {{\"O}z{\"u}lker}}, \bibinfo {author} {\bibfnamefont {E.}~\bibnamefont
  {Di~Valentino}}, \ and\ \bibinfo {author} {\bibfnamefont {W.}~\bibnamefont
  {Giar{\`e}}},\ }\href@noop {} {\  (\bibinfo {year} {2025})},\ \Eprint
  {http://arxiv.org/abs/2506.19053} {arXiv:2506.19053 [astro-ph.CO]}
  \BibitemShut {NoStop}%
\bibitem [{\citenamefont {Gialamas}\ \emph {et~al.}(2025)\citenamefont
  {Gialamas}, \citenamefont {H{\"u}tsi}, \citenamefont {Raidal}, \citenamefont
  {Urrutia}, \citenamefont {Vasar},\ and\ \citenamefont
  {Veerm{\"a}e}}]{Gialamas:2025pwv}%
  \BibitemOpen
  \bibfield  {author} {\bibinfo {author} {\bibfnamefont {I.~D.}\ \bibnamefont
  {Gialamas}}, \bibinfo {author} {\bibfnamefont {G.}~\bibnamefont {H{\"u}tsi}},
  \bibinfo {author} {\bibfnamefont {M.}~\bibnamefont {Raidal}}, \bibinfo
  {author} {\bibfnamefont {J.}~\bibnamefont {Urrutia}}, \bibinfo {author}
  {\bibfnamefont {M.}~\bibnamefont {Vasar}}, \ and\ \bibinfo {author}
  {\bibfnamefont {H.}~\bibnamefont {Veerm{\"a}e}},\ }\href@noop {} {\
  (\bibinfo {year} {2025})},\ \Eprint {http://arxiv.org/abs/2506.21542}
  {arXiv:2506.21542 [astro-ph.CO]} \BibitemShut {NoStop}%
\bibitem [{\citenamefont {Li}\ and\ \citenamefont
  {Wang}(2025{\natexlab{b}})}]{Li:2025ops}%
  \BibitemOpen
  \bibfield  {author} {\bibinfo {author} {\bibfnamefont {J.-X.}\ \bibnamefont
  {Li}}\ and\ \bibinfo {author} {\bibfnamefont {S.}~\bibnamefont {Wang}},\
  }\href@noop {} {\  (\bibinfo {year} {2025}{\natexlab{b}})},\ \Eprint
  {http://arxiv.org/abs/2506.22953} {arXiv:2506.22953 [astro-ph.CO]}
  \BibitemShut {NoStop}%
\bibitem [{\citenamefont {Liu}\ \emph {et~al.}(2025)\citenamefont {Liu},
  \citenamefont {Li}, \citenamefont {Xu}, \citenamefont {Biesiada},\ and\
  \citenamefont {Wang}}]{Liu:2025myr}%
  \BibitemOpen
  \bibfield  {author} {\bibinfo {author} {\bibfnamefont {T.}~\bibnamefont
  {Liu}}, \bibinfo {author} {\bibfnamefont {X.}~\bibnamefont {Li}}, \bibinfo
  {author} {\bibfnamefont {T.}~\bibnamefont {Xu}}, \bibinfo {author}
  {\bibfnamefont {M.}~\bibnamefont {Biesiada}}, \ and\ \bibinfo {author}
  {\bibfnamefont {J.}~\bibnamefont {Wang}},\ }\href@noop {} {\  (\bibinfo
  {year} {2025})},\ \Eprint {http://arxiv.org/abs/2507.04265} {arXiv:2507.04265
  [astro-ph.CO]} \BibitemShut {NoStop}%
\bibitem [{\citenamefont {Qiang}\ \emph {et~al.}(2025)\citenamefont {Qiang},
  \citenamefont {Jia},\ and\ \citenamefont {Wei}}]{Qiang:2025cxp}%
  \BibitemOpen
  \bibfield  {author} {\bibinfo {author} {\bibfnamefont {D.-C.}\ \bibnamefont
  {Qiang}}, \bibinfo {author} {\bibfnamefont {J.-Y.}\ \bibnamefont {Jia}}, \
  and\ \bibinfo {author} {\bibfnamefont {H.}~\bibnamefont {Wei}},\ }\href@noop
  {} {\  (\bibinfo {year} {2025})},\ \Eprint {http://arxiv.org/abs/2507.09981}
  {arXiv:2507.09981 [astro-ph.CO]} \BibitemShut {NoStop}%
\bibitem [{\citenamefont {Li}\ \emph {et~al.}(2025{\natexlab{c}})\citenamefont
  {Li}, \citenamefont {Du}, \citenamefont {Wu}, \citenamefont {Qi},
  \citenamefont {Zhang},\ and\ \citenamefont {Zhang}}]{Li:2025htp}%
  \BibitemOpen
  \bibfield  {author} {\bibinfo {author} {\bibfnamefont {T.-N.}\ \bibnamefont
  {Li}}, \bibinfo {author} {\bibfnamefont {G.-H.}\ \bibnamefont {Du}}, \bibinfo
  {author} {\bibfnamefont {P.-J.}\ \bibnamefont {Wu}}, \bibinfo {author}
  {\bibfnamefont {J.-Z.}\ \bibnamefont {Qi}}, \bibinfo {author} {\bibfnamefont
  {J.-F.}\ \bibnamefont {Zhang}}, \ and\ \bibinfo {author} {\bibfnamefont
  {X.}~\bibnamefont {Zhang}},\ }\href@noop {} {\  (\bibinfo {year}
  {2025}{\natexlab{c}})},\ \Eprint {http://arxiv.org/abs/2507.13811}
  {arXiv:2507.13811 [astro-ph.CO]} \BibitemShut {NoStop}%
\bibitem [{\citenamefont {Li}\ \emph {et~al.}(2025{\natexlab{d}})\citenamefont
  {Li}, \citenamefont {Zhang}, \citenamefont {Yao}, \citenamefont {Wu},
  \citenamefont {Zhang},\ and\ \citenamefont {Zhang}}]{Li:2025eqh}%
  \BibitemOpen
  \bibfield  {author} {\bibinfo {author} {\bibfnamefont {T.-N.}\ \bibnamefont
  {Li}}, \bibinfo {author} {\bibfnamefont {Y.-M.}\ \bibnamefont {Zhang}},
  \bibinfo {author} {\bibfnamefont {Y.-H.}\ \bibnamefont {Yao}}, \bibinfo
  {author} {\bibfnamefont {P.-J.}\ \bibnamefont {Wu}}, \bibinfo {author}
  {\bibfnamefont {J.-F.}\ \bibnamefont {Zhang}}, \ and\ \bibinfo {author}
  {\bibfnamefont {X.}~\bibnamefont {Zhang}},\ }\href@noop {} {\  (\bibinfo
  {year} {2025}{\natexlab{d}})},\ \Eprint {http://arxiv.org/abs/2506.09819}
  {arXiv:2506.09819 [astro-ph.CO]} \BibitemShut {NoStop}%
\bibitem [{\citenamefont {Li}\ \emph {et~al.}(2025{\natexlab{e}})\citenamefont
  {Li}, \citenamefont {Wu}, \citenamefont {Du}, \citenamefont {Yao},
  \citenamefont {Zhang},\ and\ \citenamefont {Zhang}}]{Li:2025dwz}%
  \BibitemOpen
  \bibfield  {author} {\bibinfo {author} {\bibfnamefont {T.-N.}\ \bibnamefont
  {Li}}, \bibinfo {author} {\bibfnamefont {P.-J.}\ \bibnamefont {Wu}}, \bibinfo
  {author} {\bibfnamefont {G.-H.}\ \bibnamefont {Du}}, \bibinfo {author}
  {\bibfnamefont {Y.-H.}\ \bibnamefont {Yao}}, \bibinfo {author} {\bibfnamefont
  {J.-F.}\ \bibnamefont {Zhang}}, \ and\ \bibinfo {author} {\bibfnamefont
  {X.}~\bibnamefont {Zhang}},\ }\href {\doibase 10.1016/j.dark.2025.102068}
  {\bibfield  {journal} {\bibinfo  {journal} {Phys. Dark Univ.}\ }\textbf
  {\bibinfo {volume} {50}},\ \bibinfo {pages} {102068} (\bibinfo {year}
  {2025}{\natexlab{e}})},\ \Eprint {http://arxiv.org/abs/2507.07798}
  {arXiv:2507.07798 [astro-ph.CO]} \BibitemShut {NoStop}%
\bibitem [{\citenamefont {Wu}\ \emph {et~al.}(2025)\citenamefont {Wu},
  \citenamefont {Li}, \citenamefont {Du},\ and\ \citenamefont
  {Zhang}}]{Wu:2025vfs}%
  \BibitemOpen
  \bibfield  {author} {\bibinfo {author} {\bibfnamefont {P.-J.}\ \bibnamefont
  {Wu}}, \bibinfo {author} {\bibfnamefont {T.-N.}\ \bibnamefont {Li}}, \bibinfo
  {author} {\bibfnamefont {G.-H.}\ \bibnamefont {Du}}, \ and\ \bibinfo {author}
  {\bibfnamefont {X.}~\bibnamefont {Zhang}},\ }\href@noop {} {\  (\bibinfo
  {year} {2025})},\ \Eprint {http://arxiv.org/abs/2509.02945} {arXiv:2509.02945
  [astro-ph.CO]} \BibitemShut {NoStop}%
\bibitem [{\citenamefont {Du}\ \emph {et~al.}(2025{\natexlab{a}})\citenamefont
  {Du}, \citenamefont {Wu}, \citenamefont {Li},\ and\ \citenamefont
  {Zhang}}]{Du:2024pai}%
  \BibitemOpen
  \bibfield  {author} {\bibinfo {author} {\bibfnamefont {G.-H.}\ \bibnamefont
  {Du}}, \bibinfo {author} {\bibfnamefont {P.-J.}\ \bibnamefont {Wu}}, \bibinfo
  {author} {\bibfnamefont {T.-N.}\ \bibnamefont {Li}}, \ and\ \bibinfo {author}
  {\bibfnamefont {X.}~\bibnamefont {Zhang}},\ }\href {\doibase
  10.1140/epjc/s10052-025-14094-0} {\bibfield  {journal} {\bibinfo  {journal}
  {Eur. Phys. J. C}\ }\textbf {\bibinfo {volume} {85}},\ \bibinfo {pages} {392}
  (\bibinfo {year} {2025}{\natexlab{a}})},\ \Eprint
  {http://arxiv.org/abs/2407.15640} {arXiv:2407.15640 [astro-ph.CO]}
  \BibitemShut {NoStop}%
\bibitem [{\citenamefont {Jiang}\ \emph {et~al.}(2025)\citenamefont {Jiang},
  \citenamefont {Giar{\`e}}, \citenamefont {Gariazzo}, \citenamefont
  {Dainotti}, \citenamefont {Di~Valentino}, \citenamefont {Mena}, \citenamefont
  {Pedrotti}, \citenamefont {da~Costa},\ and\ \citenamefont
  {Vagnozzi}}]{Jiang:2024viw}%
  \BibitemOpen
  \bibfield  {author} {\bibinfo {author} {\bibfnamefont {J.-Q.}\ \bibnamefont
  {Jiang}}, \bibinfo {author} {\bibfnamefont {W.}~\bibnamefont {Giar{\`e}}},
  \bibinfo {author} {\bibfnamefont {S.}~\bibnamefont {Gariazzo}}, \bibinfo
  {author} {\bibfnamefont {M.~G.}\ \bibnamefont {Dainotti}}, \bibinfo {author}
  {\bibfnamefont {E.}~\bibnamefont {Di~Valentino}}, \bibinfo {author}
  {\bibfnamefont {O.}~\bibnamefont {Mena}}, \bibinfo {author} {\bibfnamefont
  {D.}~\bibnamefont {Pedrotti}}, \bibinfo {author} {\bibfnamefont {S.~S.}\
  \bibnamefont {da~Costa}}, \ and\ \bibinfo {author} {\bibfnamefont
  {S.}~\bibnamefont {Vagnozzi}},\ }\href {\doibase
  10.1088/1475-7516/2025/01/153} {\bibfield  {journal} {\bibinfo  {journal}
  {JCAP}\ }\textbf {\bibinfo {volume} {01}},\ \bibinfo {pages} {153} (\bibinfo
  {year} {2025})},\ \Eprint {http://arxiv.org/abs/2407.18047} {arXiv:2407.18047
  [astro-ph.CO]} \BibitemShut {NoStop}%
\bibitem [{\citenamefont {Du}\ \emph {et~al.}(2025{\natexlab{b}})\citenamefont
  {Du}, \citenamefont {Li}, \citenamefont {Wu}, \citenamefont {Feng},
  \citenamefont {Zhou}, \citenamefont {Zhang},\ and\ \citenamefont
  {Zhang}}]{Du:2025iow}%
  \BibitemOpen
  \bibfield  {author} {\bibinfo {author} {\bibfnamefont {G.-H.}\ \bibnamefont
  {Du}}, \bibinfo {author} {\bibfnamefont {T.-N.}\ \bibnamefont {Li}}, \bibinfo
  {author} {\bibfnamefont {P.-J.}\ \bibnamefont {Wu}}, \bibinfo {author}
  {\bibfnamefont {L.}~\bibnamefont {Feng}}, \bibinfo {author} {\bibfnamefont
  {S.-H.}\ \bibnamefont {Zhou}}, \bibinfo {author} {\bibfnamefont {J.-F.}\
  \bibnamefont {Zhang}}, \ and\ \bibinfo {author} {\bibfnamefont
  {X.}~\bibnamefont {Zhang}},\ }\href@noop {} {\  (\bibinfo {year}
  {2025}{\natexlab{b}})},\ \Eprint {http://arxiv.org/abs/2501.10785}
  {arXiv:2501.10785 [astro-ph.CO]} \BibitemShut {NoStop}%
\bibitem [{\citenamefont {Feng}\ \emph
  {et~al.}(2025{\natexlab{b}})\citenamefont {Feng}, \citenamefont {Li},
  \citenamefont {Du}, \citenamefont {Zhang},\ and\ \citenamefont
  {Zhang}}]{Feng:2025mlo}%
  \BibitemOpen
  \bibfield  {author} {\bibinfo {author} {\bibfnamefont {L.}~\bibnamefont
  {Feng}}, \bibinfo {author} {\bibfnamefont {T.-N.}\ \bibnamefont {Li}},
  \bibinfo {author} {\bibfnamefont {G.-H.}\ \bibnamefont {Du}}, \bibinfo
  {author} {\bibfnamefont {J.-F.}\ \bibnamefont {Zhang}}, \ and\ \bibinfo
  {author} {\bibfnamefont {X.}~\bibnamefont {Zhang}},\ }\href {\doibase
  10.1016/j.dark.2025.101935} {\bibfield  {journal} {\bibinfo  {journal} {Phys.
  Dark Univ.}\ }\textbf {\bibinfo {volume} {48}},\ \bibinfo {pages} {101935}
  (\bibinfo {year} {2025}{\natexlab{b}})},\ \Eprint
  {http://arxiv.org/abs/2503.10423} {arXiv:2503.10423 [astro-ph.CO]}
  \BibitemShut {NoStop}%
\bibitem [{\citenamefont {Roy~Choudhury}(2025)}]{RoyChoudhury:2025dhe}%
  \BibitemOpen
  \bibfield  {author} {\bibinfo {author} {\bibfnamefont {S.}~\bibnamefont
  {Roy~Choudhury}},\ }\href {\doibase 10.3847/2041-8213/ade1cc} {\bibfield
  {journal} {\bibinfo  {journal} {Astrophys. J. Lett.}\ }\textbf {\bibinfo
  {volume} {986}},\ \bibinfo {pages} {L31} (\bibinfo {year} {2025})},\ \Eprint
  {http://arxiv.org/abs/2504.15340} {arXiv:2504.15340 [astro-ph.CO]}
  \BibitemShut {NoStop}%
\bibitem [{\citenamefont {Du}\ \emph {et~al.}(2025{\natexlab{c}})\citenamefont
  {Du}, \citenamefont {Li}, \citenamefont {Wu}, \citenamefont {Zhang},\ and\
  \citenamefont {Zhang}}]{Du:2025xes}%
  \BibitemOpen
  \bibfield  {author} {\bibinfo {author} {\bibfnamefont {G.-H.}\ \bibnamefont
  {Du}}, \bibinfo {author} {\bibfnamefont {T.-N.}\ \bibnamefont {Li}}, \bibinfo
  {author} {\bibfnamefont {P.-J.}\ \bibnamefont {Wu}}, \bibinfo {author}
  {\bibfnamefont {J.-F.}\ \bibnamefont {Zhang}}, \ and\ \bibinfo {author}
  {\bibfnamefont {X.}~\bibnamefont {Zhang}},\ }\href@noop {} {\  (\bibinfo
  {year} {2025}{\natexlab{c}})},\ \Eprint {http://arxiv.org/abs/2507.16589}
  {arXiv:2507.16589 [astro-ph.CO]} \BibitemShut {NoStop}%
\bibitem [{\citenamefont {Akita}\ and\ \citenamefont
  {Yamaguchi}(2020)}]{Akita:2020szl}%
  \BibitemOpen
  \bibfield  {author} {\bibinfo {author} {\bibfnamefont {K.}~\bibnamefont
  {Akita}}\ and\ \bibinfo {author} {\bibfnamefont {M.}~\bibnamefont
  {Yamaguchi}},\ }\href {\doibase 10.1088/1475-7516/2020/08/012} {\bibfield
  {journal} {\bibinfo  {journal} {JCAP}\ }\textbf {\bibinfo {volume} {08}},\
  \bibinfo {pages} {012} (\bibinfo {year} {2020})},\ \Eprint
  {http://arxiv.org/abs/2005.07047} {arXiv:2005.07047 [hep-ph]} \BibitemShut
  {NoStop}%
\bibitem [{\citenamefont {Froustey}\ \emph {et~al.}(2020)\citenamefont
  {Froustey}, \citenamefont {Pitrou},\ and\ \citenamefont
  {Volpe}}]{Froustey:2020mcq}%
  \BibitemOpen
  \bibfield  {author} {\bibinfo {author} {\bibfnamefont {J.}~\bibnamefont
  {Froustey}}, \bibinfo {author} {\bibfnamefont {C.}~\bibnamefont {Pitrou}}, \
  and\ \bibinfo {author} {\bibfnamefont {M.~C.}\ \bibnamefont {Volpe}},\ }\href
  {\doibase 10.1088/1475-7516/2020/12/015} {\bibfield  {journal} {\bibinfo
  {journal} {JCAP}\ }\textbf {\bibinfo {volume} {12}},\ \bibinfo {pages} {015}
  (\bibinfo {year} {2020})},\ \Eprint {http://arxiv.org/abs/2008.01074}
  {arXiv:2008.01074 [hep-ph]} \BibitemShut {NoStop}%
\bibitem [{\citenamefont {Bennett}\ \emph {et~al.}(2021)\citenamefont
  {Bennett}, \citenamefont {Buldgen}, \citenamefont {De~Salas}, \citenamefont
  {Drewes}, \citenamefont {Gariazzo}, \citenamefont {Pastor},\ and\
  \citenamefont {Wong}}]{Bennett:2020zkv}%
  \BibitemOpen
  \bibfield  {author} {\bibinfo {author} {\bibfnamefont {J.~J.}\ \bibnamefont
  {Bennett}}, \bibinfo {author} {\bibfnamefont {G.}~\bibnamefont {Buldgen}},
  \bibinfo {author} {\bibfnamefont {P.~F.}\ \bibnamefont {De~Salas}}, \bibinfo
  {author} {\bibfnamefont {M.}~\bibnamefont {Drewes}}, \bibinfo {author}
  {\bibfnamefont {S.}~\bibnamefont {Gariazzo}}, \bibinfo {author}
  {\bibfnamefont {S.}~\bibnamefont {Pastor}}, \ and\ \bibinfo {author}
  {\bibfnamefont {Y.~Y.~Y.}\ \bibnamefont {Wong}},\ }\href {\doibase
  10.1088/1475-7516/2021/04/073} {\bibfield  {journal} {\bibinfo  {journal}
  {JCAP}\ }\textbf {\bibinfo {volume} {04}},\ \bibinfo {pages} {073} (\bibinfo
  {year} {2021})},\ \Eprint {http://arxiv.org/abs/2012.02726} {arXiv:2012.02726
  [hep-ph]} \BibitemShut {NoStop}%
\bibitem [{\citenamefont {Li}(2004)}]{Li:2004rb}%
  \BibitemOpen
  \bibfield  {author} {\bibinfo {author} {\bibfnamefont {M.}~\bibnamefont
  {Li}},\ }\href {\doibase 10.1016/j.physletb.2004.10.014} {\bibfield
  {journal} {\bibinfo  {journal} {Phys. Lett. B}\ }\textbf {\bibinfo {volume}
  {603}},\ \bibinfo {pages} {1} (\bibinfo {year} {2004})},\ \Eprint
  {http://arxiv.org/abs/hep-th/0403127} {arXiv:hep-th/0403127} \BibitemShut
  {NoStop}%
\bibitem [{\citenamefont {Zhang}\ and\ \citenamefont
  {Wu}(2005)}]{Zhang:2005hs}%
  \BibitemOpen
  \bibfield  {author} {\bibinfo {author} {\bibfnamefont {X.}~\bibnamefont
  {Zhang}}\ and\ \bibinfo {author} {\bibfnamefont {F.-Q.}\ \bibnamefont {Wu}},\
  }\href {\doibase 10.1103/PhysRevD.72.043524} {\bibfield  {journal} {\bibinfo
  {journal} {Phys. Rev. D}\ }\textbf {\bibinfo {volume} {72}},\ \bibinfo
  {pages} {043524} (\bibinfo {year} {2005})},\ \Eprint
  {http://arxiv.org/abs/astro-ph/0506310} {arXiv:astro-ph/0506310} \BibitemShut
  {NoStop}%
\bibitem [{\citenamefont {Li}\ \emph {et~al.}(2009)\citenamefont {Li},
  \citenamefont {Li}, \citenamefont {Wang},\ and\ \citenamefont
  {Zhang}}]{Li:2009bn}%
  \BibitemOpen
  \bibfield  {author} {\bibinfo {author} {\bibfnamefont {M.}~\bibnamefont
  {Li}}, \bibinfo {author} {\bibfnamefont {X.-D.}\ \bibnamefont {Li}}, \bibinfo
  {author} {\bibfnamefont {S.}~\bibnamefont {Wang}}, \ and\ \bibinfo {author}
  {\bibfnamefont {X.}~\bibnamefont {Zhang}},\ }\href {\doibase
  10.1088/1475-7516/2009/06/036} {\bibfield  {journal} {\bibinfo  {journal}
  {JCAP}\ }\textbf {\bibinfo {volume} {06}},\ \bibinfo {pages} {036} (\bibinfo
  {year} {2009})},\ \Eprint {http://arxiv.org/abs/0904.0928} {arXiv:0904.0928
  [astro-ph.CO]} \BibitemShut {NoStop}%
\bibitem [{\citenamefont {Li}\ \emph {et~al.}(2013{\natexlab{b}})\citenamefont
  {Li}, \citenamefont {Li}, \citenamefont {Ma}, \citenamefont {Zhang},\ and\
  \citenamefont {Zhang}}]{Li:2013dha}%
  \BibitemOpen
  \bibfield  {author} {\bibinfo {author} {\bibfnamefont {M.}~\bibnamefont
  {Li}}, \bibinfo {author} {\bibfnamefont {X.-D.}\ \bibnamefont {Li}}, \bibinfo
  {author} {\bibfnamefont {Y.-Z.}\ \bibnamefont {Ma}}, \bibinfo {author}
  {\bibfnamefont {X.}~\bibnamefont {Zhang}}, \ and\ \bibinfo {author}
  {\bibfnamefont {Z.}~\bibnamefont {Zhang}},\ }\href {\doibase
  10.1088/1475-7516/2013/09/021} {\bibfield  {journal} {\bibinfo  {journal}
  {JCAP}\ }\textbf {\bibinfo {volume} {09}},\ \bibinfo {pages} {021} (\bibinfo
  {year} {2013}{\natexlab{b}})},\ \Eprint {http://arxiv.org/abs/1305.5302}
  {arXiv:1305.5302 [astro-ph.CO]} \BibitemShut {NoStop}%
\bibitem [{\citenamefont {Feng}\ and\ \citenamefont
  {Zhang}(2016)}]{Feng:2016djj}%
  \BibitemOpen
  \bibfield  {author} {\bibinfo {author} {\bibfnamefont {L.}~\bibnamefont
  {Feng}}\ and\ \bibinfo {author} {\bibfnamefont {X.}~\bibnamefont {Zhang}},\
  }\href {\doibase 10.1088/1475-7516/2016/08/072} {\bibfield  {journal}
  {\bibinfo  {journal} {JCAP}\ }\textbf {\bibinfo {volume} {08}},\ \bibinfo
  {pages} {072} (\bibinfo {year} {2016})},\ \Eprint
  {http://arxiv.org/abs/1607.05567} {arXiv:1607.05567 [astro-ph.CO]}
  \BibitemShut {NoStop}%
\bibitem [{\citenamefont {Chevallier}\ and\ \citenamefont
  {Polarski}(2001)}]{Chevallier:2000qy}%
  \BibitemOpen
  \bibfield  {author} {\bibinfo {author} {\bibfnamefont {M.}~\bibnamefont
  {Chevallier}}\ and\ \bibinfo {author} {\bibfnamefont {D.}~\bibnamefont
  {Polarski}},\ }\href {\doibase 10.1142/S0218271801000822} {\bibfield
  {journal} {\bibinfo  {journal} {Int. J. Mod. Phys. D}\ }\textbf {\bibinfo
  {volume} {10}},\ \bibinfo {pages} {213} (\bibinfo {year} {2001})},\ \Eprint
  {http://arxiv.org/abs/gr-qc/0009008} {arXiv:gr-qc/0009008} \BibitemShut
  {NoStop}%
\bibitem [{\citenamefont {Linder}(2003)}]{Linder:2002et}%
  \BibitemOpen
  \bibfield  {author} {\bibinfo {author} {\bibfnamefont {E.~V.}\ \bibnamefont
  {Linder}},\ }\href {\doibase 10.1103/PhysRevLett.90.091301} {\bibfield
  {journal} {\bibinfo  {journal} {Phys. Rev. Lett.}\ }\textbf {\bibinfo
  {volume} {90}},\ \bibinfo {pages} {091301} (\bibinfo {year} {2003})},\
  \Eprint {http://arxiv.org/abs/astro-ph/0208512} {arXiv:astro-ph/0208512}
  \BibitemShut {NoStop}%
\bibitem [{\citenamefont {Aghanim}\ \emph {et~al.}(2020)\citenamefont {Aghanim}
  \emph {et~al.}}]{Planck:2018vyg}%
  \BibitemOpen
  \bibfield  {author} {\bibinfo {author} {\bibfnamefont {N.}~\bibnamefont
  {Aghanim}} \emph {et~al.} (\bibinfo {collaboration} {Planck}),\ }\href
  {\doibase 10.1051/0004-6361/201833910} {\bibfield  {journal} {\bibinfo
  {journal} {Astron. Astrophys.}\ }\textbf {\bibinfo {volume} {641}},\ \bibinfo
  {pages} {A6} (\bibinfo {year} {2020})},\ \bibinfo {note} {[Erratum:
  Astron.Astrophys. 652, C4 (2021)]},\ \Eprint
  {http://arxiv.org/abs/1807.06209} {arXiv:1807.06209 [astro-ph.CO]}
  \BibitemShut {NoStop}%
\bibitem [{\citenamefont {Beutler}\ \emph {et~al.}(2011)\citenamefont
  {Beutler}, \citenamefont {Blake}, \citenamefont {Colless}, \citenamefont
  {Jones}, \citenamefont {Staveley-Smith}, \citenamefont {Campbell},
  \citenamefont {Parker}, \citenamefont {Saunders},\ and\ \citenamefont
  {Watson}}]{Beutler:2011hx}%
  \BibitemOpen
  \bibfield  {author} {\bibinfo {author} {\bibfnamefont {F.}~\bibnamefont
  {Beutler}}, \bibinfo {author} {\bibfnamefont {C.}~\bibnamefont {Blake}},
  \bibinfo {author} {\bibfnamefont {M.}~\bibnamefont {Colless}}, \bibinfo
  {author} {\bibfnamefont {D.~H.}\ \bibnamefont {Jones}}, \bibinfo {author}
  {\bibfnamefont {L.}~\bibnamefont {Staveley-Smith}}, \bibinfo {author}
  {\bibfnamefont {L.}~\bibnamefont {Campbell}}, \bibinfo {author}
  {\bibfnamefont {Q.}~\bibnamefont {Parker}}, \bibinfo {author} {\bibfnamefont
  {W.}~\bibnamefont {Saunders}}, \ and\ \bibinfo {author} {\bibfnamefont
  {F.}~\bibnamefont {Watson}},\ }\href {\doibase
  10.1111/j.1365-2966.2011.19250.x} {\bibfield  {journal} {\bibinfo  {journal}
  {Mon. Not. Roy. Astron. Soc.}\ }\textbf {\bibinfo {volume} {416}},\ \bibinfo
  {pages} {3017} (\bibinfo {year} {2011})},\ \Eprint
  {http://arxiv.org/abs/1106.3366} {arXiv:1106.3366 [astro-ph.CO]} \BibitemShut
  {NoStop}%
\bibitem [{\citenamefont {Ross}\ \emph {et~al.}(2015)\citenamefont {Ross},
  \citenamefont {Samushia}, \citenamefont {Howlett}, \citenamefont {Percival},
  \citenamefont {Burden},\ and\ \citenamefont {Manera}}]{Ross:2014qpa}%
  \BibitemOpen
  \bibfield  {author} {\bibinfo {author} {\bibfnamefont {A.~J.}\ \bibnamefont
  {Ross}}, \bibinfo {author} {\bibfnamefont {L.}~\bibnamefont {Samushia}},
  \bibinfo {author} {\bibfnamefont {C.}~\bibnamefont {Howlett}}, \bibinfo
  {author} {\bibfnamefont {W.~J.}\ \bibnamefont {Percival}}, \bibinfo {author}
  {\bibfnamefont {A.}~\bibnamefont {Burden}}, \ and\ \bibinfo {author}
  {\bibfnamefont {M.}~\bibnamefont {Manera}},\ }\href {\doibase
  10.1093/mnras/stv154} {\bibfield  {journal} {\bibinfo  {journal} {Mon. Not.
  Roy. Astron. Soc.}\ }\textbf {\bibinfo {volume} {449}},\ \bibinfo {pages}
  {835} (\bibinfo {year} {2015})},\ \Eprint {http://arxiv.org/abs/1409.3242}
  {arXiv:1409.3242 [astro-ph.CO]} \BibitemShut {NoStop}%
\bibitem [{\citenamefont {Alam}\ \emph {et~al.}(2017)\citenamefont {Alam} \emph
  {et~al.}}]{BOSS:2016wmc}%
  \BibitemOpen
  \bibfield  {author} {\bibinfo {author} {\bibfnamefont {S.}~\bibnamefont
  {Alam}} \emph {et~al.} (\bibinfo {collaboration} {BOSS}),\ }\href {\doibase
  10.1093/mnras/stx721} {\bibfield  {journal} {\bibinfo  {journal} {Mon. Not.
  Roy. Astron. Soc.}\ }\textbf {\bibinfo {volume} {470}},\ \bibinfo {pages}
  {2617} (\bibinfo {year} {2017})},\ \Eprint {http://arxiv.org/abs/1607.03155}
  {arXiv:1607.03155 [astro-ph.CO]} \BibitemShut {NoStop}%
\bibitem [{\citenamefont {Scolnic}\ \emph {et~al.}(2018)\citenamefont {Scolnic}
  \emph {et~al.}}]{Pan-STARRS1:2017jku}%
  \BibitemOpen
  \bibfield  {author} {\bibinfo {author} {\bibfnamefont {D.~M.}\ \bibnamefont
  {Scolnic}} \emph {et~al.} (\bibinfo {collaboration} {Pan-STARRS1}),\ }\href
  {\doibase 10.3847/1538-4357/aab9bb} {\bibfield  {journal} {\bibinfo
  {journal} {Astrophys. J.}\ }\textbf {\bibinfo {volume} {859}},\ \bibinfo
  {pages} {101} (\bibinfo {year} {2018})},\ \Eprint
  {http://arxiv.org/abs/1710.00845} {arXiv:1710.00845 [astro-ph.CO]}
  \BibitemShut {NoStop}%
\bibitem [{\citenamefont {Vitale}\ \emph {et~al.}(2019)\citenamefont {Vitale},
  \citenamefont {Farr}, \citenamefont {Ng},\ and\ \citenamefont
  {Rodriguez}}]{Vitale:2018yhm}%
  \BibitemOpen
  \bibfield  {author} {\bibinfo {author} {\bibfnamefont {S.}~\bibnamefont
  {Vitale}}, \bibinfo {author} {\bibfnamefont {W.~M.}\ \bibnamefont {Farr}},
  \bibinfo {author} {\bibfnamefont {K.}~\bibnamefont {Ng}}, \ and\ \bibinfo
  {author} {\bibfnamefont {C.~L.}\ \bibnamefont {Rodriguez}},\ }\href {\doibase
  10.3847/2041-8213/ab50c0} {\bibfield  {journal} {\bibinfo  {journal}
  {Astrophys. J. Lett.}\ }\textbf {\bibinfo {volume} {886}},\ \bibinfo {pages}
  {L1} (\bibinfo {year} {2019})},\ \Eprint {http://arxiv.org/abs/1808.00901}
  {arXiv:1808.00901 [astro-ph.HE]} \BibitemShut {NoStop}%
\bibitem [{\citenamefont {Yang}(2021)}]{Yang:2021qge}%
  \BibitemOpen
  \bibfield  {author} {\bibinfo {author} {\bibfnamefont {T.}~\bibnamefont
  {Yang}},\ }\href {\doibase 10.1088/1475-7516/2021/05/044} {\bibfield
  {journal} {\bibinfo  {journal} {JCAP}\ }\textbf {\bibinfo {volume} {05}},\
  \bibinfo {pages} {044} (\bibinfo {year} {2021})},\ \Eprint
  {http://arxiv.org/abs/2103.01923} {arXiv:2103.01923 [astro-ph.CO]}
  \BibitemShut {NoStop}%
\bibitem [{\citenamefont {Madau}\ and\ \citenamefont
  {Dickinson}(2014)}]{Madau:2014bja}%
  \BibitemOpen
  \bibfield  {author} {\bibinfo {author} {\bibfnamefont {P.}~\bibnamefont
  {Madau}}\ and\ \bibinfo {author} {\bibfnamefont {M.}~\bibnamefont
  {Dickinson}},\ }\href {\doibase 10.1146/annurev-astro-081811-125615}
  {\bibfield  {journal} {\bibinfo  {journal} {Ann. Rev. Astron. Astrophys.}\
  }\textbf {\bibinfo {volume} {52}},\ \bibinfo {pages} {415} (\bibinfo {year}
  {2014})},\ \Eprint {http://arxiv.org/abs/1403.0007} {arXiv:1403.0007
  [astro-ph.CO]} \BibitemShut {NoStop}%
\bibitem [{\citenamefont {Eichhorn}\ \emph {et~al.}(2019)\citenamefont
  {Eichhorn}, \citenamefont {Koslowski},\ and\ \citenamefont
  {Pereira}}]{Eichhorn:2018phj}%
  \BibitemOpen
  \bibfield  {author} {\bibinfo {author} {\bibfnamefont {A.}~\bibnamefont
  {Eichhorn}}, \bibinfo {author} {\bibfnamefont {T.}~\bibnamefont {Koslowski}},
  \ and\ \bibinfo {author} {\bibfnamefont {A.~D.}\ \bibnamefont {Pereira}},\
  }\href {\doibase 10.3390/universe5020053} {\bibfield  {journal} {\bibinfo
  {journal} {Universe}\ }\textbf {\bibinfo {volume} {5}},\ \bibinfo {pages}
  {53} (\bibinfo {year} {2019})},\ \Eprint {http://arxiv.org/abs/1811.12909}
  {arXiv:1811.12909 [gr-qc]} \BibitemShut {NoStop}%
\bibitem [{\citenamefont {Abbott}\ \emph {et~al.}(2023)\citenamefont {Abbott}
  \emph {et~al.}}]{KAGRA:2021duu}%
  \BibitemOpen
  \bibfield  {author} {\bibinfo {author} {\bibfnamefont {R.}~\bibnamefont
  {Abbott}} \emph {et~al.} (\bibinfo {collaboration} {KAGRA, VIRGO, LIGO
  Scientific}),\ }\href {\doibase 10.1103/PhysRevX.13.011048} {\bibfield
  {journal} {\bibinfo  {journal} {Phys. Rev. X}\ }\textbf {\bibinfo {volume}
  {13}},\ \bibinfo {pages} {011048} (\bibinfo {year} {2023})},\ \Eprint
  {http://arxiv.org/abs/2111.03634} {arXiv:2111.03634 [astro-ph.HE]}
  \BibitemShut {NoStop}%
\bibitem [{\citenamefont {\"Ozel}\ and\ \citenamefont
  {Freire}(2016)}]{Ozel:2016oaf}%
  \BibitemOpen
  \bibfield  {author} {\bibinfo {author} {\bibfnamefont {F.}~\bibnamefont
  {\"Ozel}}\ and\ \bibinfo {author} {\bibfnamefont {P.}~\bibnamefont
  {Freire}},\ }\href {\doibase 10.1146/annurev-astro-081915-023322} {\bibfield
  {journal} {\bibinfo  {journal} {Ann. Rev. Astron. Astrophys.}\ }\textbf
  {\bibinfo {volume} {54}},\ \bibinfo {pages} {401} (\bibinfo {year} {2016})},\
  \Eprint {http://arxiv.org/abs/1603.02698} {arXiv:1603.02698 [astro-ph.HE]}
  \BibitemShut {NoStop}%
\bibitem [{\citenamefont {Abbott}\ \emph {et~al.}(2019)\citenamefont {Abbott}
  \emph {et~al.}}]{LIGOScientific:2018mvr}%
  \BibitemOpen
  \bibfield  {author} {\bibinfo {author} {\bibfnamefont {B.~P.}\ \bibnamefont
  {Abbott}} \emph {et~al.} (\bibinfo {collaboration} {LIGO Scientific,
  Virgo}),\ }\href {\doibase 10.1103/PhysRevX.9.031040} {\bibfield  {journal}
  {\bibinfo  {journal} {Phys. Rev. X}\ }\textbf {\bibinfo {volume} {9}},\
  \bibinfo {pages} {031040} (\bibinfo {year} {2019})},\ \Eprint
  {http://arxiv.org/abs/1811.12907} {arXiv:1811.12907 [astro-ph.HE]}
  \BibitemShut {NoStop}%
\bibitem [{\citenamefont {Zhang}\ \emph {et~al.}(2017)\citenamefont {Zhang},
  \citenamefont {Liu},\ and\ \citenamefont {Zhao}}]{Zhang:2017srh}%
  \BibitemOpen
  \bibfield  {author} {\bibinfo {author} {\bibfnamefont {X.}~\bibnamefont
  {Zhang}}, \bibinfo {author} {\bibfnamefont {T.}~\bibnamefont {Liu}}, \ and\
  \bibinfo {author} {\bibfnamefont {W.}~\bibnamefont {Zhao}},\ }\href {\doibase
  10.1103/PhysRevD.95.104027} {\bibfield  {journal} {\bibinfo  {journal} {Phys.
  Rev. D}\ }\textbf {\bibinfo {volume} {95}},\ \bibinfo {pages} {104027}
  (\bibinfo {year} {2017})},\ \Eprint {http://arxiv.org/abs/1702.08752}
  {arXiv:1702.08752 [gr-qc]} \BibitemShut {NoStop}%
\bibitem [{\citenamefont {Wen}\ and\ \citenamefont {Chen}(2010)}]{Wen:2010cr}%
  \BibitemOpen
  \bibfield  {author} {\bibinfo {author} {\bibfnamefont {L.}~\bibnamefont
  {Wen}}\ and\ \bibinfo {author} {\bibfnamefont {Y.}~\bibnamefont {Chen}},\
  }\href {\doibase 10.1103/PhysRevD.81.082001} {\bibfield  {journal} {\bibinfo
  {journal} {Phys. Rev. D}\ }\textbf {\bibinfo {volume} {81}},\ \bibinfo
  {pages} {082001} (\bibinfo {year} {2010})},\ \Eprint
  {http://arxiv.org/abs/1003.2504} {arXiv:1003.2504 [astro-ph.CO]} \BibitemShut
  {NoStop}%
\bibitem [{\citenamefont {Zhao}\ and\ \citenamefont
  {Wen}(2018)}]{Zhao:2017cbb}%
  \BibitemOpen
  \bibfield  {author} {\bibinfo {author} {\bibfnamefont {W.}~\bibnamefont
  {Zhao}}\ and\ \bibinfo {author} {\bibfnamefont {L.}~\bibnamefont {Wen}},\
  }\href {\doibase 10.1103/PhysRevD.97.064031} {\bibfield  {journal} {\bibinfo
  {journal} {Phys. Rev. D}\ }\textbf {\bibinfo {volume} {97}},\ \bibinfo
  {pages} {064031} (\bibinfo {year} {2018})},\ \Eprint
  {http://arxiv.org/abs/1710.05325} {arXiv:1710.05325 [astro-ph.CO]}
  \BibitemShut {NoStop}%
\bibitem [{\citenamefont {Cutler}\ \emph {et~al.}(1993)\citenamefont {Cutler}
  \emph {et~al.}}]{Cutler:1992tc}%
  \BibitemOpen
  \bibfield  {author} {\bibinfo {author} {\bibfnamefont {C.}~\bibnamefont
  {Cutler}} \emph {et~al.},\ }\href {\doibase 10.1103/PhysRevLett.70.2984}
  {\bibfield  {journal} {\bibinfo  {journal} {Phys. Rev. Lett.}\ }\textbf
  {\bibinfo {volume} {70}},\ \bibinfo {pages} {2984} (\bibinfo {year}
  {1993})},\ \Eprint {http://arxiv.org/abs/astro-ph/9208005}
  {arXiv:astro-ph/9208005} \BibitemShut {NoStop}%
\bibitem [{\citenamefont {Sathyaprakash}\ and\ \citenamefont
  {Schutz}(2009)}]{Sathyaprakash:2009xs}%
  \BibitemOpen
  \bibfield  {author} {\bibinfo {author} {\bibfnamefont {B.~S.}\ \bibnamefont
  {Sathyaprakash}}\ and\ \bibinfo {author} {\bibfnamefont {B.~F.}\ \bibnamefont
  {Schutz}},\ }\href {\doibase 10.12942/lrr-2009-2} {\bibfield  {journal}
  {\bibinfo  {journal} {Living Rev. Rel.}\ }\textbf {\bibinfo {volume} {12}},\
  \bibinfo {pages} {2} (\bibinfo {year} {2009})},\ \Eprint
  {http://arxiv.org/abs/0903.0338} {arXiv:0903.0338 [gr-qc]} \BibitemShut
  {NoStop}%
\bibitem [{\citenamefont {Howell}\ \emph {et~al.}(2019)\citenamefont {Howell},
  \citenamefont {Ackley}, \citenamefont {Rowlinson},\ and\ \citenamefont
  {Coward}}]{Howell:2018nhu}%
  \BibitemOpen
  \bibfield  {author} {\bibinfo {author} {\bibfnamefont {E.~J.}\ \bibnamefont
  {Howell}}, \bibinfo {author} {\bibfnamefont {K.}~\bibnamefont {Ackley}},
  \bibinfo {author} {\bibfnamefont {A.}~\bibnamefont {Rowlinson}}, \ and\
  \bibinfo {author} {\bibfnamefont {D.}~\bibnamefont {Coward}},\ }\href
  {\doibase 10.1093/mnras/stz455} {\bibfield  {journal} {\bibinfo  {journal}
  {Mon. Not. Roy. Astron. Soc.}\ }\textbf {\bibinfo {volume} {485}},\ \bibinfo
  {pages} {1435} (\bibinfo {year} {2019})},\ \Eprint
  {http://arxiv.org/abs/1811.09168} {arXiv:1811.09168 [astro-ph.HE]}
  \BibitemShut {NoStop}%
\bibitem [{\citenamefont {Wanderman}\ and\ \citenamefont
  {Piran}(2015)}]{Wanderman:2014eza}%
  \BibitemOpen
  \bibfield  {author} {\bibinfo {author} {\bibfnamefont {D.}~\bibnamefont
  {Wanderman}}\ and\ \bibinfo {author} {\bibfnamefont {T.}~\bibnamefont
  {Piran}},\ }\href {\doibase 10.1093/mnras/stv123} {\bibfield  {journal}
  {\bibinfo  {journal} {Mon. Not. Roy. Astron. Soc.}\ }\textbf {\bibinfo
  {volume} {448}},\ \bibinfo {pages} {3026} (\bibinfo {year} {2015})},\ \Eprint
  {http://arxiv.org/abs/1405.5878} {arXiv:1405.5878 [astro-ph.HE]} \BibitemShut
  {NoStop}%
\bibitem [{\citenamefont {Tan}\ and\ \citenamefont {Yu}(2020)}]{Tan:2020vtc}%
  \BibitemOpen
  \bibfield  {author} {\bibinfo {author} {\bibfnamefont {W.-W.}\ \bibnamefont
  {Tan}}\ and\ \bibinfo {author} {\bibfnamefont {Y.-W.}\ \bibnamefont {Yu}},\
  }\href {\doibase 10.3847/1538-4357/abb404} {\bibfield  {journal} {\bibinfo
  {journal} {Astrophys. J.}\ }\textbf {\bibinfo {volume} {902}},\ \bibinfo
  {pages} {83} (\bibinfo {year} {2020})},\ \Eprint
  {http://arxiv.org/abs/2006.02060} {arXiv:2006.02060 [astro-ph.HE]}
  \BibitemShut {NoStop}%
\bibitem [{\citenamefont {Stratta}\ \emph {et~al.}(2018)\citenamefont
  {Stratta}, \citenamefont {Amati}, \citenamefont {Ciolfi},\ and\ \citenamefont
  {Vinciguerra}}]{Stratta:2018ldl}%
  \BibitemOpen
  \bibfield  {author} {\bibinfo {author} {\bibfnamefont {G.}~\bibnamefont
  {Stratta}}, \bibinfo {author} {\bibfnamefont {L.}~\bibnamefont {Amati}},
  \bibinfo {author} {\bibfnamefont {R.}~\bibnamefont {Ciolfi}}, \ and\ \bibinfo
  {author} {\bibfnamefont {S.}~\bibnamefont {Vinciguerra}},\ }\href@noop {}
  {\bibfield  {journal} {\bibinfo  {journal} {Mem. Soc. Ast. It.}\ }\textbf
  {\bibinfo {volume} {89}},\ \bibinfo {pages} {205} (\bibinfo {year} {2018})},\
  \Eprint {http://arxiv.org/abs/1802.01677} {arXiv:1802.01677 [astro-ph.IM]}
  \BibitemShut {NoStop}%
\bibitem [{\citenamefont {Speri}\ \emph {et~al.}(2021)\citenamefont {Speri},
  \citenamefont {Tamanini}, \citenamefont {Caldwell}, \citenamefont {Gair},\
  and\ \citenamefont {Wang}}]{Speri:2020hwc}%
  \BibitemOpen
  \bibfield  {author} {\bibinfo {author} {\bibfnamefont {L.}~\bibnamefont
  {Speri}}, \bibinfo {author} {\bibfnamefont {N.}~\bibnamefont {Tamanini}},
  \bibinfo {author} {\bibfnamefont {R.~R.}\ \bibnamefont {Caldwell}}, \bibinfo
  {author} {\bibfnamefont {J.~R.}\ \bibnamefont {Gair}}, \ and\ \bibinfo
  {author} {\bibfnamefont {B.}~\bibnamefont {Wang}},\ }\href {\doibase
  10.1103/PhysRevD.103.083526} {\bibfield  {journal} {\bibinfo  {journal}
  {Phys. Rev. D}\ }\textbf {\bibinfo {volume} {103}},\ \bibinfo {pages}
  {083526} (\bibinfo {year} {2021})},\ \Eprint
  {http://arxiv.org/abs/2010.09049} {arXiv:2010.09049 [astro-ph.CO]}
  \BibitemShut {NoStop}%
\bibitem [{\citenamefont {Hirata}\ \emph {et~al.}(2010)\citenamefont {Hirata},
  \citenamefont {Holz},\ and\ \citenamefont {Cutler}}]{Hirata:2010ba}%
  \BibitemOpen
  \bibfield  {author} {\bibinfo {author} {\bibfnamefont {C.~M.}\ \bibnamefont
  {Hirata}}, \bibinfo {author} {\bibfnamefont {D.~E.}\ \bibnamefont {Holz}}, \
  and\ \bibinfo {author} {\bibfnamefont {C.}~\bibnamefont {Cutler}},\ }\href
  {\doibase 10.1103/PhysRevD.81.124046} {\bibfield  {journal} {\bibinfo
  {journal} {Phys. Rev. D}\ }\textbf {\bibinfo {volume} {81}},\ \bibinfo
  {pages} {124046} (\bibinfo {year} {2010})},\ \Eprint
  {http://arxiv.org/abs/1004.3988} {arXiv:1004.3988 [astro-ph.CO]} \BibitemShut
  {NoStop}%
\bibitem [{\citenamefont {Kocsis}\ \emph {et~al.}(2006)\citenamefont {Kocsis},
  \citenamefont {Frei}, \citenamefont {Haiman},\ and\ \citenamefont
  {Menou}}]{Kocsis:2005vv}%
  \BibitemOpen
  \bibfield  {author} {\bibinfo {author} {\bibfnamefont {B.}~\bibnamefont
  {Kocsis}}, \bibinfo {author} {\bibfnamefont {Z.}~\bibnamefont {Frei}},
  \bibinfo {author} {\bibfnamefont {Z.}~\bibnamefont {Haiman}}, \ and\ \bibinfo
  {author} {\bibfnamefont {K.}~\bibnamefont {Menou}},\ }\href {\doibase
  10.1086/498236} {\bibfield  {journal} {\bibinfo  {journal} {Astrophys. J.}\
  }\textbf {\bibinfo {volume} {637}},\ \bibinfo {pages} {27} (\bibinfo {year}
  {2006})},\ \Eprint {http://arxiv.org/abs/astro-ph/0505394}
  {arXiv:astro-ph/0505394} \BibitemShut {NoStop}%
\bibitem [{\citenamefont {Lewis}\ and\ \citenamefont
  {Bridle}(2002)}]{Lewis:2002ah}%
  \BibitemOpen
  \bibfield  {author} {\bibinfo {author} {\bibfnamefont {A.}~\bibnamefont
  {Lewis}}\ and\ \bibinfo {author} {\bibfnamefont {S.}~\bibnamefont {Bridle}},\
  }\href {\doibase 10.1103/PhysRevD.66.103511} {\bibfield  {journal} {\bibinfo
  {journal} {Phys. Rev. D}\ }\textbf {\bibinfo {volume} {66}},\ \bibinfo
  {pages} {103511} (\bibinfo {year} {2002})},\ \Eprint
  {http://arxiv.org/abs/astro-ph/0205436} {arXiv:astro-ph/0205436} \BibitemShut
  {NoStop}%
\bibitem [{ETc()}]{ETcurve-web}%
  \BibitemOpen
  \href@noop {} {}\bibinfo {howpublished}
  {\url{https://www.et-gw.eu/index.php/etsensitivities/}}\BibitemShut {NoStop}%
\bibitem [{CEc()}]{CEcurve-web}%
  \BibitemOpen
  \href@noop {} {}\bibinfo {howpublished}
  {\url{https://cosmicexplorer.org/sensitivity.html}}\BibitemShut {NoStop}%
\bibitem [{\citenamefont {Zhu}\ \emph {et~al.}(2023)\citenamefont {Zhu} \emph
  {et~al.}}]{Zhu:2021ram}%
  \BibitemOpen
  \bibfield  {author} {\bibinfo {author} {\bibfnamefont {J.-P.}\ \bibnamefont
  {Zhu}} \emph {et~al.},\ }\href {\doibase 10.3847/1538-4357/aca527} {\bibfield
   {journal} {\bibinfo  {journal} {Astrophys. J.}\ }\textbf {\bibinfo {volume}
  {942}},\ \bibinfo {pages} {88} (\bibinfo {year} {2023})},\ \Eprint
  {http://arxiv.org/abs/2110.10469} {arXiv:2110.10469 [astro-ph.HE]}
  \BibitemShut {NoStop}%
\end{thebibliography}%

\end{document}